\documentstyle[emulateapj,onecolfloat]{article}

\def\etal{{\rm et~al.\ }}
\def\hmpc{\;h^{-1}{\rm Mpc}}
\def\invhmpc{\;h\;{\rm Mpc}^{-1}}

\def\kms{{\rm \;km\;s^{-1}}}
\def\invkms{({\rm km\;s}^{-1})^{-1}}

\def\lya{Ly$\alpha$}
\def\lyb{Ly$\beta$}

\def\nh1{n_{\rm HI}}
\def\taueff{\overline{\tau}_{\rm eff}}
\def\deltaf{\Delta^{2}_{\rm F}(k)}

\def\overden{(\rho/{\overline\rho})}

\def\p1dk{P_{\rm 1D}(k)}
\def\gprime{{\Gamma^{\prime}}}
\newcommand{\PSbox}[3]{\mbox{\rule{0in}{#3}\includegraphics{#1}\hspace{#2}}}

\begin{document}

\twocolumn[
 
\title{
Towards a Precise Measurement of Matter Clustering:
Lyman-alpha Forest Data at Redshifts 2-4
}
  
\author{
Rupert A.C. Croft$^{1,2,3}$,
David H. Weinberg$^{4}$
Mike Bolte$^{1,5}$,
Scott Burles$^{1,6,7,8}$,
Lars Hernquist$^{1,2}$,
Neal Katz$^{9}$, 
David Kirkman$^{1,6,10}$,
David Tytler$^{1,6}$, 
}  

\begin{abstract}
We measure the filling factor, correlation function, and power spectrum of
transmitted flux in a large sample of \lya\ forest spectra, comprised of 30
Keck HIRES spectra and 23 Keck LRIS spectra.
We infer the linear matter power spectrum $P(k)$ from the
flux power spectrum $P_F(k)$, using an improved version of the method of
Croft et al.\ (1998) that accounts for the influence of redshift-space
distortions, non-linearity, and thermal broadening on the shape of $P_F(k)$.
The evolution of the shape and amplitude of $P(k)$ over the redshift range
of the sample ($z \approx 2-4$) is consistent with the predictions of 
gravitational instability, implying that non-gravitational fluctuations do not
make a large contribution to structure in the \lya\ forest.  Our fiducial
measurement of $P(k)$ comes from a subset of the data with $2.3<z<2.9$, mean
absorption redshift $\langle z \rangle = 2.72$, and total path length
$\Delta z \approx 25$.  It has a dimensionless amplitude
$\Delta^2(k_p)=0.74^{+0.20}_{-0.16}$ at wavenumber 
$k_p=0.03\invkms$ and is well described by a power-law of index 
$\nu = -2.43 \pm 0.06$ or by a CDM-like power spectrum with shape parameter
$\gprime=1.3^{+0.7}_{-0.5}\times 10^{-3}\invkms$ at $z=2.72$ (all error bars
$1\sigma$).  The correspondence to present day $P(k)$ parameters depends on the
adopted cosmology. For $\Omega_m=0.4$, $\Omega_{\Lambda}=0.6$, the best-fit 
shape parameter is $\Gamma=0.16 h\,{\rm Mpc}^{-1}$, 
in good agreement with measurements from the 2dF Galaxy Redshift Survey, 
and the best-fit normalization is $\sigma_8=0.82 (\Gamma/0.15)^{-0.44}$.  
Matching the observed cluster mass function and our measured $\Delta^2(k_p)$ in
spatially flat cosmological models requires 
$\Omega_m=0.38^{+0.10}_{-0.08} + 2.2 (\Gamma-0.15)$.  Matching $\Delta^2(k_p)$
in COBE-normalized, flat CDM models with no tensor fluctuations requires
$\Omega_m = (0.29 \pm 0.04)n^{-2.89} h_{65}^{-1.9}$, and models that 
satisfy this constraint are also consistent with our measured logarithmic
slope.  The \lya\ forest complements other observational probes of the linear
matter power spectrum by constraining a regime of redshift and lengthscale
not accessible by other means, and the consistency of these inferred parameters
with independent estimates provides further support for a cosmological model
based on inflation, cold dark matter, and vacuum energy.
\end{abstract}
 

]

\footnotetext[1] {Visiting Astronomer, W.M. Keck Observatory,
a joint facility of the University
of California, the California Institute of Technology, and the
National Aeronautics and Space Administration.}
\footnotetext[2]{Harvard-Smithsonian Center for Astrophysics, 
Cambridge, MA 02138; lars@cfa.harvard.edu}
\footnotetext[3]{Present address: Department of Physics,
Carnegie Mellon University, Pittsburgh, PA 15213;
rcroft@cmu.edu}
\footnotetext[4]{Department of Astronomy, The Ohio State University,
Columbus, OH 43210; dhw@astronomy.ohio-state.edu}
\footnotetext[5]{UCO/Lick Observatory,
Board of Studies in Astronomy and Astrophysics,
University of California, Santa Cruz, CA 95064; bolte@ucolick.org}
\footnotetext[6]{Center for Astrophysics and Space Sciences,
University of California, San Diego, MS0424, 
La Jolla, CA 92093; tytler@ucsd.edu}
\footnotetext[7]{
University of Chicago,
Astronomy and Astrophysics Center, Chicago, IL 60615}
\footnotetext[8]{
Present address: Department of Physics,
Massachusetts Institute of Technology, Cambridge, MA, 02139-4307
burles@mit.edu}
\footnotetext[9]{Department of Physics and Astronomy, 
University of Massachusetts, Amherst, MA, 01003;
nsk@kaka.phast.umass.edu}
\footnotetext[10]{Present address: Bell Laboratories,
Lucent Technologies, Murray Hill, NJ 07974;
dkirkman@physics.bell-labs.com}

\section{Introduction}
\label{sec:intro}

Over the last few years, the study of the \lya\ forest
has been revolutionized by high-resolution spectra
(mostly using the HIRES spectrograph [Vogt \etal\ 1994] on the
Keck telescope), by measurements of coherent absorption along
lines of sight to quasar pairs (Bechtold \etal 1994;
Dinshaw \etal 1994; Crotts \& Fang 1998), and by 
a new theoretical understanding
made possible by cosmological hydrodynamic simulations
(e.g., Cen \etal 1994; Zhang, Anninos, \& Norman 1995;
Petitjean, M\"ucket, \& Kates 1995;
Hernquist \etal 1996; Wadsley \& Bond 1996; Theuns \etal 1998).
In these simulations the physical state of the diffuse
intergalactic gas responsible for the \lya\ forest is relatively
simple, implying a direct connection between \lya\ absorption and the
underlying density and velocity fields similar to that in some analytic
models of the forest (e.g., McGill 1990; Bi 1993; Bi \& Davidsen 1997;
Hui, Gnedin, \& Zhang 1997).
Several methods have been proposed for using the forest to
measure the amplitude
of mass fluctuations (Gnedin 1998; Gazta\~{n}aga \& Croft 1999;
Nusser \& Haehnelt 1999), 
the matter power spectrum (Croft \etal 1998,
hereafter CWKH; Hui 1999;
McDonald \& Miralda-Escud\'{e} 1999; Feng \& Fang 2000; Hui \etal 2000),
and the geometry of the universe (Hui 1999; 
McDonald \& Miralda-Escud\'{e} 1999).
Observational \lya\ forest data have been used to 
constrain these quantities by Croft \etal (1999b, hereafter CWPHK),
McDonald \etal (2000, hereafter M00), and Nusser \&
Haehnelt (2000). In this paper we present flux
statistics measured from two large samples of Keck \lya\ 
forest spectra, and we use them to measure the linear matter
power spectrum with an improved version of the method presented in
CWKH. The data sample represents a fourfold increase in 
total length of spectra compared to that in CWPHK, and it includes
high resolution spectra that enable us to extend our measurement
to smaller scales.  The higher statistical precision of this
data set requires that we address systematic errors that were 
not significant in our earlier measurement, and much of this paper is 
devoted to assessing and correcting for these systematic effects.

Our analysis breaks into two parts: a measurement of the power
spectrum of transmitted flux in the \lya\ forest, and an inference
of the matter power spectrum from the flux power spectrum.
The theoretical model that motivates our method for the second
step is the ``Fluctuating Gunn-Peterson Approximation'' 
(FGPA; see Rauch \etal 1997; CWKH; Weinberg \etal 1998b),
which describes the relation between \lya\ opacity and matter
density for the diffuse intergalactic gas that produces most 
of the \lya\ forest absorption at high redshift.
Photoionization heating by the ultraviolet (UV) background radiation
and adiabatic cooling by the expansion of the universe 
combine to drive most of the gas with $\rho_b<10$ onto a power-law 
temperature-density relation, 
\begin{equation}
T = T_{0} \rho_{b}^\alpha,
\label{eqn:td}
\end{equation}
where $\rho_{b}$ is the baryon overdensity in units of the cosmic mean
(Katz, Weinberg \& Hernquist 1996; Hui \& Gnedin 1997). 
The parameters $T_{0}$ and $\alpha$ depend on the reionization
history and on the spectral shape of the UV background; they can
be predicted theoretically with the formalism of Hui \& Gnedin (1997)
and constrained observationally with techniques described by
Schaye et al. (1999, 2000), Bryan \& Machacek (2000), 
Ricotti et al. (2000), and McDonald et al. (2000b).
The optical depth for \lya\ absorption is proportional to
the neutral hydrogen density (Gunn \& Peterson 1965),
which  for this gas in photoionization
equilibrium is proportional to the density times the recombination rate. 
This leads to a power-law relation between the 
fluctuating gas density and the \lya\ optical depth,
\begin{eqnarray}
\tau &\propto & \rho_b^2 T^{-0.7} ~=~ A\rho_{b}^\beta, 
\label{eqn:tau} \\
A & = & 0.433
\left(\frac{1+z}{3.5}\right)^6 
\left(\frac{\Omega_b h^2}{0.02}\right)^2
\left(\frac{T_0}{6000\;{\rm K}}\right)^{-0.7} \;\times \nonumber \\
& & \left(\frac{h}{0.65}\right)^{-1}
\left(\frac{H(z)/H_0}{3.68}\right)^{-1} 
\left(\frac{\Gamma_{\rm HI}}
{1.5\times 10^{-12}\;{\rm s}^{-1}}\right)^{-1}\; ,\nonumber
\end{eqnarray}
with $\beta \equiv 2 - 0.7\alpha$ in the range $1.6-1.8$.
Here $\Gamma_{\rm HI}$ is the HI photoionization rate
and $\rho_b$ is in units of the mean cosmic baryon density.
This relation holds to a good approximation on a pixel-by-pixel
basis in spectra extracted from hydrodynamic simulations, even
when the effects of peculiar velocities and thermal broadening
are included (Croft et al.\ 1997b; Weinberg et al.\ 1999b).

The FGPA implies a tight correlation between the observable
quantity $F=e^{-\tau}$ and the underlying gas density $\rho_b$,
which in turn follows the dark matter density because pressure
gradients are weak in the diffuse, cool gas.  CWKH show that
the matter power spectrum $P(k)$ is proportional to the flux
power spectrum $P_F(k)$ on large scales.  The constant of proportionality
depends on the parameter $A$, but this can be fixed empirically by
matching a single observational constraint, such as the mean opacity
of the forest at the redshift under consideration.
In practice, we do not use equation~(\ref{eqn:tau}) itself but
a closely related numerical approximation.  To predict properties
of the \lya\ forest for a given matter power spectrum, we evolve
a collisionless N-body simulation, assign optical depths in real space
via equation~(\ref{eqn:tau}), then extract absorption spectra including
the effects of peculiar velocities and thermal broadening.
Throughout this paper, we will use the term FGPA to refer to this
full numerical procedure, not just to equation~(\ref{eqn:tau}).

Relative to CWPHK, the most important change in our methodology here
is to replace the assumption that $P(k)=b^2 P_F(k)$ with the
more general assumption that $P(k)=b^2(k) P_F(k)$, where $b(k)$ is
a function calculated from simulations constrained to match the
observed flux power spectrum.  This change allows a more accurate
treatment of the effects of non-linear evolution, peculiar velocities,
and thermal broadening.  We have titled our paper ``Towards a Precise
Measurement of Matter Clustering ...'' because, while the measurement
of $P_F(k)$ is quite precise, the determination of $b(k)$ still 
suffers from some systematic uncertainties.  The most important of these
is the uncertainty in the mean opacity of the forest, measurement
of which requires careful attention to continuum fitting.
Other sources of uncertainty are the values of $T_0$ and $\alpha$
and the numerical limitations of our simulations.  We have tried to
provide the information needed to convert our flux power spectrum
into a more accurate and more precise matter power spectrum as
observational parameter determinations improve.

In \S 2 we present the two observational data samples,
describing the sample selection, the observations, and
the data reduction. In \S 3 we present statistics of the transmitted flux
measured from the data: the flux filling factor, the flux two-point correlation
function, and the flux power spectrum. In \S 4 we describe the 
method used to recover the matter power spectrum from the flux
power spectrum, and in \S5 we enumerate and quantify systematic uncertainties
in this procedure. We present our main results in \S6, giving a table of
matter power spectrum values as well as power law fits to the data at 
several different redshifts. In \S7 we discuss some implications of our
results for cosmological parameter values and structure formation scenarios.
We provide a fairly comprehensive summary of our results in \S8, and the
reader who does not wish to dive immediately into a long paper may prefer
to start with this summary.

\section{Data}

We use data from two sets of quasar spectra. One is a moderate resolution
and signal-to-noise \lya\ forest survey, carried out with
the Low Resolution Imaging
Spectrometer (LRIS; Oke \etal 1995) on the Keck II telescope and 
specifically designed for the task of measuring mass fluctuations
and constraining cosmology. The exposure times for these observations are 
quite short,  the aim of the survey being to provide as many quasar spectra as 
possible to cut down cosmic variance.
The other data set consists of high
resolution quasar spectra taken with the Keck HIRES spectrometer.
The HIRES data enable us to probe smaller scales, 
they allow more accurate continuum fitting because they have
more data points close to the unabsorbed continuum,
and they avoid potential biases 
associated with highly smoothed data (see \S 3.3). 
We now describe both data sets in more detail.

\begin{figure}[t]
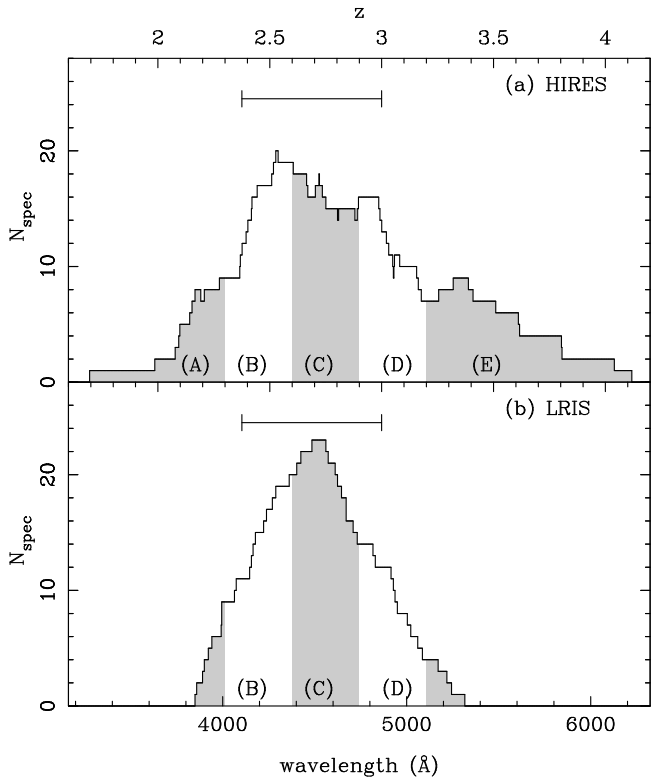

\centering
\PSbox{hist.ps angle=-90 voffset=330 hoffset=-100 vscale=58 hscale=58}
{3.5in}{4.2in} 
\caption[hist]{
Histograms of the data. The horizontal bar above the histograms shows
the length of a \lya\ forest spectrum with a midpoint at $z=2.7$.
Shading marks the boundaries of the different redshift subsamples
(see Table~\ref{subtab}).
\label{hist}
}
\end{figure}

\begin{table}[ht]
\centering
\caption[subtab]{\label{subtab} 
The data subsamples.}
\begin{tabular}{ccccc}
\hline &&\\
Subsample &  HIRES  & HIRES  & LRIS &
LRIS  \\
& $\langle z \rangle $ & length  & $\langle z \rangle$ &
length   \\
&   & ($\kms$) &  &
 ($\kms$)  \\
\hline &&\\
A &     2.13 &   180000 &       &         \\
B &     2.47 &   389000 &  2.47 & 367000  \\
C &     2.74 &   375000 &  2.74 & 469000  \\ 
D &     3.03 &   255000 &  3.02 & 199000  \\ 
E &     3.51 &   278000 &       &         \\
Fiducial &   2.72  &  1016000 & 2.72 & 1049000  \\
\hline &&\\
\end{tabular}
\end{table}

\subsection{HIRES sample}

The HIRES data were obtained as part of the effort to 
measure the mean cosmic baryon density $\Omega_{b}h^2$
by comparing the ratio of deuterium to hydrogen with the predictions
of primordial nucleosynthesis (e.g., Tytler, Fan \& Burles 1996;
Burles \& Tytler 1997, 1998, Burles, Kirkman \& Tytler 1999).
 The reader is referred
to these papers, as well as 
Kirkman \& Tytler (1997), for
details of observational techniques
and data reduction procedures.

The sample used here consists of 30 Keck HIRES spectra,
with QSO emission redshifts ranging from
2.19 to 4.11. The resolution is $8 \kms$ FWHM, and the spectra  have been
rebinned onto $2.1 \kms$ pixels. 
A Legendre polynomial continuum was fit to the echelle orders,
using the IRAF task CONTINUUM, before they were combined to form a
continuous spectrum. 
Averaging over all spectra, the mean $1\sigma$ uncertainty in the
flux values of pixels relative to the continuum in the \lya\ forest
region is $4\%$. 

The region that we use from each spectrum 
spans the wavelength range from \lyb\ to 5000 $\kms$ blueward of  
\lya\ (to avoid the effect of any ionizing radiation arising from
close proximity
to the quasar). We show the redshift distribution of these \lya\ forest data 
in Figure~\ref{hist}a, where we can see that redshifts from 1.6 to 4.1
are represented and that the peak in the histogram lies between $z=2.5$
and $z=3$. We have divided the dataset into different redshift
subsamples, which are marked on Figure ~\ref{hist}a and summarized in 
Table ~\ref{subtab}. The boundaries between the subsamples are
at $z=2.3, 2.6, 2.9$, and $3.2$. The bulk of our analysis will be performed on
a fiducial sample, which is the sum of data in subsamples B,C and D.
This fiducial sample comprises the bulk of the data, and it has a relatively 
small redshift range in order to minimize the effects of
evolution internal to the sample. 
It is centered roughly on the peak of the $z$ distribution, 
and the fiducial HIRES sample is comparable in size and mean 
redshift to the full LRIS sample discussed below.

\begin{figure*}[t]
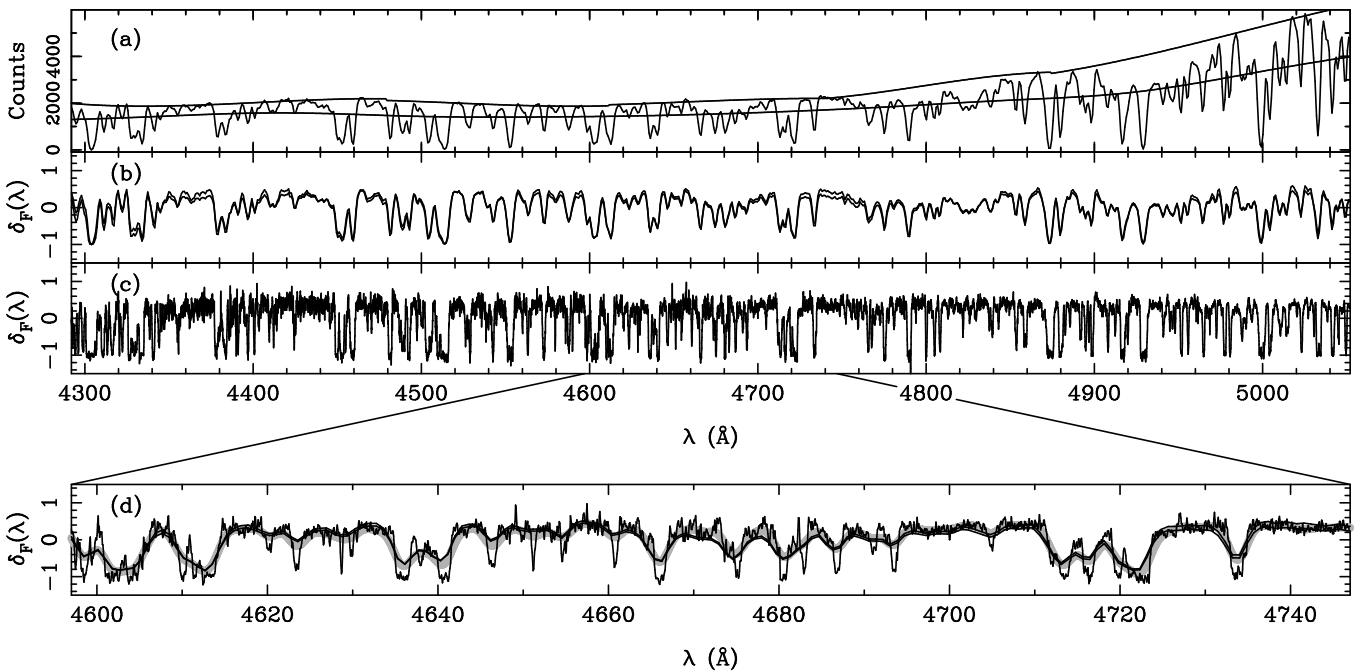

\centering
\PSbox{spec.ps angle=-90 voffset=280 hoffset=-215 vscale=85 hscale=85}
{3.5in}{4.0in}
\caption[spec]{
Determination of $\delta_F(\lambda)$, for the quasar Q1017+1055. 
({\it a}) LRIS spectrum (wiggly line), the continuum fitted over
100\AA\ regions (upper smooth curve), and the spectrum 
smoothed with a 50\AA\ Gaussian (lower smooth curve).
({\it b}) Fluctuations $\delta_F(\lambda)$ derived using the 
continuum fitted spectrum and the smoothed spectrum. The 
continuum fitted curve is slightly higher where the two are distinguishable.
({\it c}) Fluctuations $\delta_F(\lambda)$ from
the HIRES spectrum of Q1077+1055.
({\it d}) A zoom of the central 150\AA\ showing
the two variations of the LRIS spectrum,
the HIRES spectrum, and (grey curve) the HIRES spectrum smoothed
to the resolution of the LRIS data.
\label{spec}}
\end{figure*}

\subsection{LRIS sample}

Large aperture telescopes are commonly used to probe to faint
magnitudes with long exposures. Another possible application of
their light gathering 
power is to build up a large sample of shallower exposures. This sort of 
approach is useful for the statistical study of large-scale structure,
where it is important to minimize ``cosmic variance.''
In choosing the present 
LRIS data sample, our intention was to use the Keck telescope
to carry out a quick survey comprising a relatively large number
of quasar spectra. Because a high signal-to-noise ratio is not necessary
for flux clustering measurements [at least for scales $\la 0.1 \invkms$;
see CWKH, and \S3.3], it is possible to obtain useful spectra of 
bright $z \sim3$ quasars with a few minutes integration time.

Because of the limited blue response of the instrument detectors, 
we were restricted to  quasar targets with $z > 2.7$. These were drawn from 
the quasar catalogue of Veron-Cetty \& Veron (1998), and chosen to be as bright
as possible. Of our targets, we were able to obtain spectra for 23,
objects with V magnitudes ranging from  15.8 to 18.7, with the majority
between 17.0 and 18.0. The highest redshift quasar in our sample is at $z=3.37$
and the lowest at $z=2.75$. The histogram of regions
in the \lya\ forest (running from \lyb\ emission to 
20 \AA\ blueward of \lya\ emission) is shown in Figure ~\ref{hist}b. 
Our fiducial LRIS \lya\ forest sample comprises
all the LRIS data from $z=2.3$ to $z=3.2$. 
Our fiducial flux statistics will be measured
by combining results from this sample and the HIRES data sample
that covers the same redshift range. 
We have also subdivided the
data  to make subsamples with the same redshift boundaries
as those of HIRES subsamples B,C, and D. Details are given in  
Table ~\ref{subtab}. Four of the quasars are also in the HIRES dataset
and were used for comparison purposes (see below and \S3.3).

The grating used for the LRIS
observations  was ruled with 900 lines/mm, with the blaze at 5500 \AA.
The  FWHM resolution of the data was 
2 \AA, sampled with 0.85 \AA\  pixels.
The data were taken on Dec. 10, 1998, with some
spectra also taken during director's observing time in Jan. 1999 and Feb. 1999.
The integration times were between 10 and 30 minutes per object.
Data reduction was carried out using standard IRAF packages for longslit 
spectroscopy. The signal-to-noise ratio of the resulting spectra
varies between 10 and 50, with the majority having S/N of $\sim40$
per pixel.
Figure~\ref{spec} (discussed below) displays a typical example  
from our LRIS spectra.

In our analysis of flux statistics, we will be interested in the 
mean flux $\langle F \rangle$ and the fluctuations about the mean
$\delta_{F}(\lambda) \equiv F(\lambda)/\langle F \rangle-1$.
We use $F(\lambda)$ to denote
the transmitted flux, i.e., the ratio of the flux at a given 
wavelength $\lambda$ to the unabsorbed quasar continuum flux
at $\lambda$.
In order to find 
$F(\lambda)$, and hence $\langle F \rangle$, it is necessary to make an
estimate of the unabsorbed continuum level. The quantity $\delta_{F}(\lambda)$
is  much less sensitive to the exact assumed continuum level, as 
$\langle F \rangle$ has already been divided out. In the present paper,
we will not attempt to make accurate determinations of $\langle F \rangle$
from our data.  Instead, we will use $\langle F \rangle$ results 
from the literature and show
how our results (for example for the amplitude of the 
matter power spectrum) would change for given future determinations of 
$\langle F \rangle$.

In order to calculate $\delta_{F}(\lambda)$, we have two choices. The first
is to estimate a continuum level by fitting a line that passes
through apparently unabsorbed regions of the spectrum. This
has already been  done in a semi-automated way
for the HIRES data as described in \S2.1 (see also Burles \& Tytler 1998).
For the LRIS data, which has much lower spectral resolution,
our $\delta_{F}(\lambda)$ results are more likely to be sensitive to
the continuum fitting technique used.
We therefore compare two techniques applied to the LRIS data. The first is the 
automated technique described in CWPHK.
This involves fitting a third-order
polynomial through the datapoints in a given length of 
spectrum, rejecting points that lie 2$\sigma$ below the fit line, and 
iterating until convergence is reached. 
We implement this procedure using 100 \AA\ fitting segments.

The second method for estimating $\delta_{F}(\lambda)$ is
to calculate the mean flux level of the spectrum directly, rather than 
first fitting the continuum to scale unabsorbed flux to $F=1$.
The mean level must be estimated from a region much larger 
than the length scales for which we   
are interested in measuring variations in $\delta_{F}(\lambda)$. This
can be done either by fitting a low order polynomial to the spectrum itself
(Hui et al.\ 2000) or by smoothing the spectrum with a
large radius filter. We do the latter, using a 50 \AA\ Gaussian filter.  
The value of $\delta_{F}(\lambda)$
is then given by $C(\lambda)/C_{S}(\lambda)-1$, where 
$C(\lambda)$ is the number of counts in the spectrum at a
wavelength $\lambda$ and $C_{S}(\lambda)$ is
the smoothed number of counts.

Figure~\ref{spec} illustrates these two methods of determining 
$\delta_F(\lambda)$.  Figure~\ref{spec}a shows the LRIS spectrum
of the $z=3.16$ quasar Q107+1055, along with the fitted continuum
(upper smooth curve) and the 50\AA\ smoothed spectrum (lower smooth curve).
Figure~\ref{spec}b compares $\delta_F(\lambda)$ estimated using the
fitted continuum and using the smoothed spectrum.
The two methods yield nearly indistinguishable results, with small
differences appearing in regions where the spectrum is apparently
close to the unabsorbed continuum.
Figure~\ref{spec}c shows $\delta_F(\lambda)$ from the
(continuum-fitted) HIRES spectrum of Q107+1055.
Figure~\ref{spec}d blows up the central 150\AA\ of the spectrum,
superposing the two LRIS $\delta_F(\lambda)$, the HIRES $\delta_F(\lambda)$,
and the HIRES $\delta_F(\lambda)$ smoothed to the spatial resolution
of the LRIS data.  The smoothed HIRES spectrum matches the LRIS
spectrum almost perfectly, providing further evidence of the
robustness of the $\delta_F(\lambda)$ determination.
In \S 3.3 we will compare the HIRES and LRIS flux power spectra
for the four quasars common to both samples.
We will also show that the two methods of determining $\delta_F(\lambda)$
from the LRIS spectra yield similar power spectrum results.
We will adopt the smoothed spectrum method as our standard, since
it does not involve splitting a spectrum into discrete segments
and is simpler to implement in a robust manner.

As in CWPHK, we scale the individual
pixel widths in the spectra to 
the size they would have 
at the mean redshift of the sample in question. In the present
work, we do this assuming that 
the evolution of $H(z)$ follows that in an EdS universe,
which should be a good approximation at these high redshifts
(see also M00).
We also follow Rauch \etal (1997) and CWPHK  in scaling the optical 
of pixels depths by a factor of $(1+z)^{4.5}$ to the mean $z$ of the sample,
in order to mitigate the effects of evolution.
We have done this only for the HIRES data, since for the LRIS data
we are already calculating $\delta_{F}(\lambda)$ with respect to the
local mean. The effects of both of these rescalings on the flux power 
spectrum are investigated in \S3.3.

\section{Statistics of the transmitted flux}
\label{sec:statistics}

In the fluctuating intergalactic medium (IGM)
view of the \lya\ forest described in \S 1,
the most natural statistical descriptors of the forest are those
that treat each spectrum of transmitted flux as a continuous
one-dimensional field, rather than a collection of lines.
Many such statistics have been discussed in the literature,
including the one-point flux probability distribution, the
threshold-crossing frequency, and the filling factor of 
saturated regions (e.g., Miralda-Escud\'e et al.\ 1996, 1997;
Cen 1997; Croft et al.\ 1997a; Rauch et al.\ 1997; 
Croft \& Gazta\~naga 1999; Theuns, Schaye, \& Haehnelt 1999;
Weinberg et al.\ 1999b; M00).
These measures are analogous to, and in some cases borrowed from,
those used to study large-scale structure in the galaxy distribution.
They encode information about the underlying matter distribution
and about the temperature and physical state of the IGM.

In this section we focus on the statistics that are most
relevant to determination of the matter power spectrum, namely the
flux power spectrum and its Fourier transform, the flux correlation
function.  As mentioned in \S 1, the inference of the matter
power spectrum depends on the value of the mean transmitted flux 
$\langle F\rangle$, or, equivalently, the effective mean optical depth
\begin{equation}
\taueff \equiv -\ln \langle F\rangle .
\label{eqn:taueff}
\end{equation}
We will not undertake a direct measurement of $\taueff$ here, but
we will examine a statistic, the flux filling factor, that 
provides an indirect handle on $\taueff$.  Although determination
of the matter power spectrum is our long-term goal, the measurements
of flux statistics in this section can also stand as direct tests of
theoretical predictions for the \lya\ forest, derived from numerical 
simulations or analytic approximations.  The flux correlation function
and flux power spectrum have been measured for an independent
sample of Keck HIRES spectra by M00.

\begin{figure}[t]
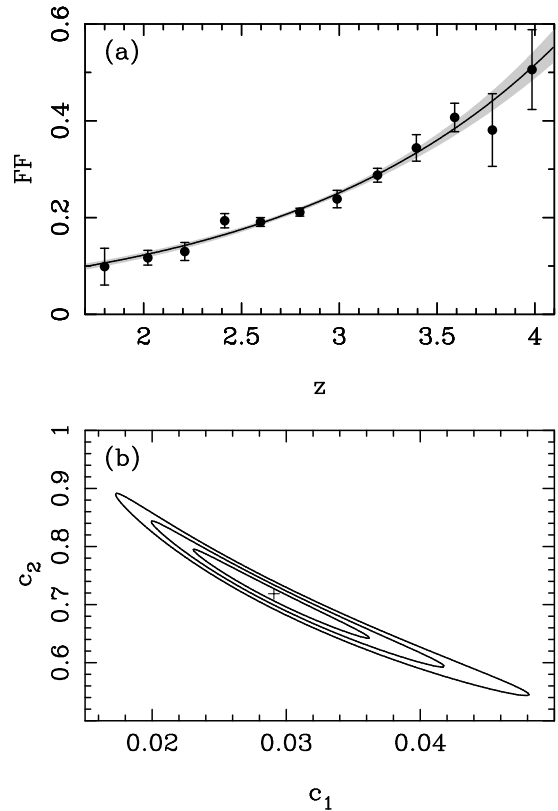

\centering
\PSbox{fillfac.ps angle=-90 voffset=410 hoffset=-185 vscale=78 hscale=78}
{3.5in}{4.3in} 
\caption[fillfac]{
{\it (a)} The filling factor FF of regions below a flux threshold
of 0.5, computed as a function of redshift from the HIRES data.
Error bars ($1\sigma$) are calculated using a jackknife estimator.
{\it (b)} Confidence intervals (68\%, 95\%, 99.7\%) on parameters
$c_1$ and $c_2$, determined by fitting the functional form
FF$=c_{1}\exp{(c_{2}z)}$ to the data points in {\it (a)}.
Best fit values (cross) are $c_{1}=0.0291$, $c_{2}=0.719$. 
In {\it (a)}, the shaded bands show the region corresponding 
to the 68\% confidence interval on $(c_1,c_2)$.
\label{fillfac}
}
\end{figure}

\subsection{Filling factor}
\label{sec:ff}

In terms of the discussion in \S 1, the role of $\taueff$ in our
analysis is to allow us to determine the constant $A$ in 
equation~(\ref{eqn:tau}), given a model of the density fluctuations
$\overden$ obtained from N-body simulations.  However, observational
determinations of $\taueff$ are sensitive to the details of continuum
fitting because a significant fraction of the mean opacity arises
in long stretches of weak absorption close to the unabsorbed
continuum.  The filling factor FF, the fraction of a spectrum
that lies below a specified flux threshold, offers another diagnostic
for $A$ that is less sensitive to the continuum level, and we
will use it as a consistency check on our adopted value of $\taueff$.
While the filling factor of saturated regions ($F \approx 0$) would
be the least sensitive to continuum determination, it is also not
sufficiently sensitive to $A$ to be useful for our purposes.
Instead, we measure the filling factor of regions with $F \leq 0.5$, 
which is likely to be almost as insensitive to continuum
uncertainty (see, e.g., figure~\ref{spec}b, and figure 5 of Nusser \& Haehnelt
2000).

For this statistic, we use the HIRES data only, as we are interested
in unsmoothed spectra. We split the data into 12 redshift bins of width
$\Delta z=0.2$, spanning $z=1.7$ to $z=4.1$. We calculate the FF for $F=0.5$
for each redshift range and plot the results in Figure~\ref{fillfac}a,
as a function of redshift. The error bars are computed using 
a jackknife estimator (Bradley 1982), which we will employ for
error estimation on other statistics as well.
For a statistic $X$ estimated from a data sample, the jackknife
estimate of the $1\sigma$ uncertainty on $X$ is obtained by dividing
the sample into $N$ subsamples and computing
$\sigma = \left[\sum_{i=1}^N\left(X_i-\widehat{X}\right)^2\right]^{1/2}$,
where $\widehat{X}$ is the estimate from the full data sample and
$X_i$ is the value estimated by leaving out subsample $i$.
For Figure~\ref{fillfac}a, we split the data for each redshift bin into
$N=5$ subsamples to estimate the error bars.

A rapid increase in FF with redshift can be seen from the plot. By 
$z\simeq4.0$, half of each spectrum lies below the $F=0.5$ level, as opposed
to $\simeq 10 \%$ at $z=1.8$. In order to present the results in a form which 
is easy to use, we carry out a $\chi^{2}$ fit to the function
FF$=c_{1}\exp{(c_{2}z)}$. The best fit line is shown on 
Figure~\ref{fillfac}a. We find that $c_{1}=0.0291\pm0.0006$ and
$c_{2}=0.719\pm0.008$ (these are $1 \sigma$ errors for the parameters
taken individually). The $\chi^{2}$ for these parameters
is 6.9 (for 10 degrees of freedom). The errors on $c_{1}$ and $c_{2}$
are highly correlated, as can be seen in panel (b) of 
Figure ~\ref{fillfac}.
The shaded band of panel (a) shows the values of FF$(z)$ that result
from varying the parameters $c_1$ and $c_2$ so that they sample the entire 
joint $68\%$ confidence interval on their values. 
Using this fit information  gives 
fractional errors on FF of $4.9\%$ at $z=2$, $2.0\%$ at $z=3$ and
$5.8\%$ at $z=4$. At the redshift of the fiducial sample ($z=2.72$), 
the error is $2.1\%$ and the FF is 0.205. Of course these error values 
are derived from the fit, and their use implies the assumption that the 
FF is changing smoothly with $z$ in accordance with the
shape given by the fit. This assumption has been used by others for 
studying the
evolution of the mean flux level with $z$ (e.g., Press, Rybicki \& Schneider
1993, hereafter PRS). 

\subsection{Flux correlation function}

The flux correlation function, $\xi_{F}(r)$, is a simple 
statistic to calculate. Its
usefulness has been emphasized by Zuo \& Bond (1994)
and Cen \etal (1998), amongst 
others, and $\xi_{F}(r)$ has been measured from a sample of
eight Keck HIRES spectra 
by M00. 
We will also use it for a consistency check on our 3-d $P(k)$ inversion,
in \S 3.3.3 below.
We estimate $\xi_{F}(r)$ from our  quasar
data using the estimator 
$\xi_{F}(r)=\langle\delta_{F}(x) \delta_{F}(x+r) \rangle $.
We present results for the fiducial sample in Table~\ref{xitabfid}.
The errors were again calculated using a jackknife estimator, and 
although we only give the diagonal terms in the covariance matrix,
the full matrix (which has large off-diagonal terms) is available from
the authors on request. In Table~\ref{xitabfid}, we have 
averaged the results
from the LRIS and HIRES samples on large scales
($r > 250\kms$), where the finite LRIS resolution is not important.
On smaller scales,
only the  HIRES results are used.
Table~\ref{xitabz} (see Appendix) gives  $\xi_{F}(r)$ for the different
redshift subsamples.

\begin{table}[ht]
\centering
\caption[xitabfid]{\label{xitabfid} 
The flux correlation function, $\xi_{F}(r)$,
for the fiducial sample ($\langle z \rangle =2.72$).}
\begin{tabular}{cc}
\hline &\\
r  &  $\xi_{F}(r)$ \\
    ($\kms$) & \\
\hline &\\
 11.4 & $0.184 \pm 0.007$  \\
 14.9 & $0.174 \pm 0.007$ \\
 19.4 &  $0.172 \pm 0.008$  \\
 25.3 & $ 0.157 \pm 0.007$ \\
 32.9 & $0.144 \pm 0.007$  \\
 42.9 & $0.127 \pm 0.007$  \\
 56.0 & $0.104 \pm 0.006$  \\
 72.9 & $(8.5 \pm 0.6) \times 10^{-2}$ \\
 95.0 &$(6.6 \pm 0.5) \times 10^{-2}$ \\
 124  &$(5.0 \pm 0.5) \times 10^{-2}$ \\
 161  &$(3.7 \pm 0.4) \times 10^{-2}$ \\
 210  &$(2.7 \pm 0.4) \times 10^{-2}$ \\
 274  &$(1.9 \pm 0.3) \times 10^{-2}$ \\
 357  &$(1.4 \pm 0.3) \times 10^{-2}$ \\
 466  &$(9.4 \pm 2.1) \times 10^{-3}$ \\
 607  &$(6.2 \pm 2.2) \times 10^{-3}$ \\ 
 791  &$(4.1 \pm 2.1) \times 10^{-3}$  \\
 1030 &$(2.6 \pm 2.0) \times 10^{-3}$ \\
 1340 &$(-7.9 \pm 17.7) \times 10^{-4}$ \\
 1750 &$(-1.8 \pm 1.6) \times 10^{-3}$ \\
\hline &\\
\end{tabular}
\end{table}

\begin{figure*}[t]
\centering
\vspace{14cm}
\includegraphics{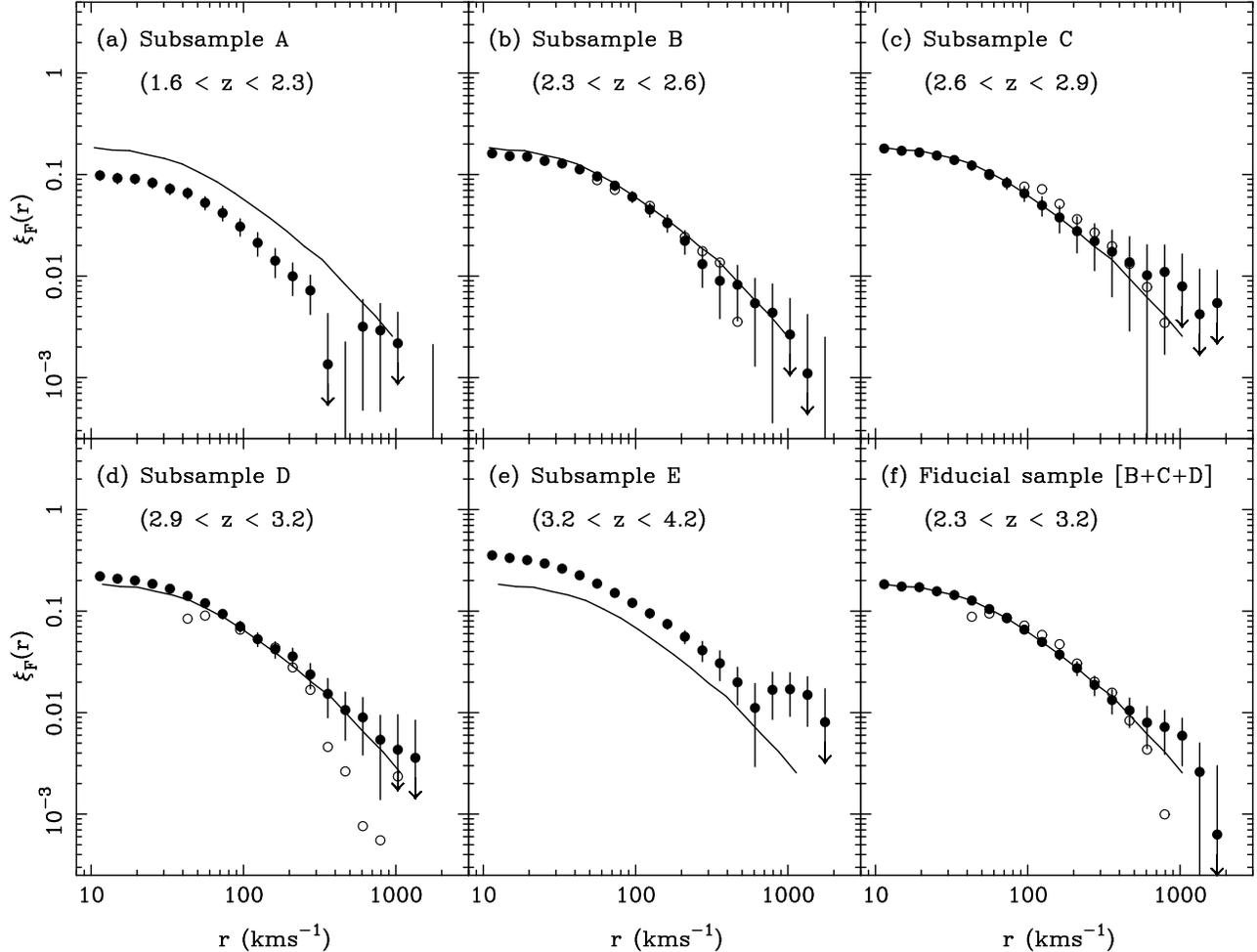}
\caption[fluxxi]{
The flux correlation function of the various data subsamples.
The solid line in each case is the fiducial combined sample, with the
length scaled so that the comoving lengths stay the same in an
EdS model. The multiplicative factor used is
$[(1+z_{i})/(1+z_{F})]^{1/2}$, where $z_{i}$, $z_{F}$ are the mean
redshifts of subsample $i$ and the fiducial sample respectively.
Filled circles represent the HIRES data and open circles
the LRIS data. Error bars have been omitted from the latter, for
clarity.
\label{fluxxi}}
\end{figure*}

We plot $\xi_{F}(r)$ for the different redshift subsamples 
in Figure~\ref{fluxxi}, with results for the fiducial sample
shown as the solid curve in each panel.
There is a measurable clustering signal out to $r \sim 2000 \kms$ (note that we
are plotting both axes with a log scale). In the same way that the flux
power spectrum in simulations appears to have much the same
shape as the (linear) matter power spectrum (e.g., CWKH), we
expect the shape of  $\xi_{F}(r)$  to reflect that of the underlying
matter correlation function, at least over
some range of scales.
A comparison of $\xi(r)$ for flux and mass has
been carried out by Cen \etal (1998). 

On the largest scales, and particularly at
high redshift, $\xi_{F}(r)$ could in principle be influenced
by UV background fluctuations or by continuum fitting errors.
However, Figure~\ref{fluxxi} implies that any such effects are not strong,
since the shape of $\xi_{F}(r)$
in the different panels does not appear to change significantly
from one redshift to the next. This consistency is expected if
$\xi_{F}(r)$ is mainly determined
by the underlying matter distribution, but it seems coincidental
 if the fluctuations that are quantified by $\xi_{F}(r)$
are generated by some other mechanism. However, the possibility that 
UV background fluctuations could reproduce this behaviour merits further study,
since, for example, clustered sources and absorbing material could conceivably
yield a $\xi_{F}(r)$ related to that of the matter distribution.
For work that shows that this is unlikely to occur on the scales
of interest to us here, see, e.g., Zuo (1992) and CWPHK.

Predicting the evolution of $\xi_{F}(r)$ 
in a given cosmological model involves a combination of
change in length units,
growth of matter clustering, and evolution of the mean opacity.
We leave such predictions to future work, which should also investigate
the consistency of higher-order statistics of the flux with the
FGPA predictions.
For the time being, we note that
the amplitude of $\xi_{F}(r)$ decreases as we move to 
lower redshift, as the rapidly decreasing value of $\taueff$ 
counteracts the effect of gravitational clustering. This means that a good 
knowledge of $\taueff$ is needed to make measurements of the amplitude
of matter fluctuations (see \S\ref{sec:taueff}).
On small scales  ($r \la 200 \kms$), the shape of 
$\xi_{F}(r)$ reflects the broadening of individual absorption
features by Hubble flow, peculiar velocities, and thermal motions
(Zuo \& Bond 1994; Hernquist et al.\ 1996).

\subsection{Flux power spectrum}

\subsubsection{Definitions}

To compute the one-dimensional flux power spectrum, $P_{F,1D}(k)$,
we must decompose the absorption spectra into Fourier modes and measure
their variance as a function of wavenumber.
In CWKH and CWPHK, we accomplished this task using
a Fast Fourier Transform (FFT).
This approach requires mapping the spectra
onto equally spaced bins using spline interpolation.
M00 calculated $P_{F,1D}(k)$ by an alternative technique,
the Lomb periodogram, which does not require the assumption
of periodic boundary conditions and which works for unequally spaced bins
(see Press \etal 1992).
In the present paper we also adopt this approach to derive our
fiducial results (using the Lomb code 
from Press \etal 1992). We compare results obtained using this
method and the FFT below.

The power spectrum along a line of sight is an integral over the power
spectrum of the corresponding 3-dimensional field
(Kaiser \& Peacock 1991).  Since we are ultimately interested in the
3-dimensional matter power spectrum, we want to work with the corresponding
property of the flux.  We will {\it define} the 3D flux power spectrum
$P_F(k)$ by the relation
\begin{equation}
P_{F}(k)=-\frac{2\pi}{k} \frac{d}{dk}P_{F,1D}(k),
\label{eqn:invert}
\end{equation}
so that $P_F(k)$ is the power spectrum of the 3D ``flux field'' that would
have a line-of-sight power spectrum $P_{F,1D}(k)$ if it were isotropic.
In practice, peculiar velocities and thermal motions make the flux field
anisotropic (Hui 1999; McDonald \& Miralda-Escud\'e 1999), 
but this anisotropy has relatively little impact on the inferred matter
power spectrum over the scales provided by our analysis, and our
procedure for estimating the matter $P(k)$ will account for it 
automatically.
In place of $P_F(k)$, we will show in our plots the quantity
\begin{equation}
\deltaf\equiv\frac{1}{2\pi^{2}}k^{3}P_{F}(k),
\label{eqn:deltaf}
\end{equation}
which is the contribution
to the variance of the flux from an interval $d\ln k =1$.
The reader should note that this definition of $\deltaf$ 
differs from that in CWKH and CWPHK by the
factor $(1/2\pi^{2})$.  Also,
with this convention our definition of $P_{F,1D}(k)$ is larger than that 
given in M00 by a factor of 2. 

\subsubsection{Tests for systematic errors}
\label{sec:fluxpksystematics}

\begin{figure}[t]
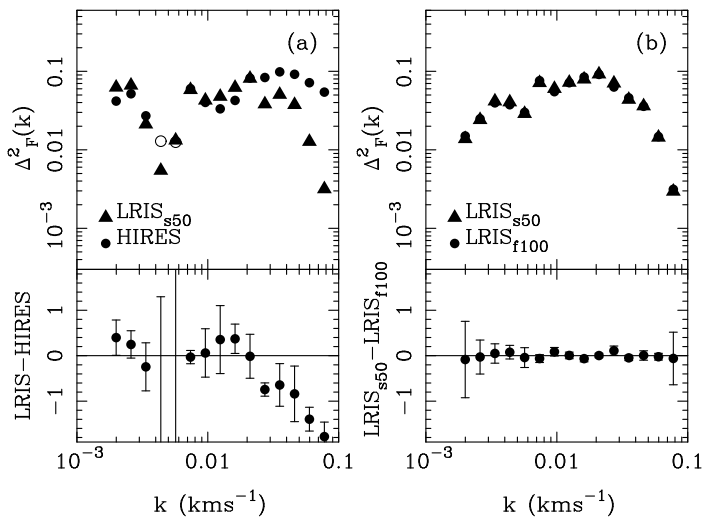

\centering
\PSbox{compmodes.ps angle=-90 voffset=240 hoffset=-110 vscale=58 hscale=58}
{3.5in}{3.0in} 
\caption[compmodes]
{Dependence of the flux power spectra $\deltaf$ on spectral resolution
and continuum fitting method.
({\it a})
We use the four quasar spectra common to the
LRIS and HIRES samples
and show the
 power spectrum derived from the HIRES observations (circles)
and the LRIS observations with 50\AA\ Gaussian smoothing for determination
of the mean level (triangles).  Open circles correspond to negative
values of $\deltaf$. 
Lower panel shows the difference from the mean of the two power
spectra in units of the mean.  Error bars represent $1\sigma$ jackknife
errors derived from the four spectra.
({\it b}) Comparison of LRIS results using
two methods of continuum fitting, 
smoothing with a 50\AA\ Gaussian (triangles)
and fitting over a 100\AA\ region with a 
3rd order polynomial (circles). Here we use all 23 LRIS spectra in
the LRIS sample, and the error bars represent jackknife errors
derived by partitioning those 23 spectra into 50 subsets of equal length
(see \S3.3.2).
 \label{compmodes}}
\end{figure}

We first compare $\deltaf$ measured from the four quasars for which we 
have both LRIS and HIRES spectra. These are Q0636+6801,
Q0940-1050, Q1017+1055, and Q1107+4847. In the top panel
of Figure~\ref{compmodes}a,
we show  $\deltaf$ measured from the LRIS spectra, where
$\delta_{F}(r)$ was calculated using the ($50$ \AA) Gaussian smoothing
method for finding the local mean flux. We also show the HIRES results.
In the lower panel, we plot the difference between the two measurements of 
$\deltaf$, in units of the mean  $\deltaf$ for the two sets of spectra.
The error bars were calculated by applying  a jackknife estimator to the
four spectra in each set. We can see that there are some differences
in detail between the two sets of $\deltaf$ results, but they appear
to be consistent within the errors, at least on large scales. On small
scales, the LRIS results are systematically lower, as we expect 
because of the lower resolution of the spectra.
We shall see later (\S\ref{sec:smoothingbias}) 
that smoothing the spectra can potentially
have effects on the inversion from 1D to 3D clustering on fairly large scales.
As a test of any systematic differences,
we have tried a least-squares fit of a horizontal 
straight line to the first ten points
in the lower panel of Figure~\ref{compmodes}a.
 We find that any uniform bias on these scales 
is consistent with zero within the
errors (we find $0.08\pm 0.11$ at 1 $\sigma$).

In Figure~\ref{compmodes}b, we carry out a similar test of the two 
different ways of analyzing the LRIS spectra, measuring $\delta_{F}(r)$ using
a fitted continuum versus smoothing the spectrum to define the mean.  
Because we are just using the LRIS data for this, we carry out the test 
using all 23 spectra. 
A horizontal fit to the 
first 10 points yields a mean bias between the two 
of  $-0.005 \pm 0.030 $ at $1\sigma$.

\begin{figure}[b]
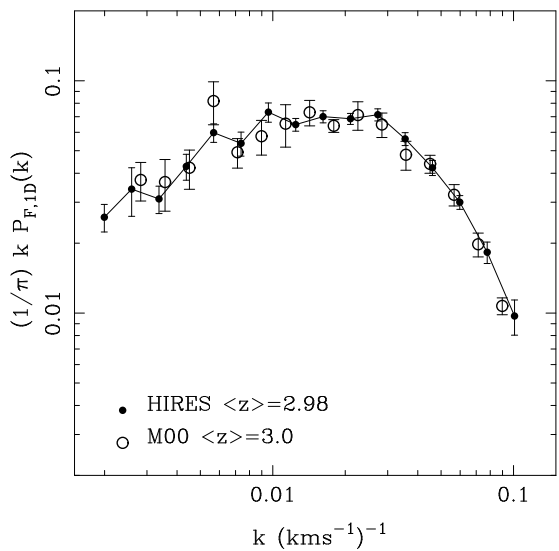

\centering
\PSbox{1dpk.ps angle=-90 voffset=270 hoffset=-60 vscale=48 hscale=48}
{3.5in}{3.2in} 
\caption[1dpk]{
Comparison of the 1D flux power spectrum from M00
(open circles) and from our HIRES spectra (filled circles),
for subsamples covering
the redshift range $z=2.67$ to $z=3.39$.
The mean redshift of the M00 data is 3.0 and that of
our subsample 2.98.
Because of
differing Fourier conventions, we have multiplied 
the M00 results by a factor of 2.
\label{1dpk}
}
\end{figure}

Previously, M00 presented $P_{F,1D}(k)$ measured from a sample of
Keck HIRES spectra. In order to compare our results to theirs, we have prepared
a sample of our HIRES data that has the same redshift boundaries
as one of the data samples in M00, $z=2.67$ to $z=3.39$.
The M00 sample with these boundaries has $\langle z \rangle=3.0$, and ours
has $\langle z \rangle=2.98$. 
The M00 spectra are a sample of eight 
with extremely high signal-to-noise ratio, and which have a FWHM
of $6.6 \kms$, binned into $2.4 \kms$ pixels.
The equivalent of $\sim 2.5$ full spectra contribute to the M00 results
for the redshift range we
use here, compared to $\sim 13$ (total length $6.4 \times 10^{5} \kms$) for 
the comparison sample of our data.
 
In Figure~\ref{1dpk}, we show $P_{F,1D}(k)$ from 
M00 and our comparison sample. 
We can see that on scales $k < 0.1 \invkms$,
there is good agreement between the two measurements,
and the larger number of spectra in our sample
is reflected in a smoother curve and smaller error bars. 
We have calculated these error bars with
a jackknife estimator, as we did for the
$\xi_{F}$ results (see \S3.2), except that when partitioning the
data we use 50 subsets. The error bars on the M00 points come from
bootstrap resampling, which should yield similar results to the 
jackknife technique.
In  Figure~\ref{1dpk}, we show only scales
 $k < 0.1 \invkms$, as on smaller scales,
we find that the results diverge. This is likely to be due to the
lower S/N of our data. We investigate
the effects of S/N on $\deltaf$ below. 
The M00 data are also of slightly higher resolution (our
data have FWHM of $8.0 \kms$, $20\%$ broader than M00).

Although we would like to have as large a dynamic range possible in our
flux power spectrum measurements, $k = 0.1 \invkms$
represents the scale  below which M00
have found that their results are sensitive to whether known metal lines
are removed or not. Since we do not attempt this
procedure, which would introduce additional uncertainties,
we are limited to points with $k < 0.1 \invkms$ even without
consideration of S/N and spectral resolution.

\begin{figure}[t]
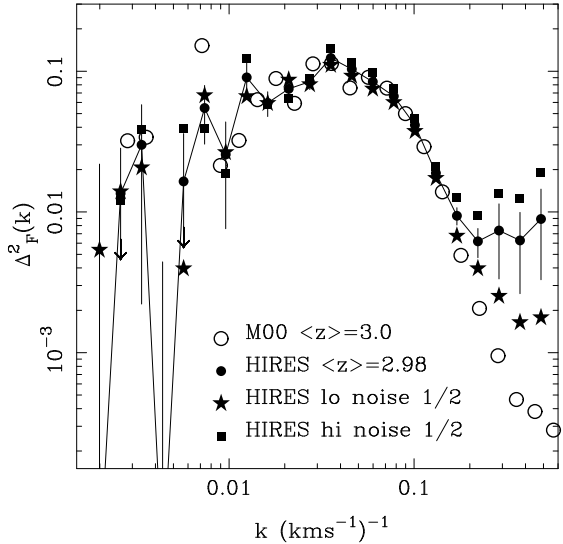

\centering
\PSbox{noisepk.ps angle=-90 voffset=270 hoffset=-60 vscale=48 hscale=48}
{3.5in}{3.2in} 
\caption[noisepk]{
Dependence of $\deltaf$ on S/N ratio of the quasar spectra.
Open circles show the 3D flux power spectrum from M00, and filled
circles show the measurement from our HIRES subsample with the
same redshift boundaries.  Stars and squares show $\deltaf$ derived
from the half of the HIRES subsample with highest and lowest S/N,
respectively.
\label{noisepk}
}
\end{figure}

In order to test the effect of noise on $\deltaf$, we split the M00 
comparison sample into two.  Spectra with a mean $1\sigma$ error in 
$F$ per pixel $>0.040$ are in the high
noise subsample. This subsample has a mean error per pixel of $0.074$ and 
$\langle z \rangle= 3.0$.
The low noise subsample, comprising the rest of the data,
has a mean noise per pixel
of $0.026$ and  $\langle z \rangle= 2.96$. The total  lengths
of spectra in the high and low noise subsamples are approximately
equal. We show the $\deltaf$ results in Figure~\ref{noisepk}, together with
$\deltaf$ for the M00 data. The noise level of the data (which is
dominated by
Poisson distributed photon noise) does  affect
the level of power on scales $k \ga 0.15 \invkms$.
However, we will limit our measurement of the matter $P(k)$ to 
scales $k < 0.05 \invkms$ anyway 
because of separate uncertainties
related to the flux to mass reconstruction.
On these scales,
any systematic bias associated with 
S/N is small,  with low S/N points
being slightly lower for $k \ga 0.03 \invkms$.

\begin{figure}[t]
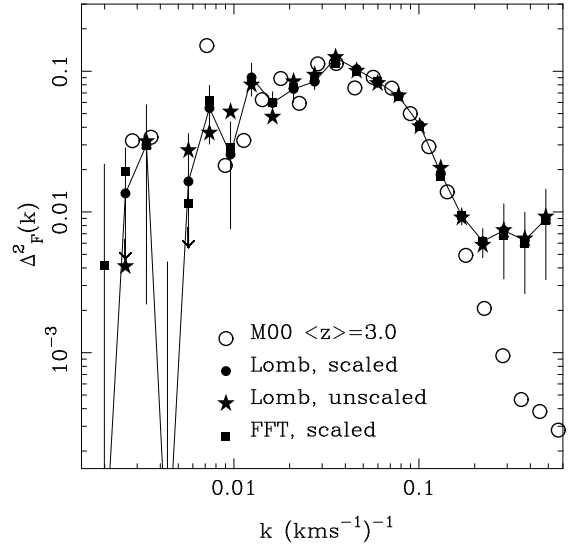

\centering
\PSbox{lombscale.ps angle=-90 voffset=270 hoffset=-60 vscale=48 hscale=48}
{3.5in}{3.2in} 
\caption[lombscale]{
Tests of different ways of computing the power spectrum and of
scaling the data.
Open circles show the 3D flux power spectrum of M00 and filled
circles the results of our standard treatment, for the comparison
HIRES subsample with the same redshift boundaries.
For these data points, and for those in the rest of the paper,
we scale pixel sizes and optical depths to the sample mean redshift
and estimate $P_F(k)$ using the Lomb periodogram.
Stars show the result with no scaling to the mean redshift.
Squares show the result of scaling the data but estimating $P_F(k)$
with an FFT instead of the Lomb periodogram.
\label{lombscale}
}
\end{figure}

In Figure~\ref{lombscale} we test details of the power spectrum 
estimation method, again using the HIRES subsample designed for
comparison to M00.  Filled circles, repeated from Figure~\ref{noisepk},
show results of our standard treatment.  Stars show
$\deltaf$ computed using
the same power spectrum estimator (the Lomb periodogram) but no scaling of
pixel sizes or optical depths to the mean redshift
(see \S2.2). There appear
to be only small and non-systematic differences between these two treatments.
The differences are even smaller
when we compare the fiducial results to those obtained from 
FFT measurements of the power spectrum (squares). We are therefore 
confident that no problems have been introduced by the use of an FFT in 
previous papers. However, the Lomb periodogram is better motivated,
so we adopt it here.

\begin{table}[ht]
\centering
\caption[pktabfid]{\label{pktabfid} 
The flux power spectrum, 
for the fiducial sample ($\langle z \rangle =2.72$).}
\begin{tabular}{ccc}
\hline &\\
k  &  $P_{F,1D}(k)$  & $P_{F}(k)$    \\
 $(\kms)^{-1}$  &  $(\kms)^{-1}$  & $(\kms)^{-3}$ \\
\hline &\\
 0.00199 & $41.1 \pm 4.4$ & $(2.50 \pm 3.19) \times 10^{7}$ \\
 0.00259 & $36.3 \pm 3.6$ & $(1.97 \pm 1.15 ) \times 10^{7}$ \\
 0.00337 &  $29.7 \pm 2.8$ & $(1.26 \pm 0.54 ) \times 10^{7}$ \\
 0.00437 &  $24.8 \pm 2.1$ &  $(3.45 \pm 2.32 ) \times 10^{6}$ \\
 0.00568&  $25.5 \pm 1.6 $ &  $(1.45 \pm 0.96 ) \times 10^{7}$ \\
 0.00738& $21.8 \pm 1.6 $ &  $(2.31 \pm 0.43 ) \times 10^{7}$ \\
 0.00958& $18.7 \pm 1.2 $ &  $(8.79 \pm  1.57) \times 10^{5}$ \\
 0.0124& $15.1 \pm 0.9 $ &  $(5.17 \pm 0.75 ) \times 10^{5}$ \\
 0.0162& $12.0 \pm 0.6 $ &  $(3.01 \pm 0.53 ) \times 10^{5}$ \\
 0.0210& $8.69 \pm 0.33 $ &  $(1.64 \pm 0.23 ) \times 10^{5}$ \\
 0.0272& $ 6.30 \pm 0.27 $ &  $(7.77 \pm 0.68 ) \times 10^{4}$ \\
 0.0355& $ 4.00\pm 0.14 $ &  $(4.07 \pm 0.27 ) \times 10^{4}$ \\
 0.0461& $2.28 \pm 0.12 $ &  $(1.72 \pm 0.08 ) \times 10^{4}$ \\
 0.0598& $1.17 \pm 0.06 $ &  $ 6340\pm 410$ \\
 0.0777& $0.561 \pm 0.036 $ &  $ 2050 \pm 110$ \\ 
 0.101&  $ 0.239\pm 0.021 $ &  $ 600 \pm 35$ \\ 
 0.131&  $ 0.114\pm 0.015 $ & $ 135 \pm 9$ \\ 
 0.170& $ 0.0712\pm 0.013 $ &  $28.5 \pm 2.7$ \\
 0.221& $ 0.0533 \pm 0.013 $ & $8.14 \pm 0.81$ \\ 
 0.287& $ 0.0400 \pm 0.010 $ & $3.74 \pm 0.86$ \\ 
\hline &\\
\end{tabular}
\end{table}

\subsubsection{
Test of inversion from the 1D to the
3D flux power spectrum}

\begin{figure}[t]
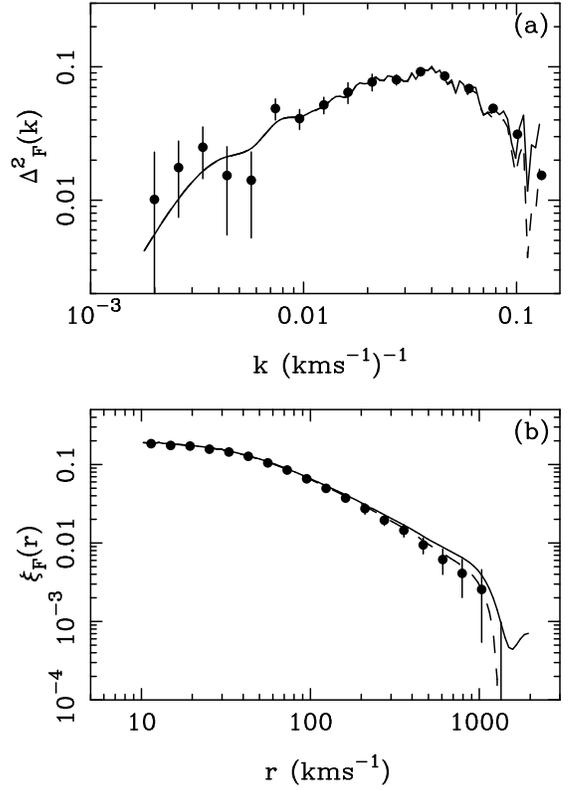

\centering
\PSbox{fft.ps angle=-90 voffset=410 hoffset=-175 vscale=78 hscale=78}
{3.5in}{4.5in} 
\caption[fft]{
(a) Comparison of the flux power spectrum measured directly
from the fiducial (B+C+D, $\langle z \rangle=2.72$)
 sample (points) with that estimated by carrying out a Fourier
transform of the flux correlation function of the fiducial sample (lines).
The different lines show two different ways of extrapolating
the measured $\xi_{F}(r)$ when carrying out the numerical integral.
The solid line shows an extrapolation which uses the LCDM shape, and
the dashed line uses a truncated form for $\xi_{F}(r)$ (see text).
(b) Comparison of the flux correlation
function $\xi_{F}(r)$ measured  directly
from the fiducial sample (points) with that estimated by carrying out a Fourier
transform of the flux power spectrum of the fiducial sample (lines). Again the
lines show two different ways of carrying out the extrapolation, solid 
being LCDM and dashed a truncated $P_{F}(k)$  (see text).
\label{fft}
}
\end{figure}

We would also like to test our method of deriving the 3D flux 
power spectrum from the 1D flux power spectrum, since one might
worry that the differentiation required by equation~(\ref{eqn:invert})
leads to biases in the presence of noise.
A simple test, illustrated in Figure~\ref{fft}, is to check that
$P_F(k)$ and $\xi_F(r)$ form the expected Fourier transform pair:
\begin{equation}
\xi_{F}(r)=\frac{1}{(2\pi)^3}\int^{\infty}_{0} P_{F}(k) 
           \frac{\sin{(kr)}}{kr} 4\pi k^2 dk,
\label{eqn:xitopk}
\end{equation}
\begin{equation}
P_{F}(k)=\int^{\infty}_{0} \xi_F(r) \frac{\sin{(kr)}}{kr} 4\pi r^2 dr.
\label{eqn:pktoxi}
\end{equation}
Points in the upper panel show $\deltaf$ estimated from our full fiducial
sample (B+C+D, HIRES and LRIS), with our standard methodology.
The solid curve shows the Fourier transform of the flux correlation
function (eq.~\ref{eqn:pktoxi}), where we have used the linear power
spectrum of the LCDM model described in \S 4.2 to extrapolate 
$\xi_F(r)$ beyond the observed $r$ limits.  The dashed curve shows the 
Fourier transform when the integral is simply truncated at the
observational limits $r_{\rm min}$ and $r_{\rm max}$.  The lower
panel of Figure~\ref{fft} displays an analogous comparison between
the directly measured $\xi_F(r)$ and the Fourier transform
(eq.~\ref{eqn:xitopk}) of $P_F(k)$.

These comparisons show that the two totally 
different methods for inferring three-dimensional clustering give 
very similar results. There is also little 
impact on Fourier transform estimates of $P_F(k)$ or $\xi_F(r)$
from scales where we have no direct measurements.
The agreement found in Figure~\ref{fft} justifies our earlier
assertion that equation~(\ref{eqn:invert}) defines a quantity
close to the power spectrum of the three-dimensional ``flux field,''
despite the presence of some redshift-space anisotropy
(see Hui 1999; McDonald \& Miralda-Escud\'e 1999).
We adopt this approach in preference to the inversion of $\xi_F(r)$,
which is rather difficult to handle numerically, particularly on the smallest
scales. 

\subsubsection{Smoothing bias}
\label{sec:smoothingbias}

There is another, somewhat subtle effect that influences
the inversion from 1D to 3D when using low resolution spectra.
Finite spectral resolution smooths the 1D power spectrum by convolution
with the square of the instrument response function.
Because the 3D power spectrum is obtained by differentiation
(eq.~\ref{eqn:invert}), this steepening of the 1D power spectrum
artificially boosts the amplitude of the 3D power spectrum, even
on scales that are significantly larger than the smoothing scale.

\begin{figure}[t]
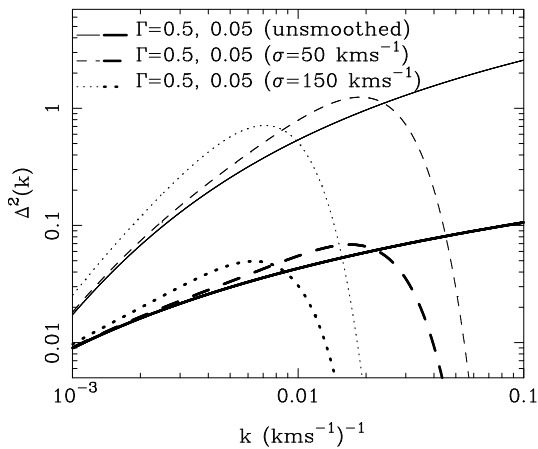

\PSbox{smoothbias.ps angle=-90 voffset=235 hoffset=-55 vscale=45 hscale=45}
{3.5in}{2.8in} 
\caption[smoothbias]{
Test of ``smoothing bias'', the effect of smoothing a spectrum 
(for example due to finite spectral resolution) on the inversion from 1D to 3D.
Solid lines show the linear theory power spectra $\Delta^2(k)$ of
CDM models with shape parameter (see \S 4.2) $\Gamma=0.05$ (thick)
and $\Gamma=0.5$ (thin).  
Dashed lines show the result of deriving the 3D power spectrum from
the corresponding 1D power spectra smoothed with a Gaussian filter
of $\sigma=50 \kms$, approximately equivalent to 2\AA\ FWHM spectral
resolution.
Dotted lines show the result for $\sigma=150\kms$ (6\AA\ FWHM) smoothing.
The derived $\Delta^2(k)$ is suppressed on small scales by smoothing,
but it is boosted on intermediate scales because of the differentiation
required to go from 1D to 3D.
\label{smoothbias}
}
\end{figure}

We show the
effect of this ``smoothing bias'' on linear theory power spectra in 
Figure~\ref{smoothbias}. Here we have multiplied $P_{F,1D}$ by a Gaussian
filter, to simulate observational smoothing, then used
equations~(\ref{eqn:invert}) and~(\ref{eqn:deltaf}) to find
$\deltaf$. We show results for two different power spectrum shapes,
characterized by the shape parameter $\Gamma$ (see \S 4.2).
The amount of bias depends on the shape, but the two we show are
fairly close to the observed shape, at least on large scales,
and the bias seen in both should be representative.
We find that on the largest scale we observe, $2 \times 10^{-3} \invkms$,
the boost in $\deltaf$ with the
2 \AA\ spectral resolution typical of our LRIS data is $3 \%$, 
while for much lower resolution of 6 \AA\ it would be $20\%$. 
This artificial amplification increases to a maximum of
14\% for the smallest scale we make use of for the 2 \AA\ case,
and would be 32\% for 6\AA. On the very smallest scales, smoothing 
suppresses $\deltaf$.
On the scales where we use the LRIS data, $k<0.014 \invkms$, they 
contribute about half the signal of the fiducial sample, so with 
no correction we would expect a bias of $1.5\%-7\%$ on these scales.
To remove this effect, we adjust the LRIS contributions to $P_F(k)$ using 
correction factors derived from 
the fractional differences between the lower curves in Figure~\ref{smoothbias};
however, we limit the maximum correction to 10\%, since the value close
to the smoothing scale is sensitive to the assumed form of the input spectrum.
Since the maximum corrections to $P_F(k)$ are only a few percent,
the {\it uncertainties} in the
corrections are much smaller than the $1\sigma$ error bars on the
affected data points, which are $\ga 15\%$.
The agreement of $\deltaf$ derived from HIRES and LRIS spectra of
the same quasars (Figure~\ref{compmodes}), for which we applied no
correction to the LRIS $\deltaf$,
is further evidence that smoothing bias is 
a minor issue in the context of this data set. 
However, it could be important for data of substantially lower spectral
resolution.  Because the degree of bias depends on the precise form
of the spectral response function and on the shape of the underlying
power spectrum, a rigorously accurate correction is difficult,
and alternative analysis methods
should be considered for lower resolution data.

\subsubsection{The flux power spectrum and its covariance matrix}
\label{sec:fluxpk}

\begin{figure*}[t]
\centering
\vspace{14cm}
\includegraphics{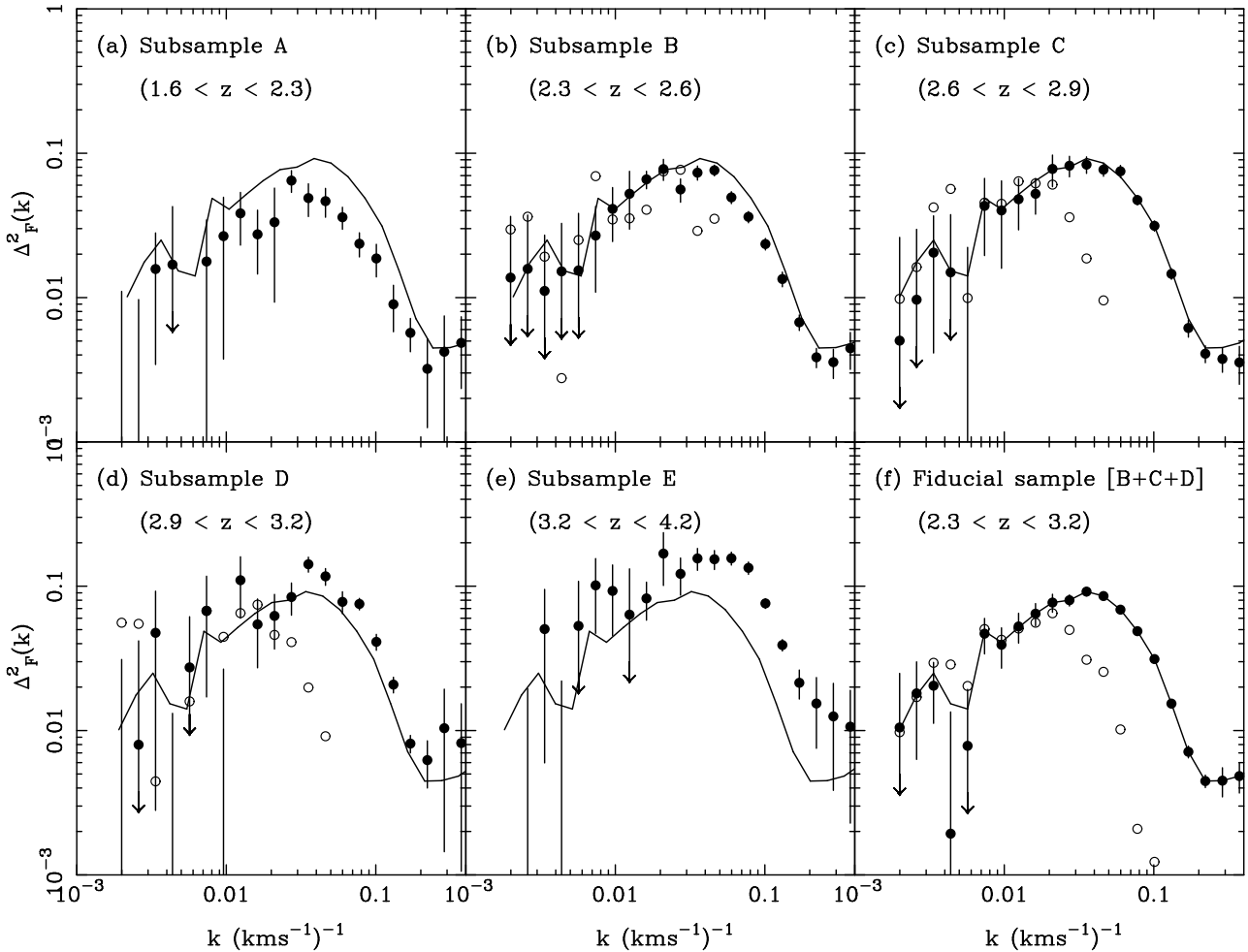}
\caption[fluxpk]{
The flux power spectrum of the various data subsamples.
The solid line in each case is the fiducial combined sample, with the
length scaled so that the comoving lengths stay the same in an
EdS model. The filled points are the HIRES data and the open circles
the LRIS data. Error bars have been omitted from the latter, for
clarity.
\label{fluxpk}}
\end{figure*}

Figure~\ref{fluxpk} presents the principal results of this section,
the flux power spectrum $\deltaf$ of the fiducial sample and the
various redshift subsamples.  The values of $P_{F,1D}(k)$ and
$P_F(k)$ for the fiducial sample are listed in Table~\ref{pktabfid},
and the values of $P_{F,1D}(k)$ for the redshift subsamples are
listed in Table~\ref{pktabz} of the Appendix.  We average the
contributions of the LRIS and HIRES data on scales
$k<0.014 \invkms$ for $\deltaf$ and $k<0.006\invkms$ for
$P_{F,1D}(k)$, which is more strongly affected by the spectral resolution.
We also average the $1\sigma$ error bars 
and divide them by $\sqrt{2}$. 
We do not account for the fact that four 
spectra appear in both samples, so our error bars on these
large-scale datapoints may be systematically underestimated by as
much as $\sim1-\sqrt{49/53}=4\%$.
We use only the HIRES data on smaller scales.

\begin{figure}[b]
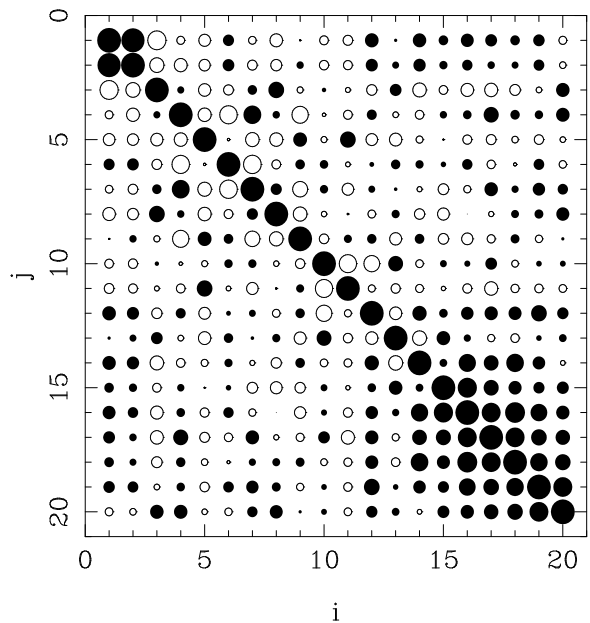

\centering
\PSbox{covhires.ps angle=-90 voffset=295 hoffset=-65 vscale=50 hscale=50}
{3.5in}{3.4in} 
\caption[covhires]{
The covariance matrix of the flux power spectrum of
the fiducial HIRES sample. The symbol area is proportional to
$C_{ij}/(C_{ii}C_{jj})^{1/2}$, with negative elements shown 
by open symbols. The flux power spectrum values which correspond
to this covariance matrix are given in 
Table~\ref{pktabfid}. The elements span scales of
0.00199 $\invkms$ (element 1) to 0.287  $\invkms$ (element 20).
\label{covhires}
}
\end{figure}

The error bars in Figure~\ref{fluxpk} and Tables~\ref{pktabfid}
and~\ref{pktabz} are computed using a jackknife estimator,
with 50 data subsets in each case.  Although
the flux correlation function
$\xi_{F}(r)$ has strongly covariant errors, we
might expect the errors on the $\deltaf$ data points to be
close to independent, at least if they reflect the
behavior of the linear matter power spectrum. 
M00 found that the
covariance matrix of $P_{F,1D} (k)$ measured from their data is extremely
noisy but consistent with the off-diagonal elements being zero.

Figure~\ref{covhires} illustrates the 
covariance matrix $C_{ij}$ of $\deltaf$ for
our fiducial HIRES data (with $\langle z \rangle=2.72$),
again estimated by the jackknife technique.
We have divided out the diagonal elements,
so that the symbol area is proportional to $C_{ij}/(C_{ii}C_{jj})^{1/2}$. 
It is obvious from the plot that $C_{ij}$ is fairly close to 
diagonal, at least for the elements with
 $i$ and $j$ $\la 13$.  The matrix is also quite noisy,
with the uncertainty on $C_{ij}$ increasing as we move towards 
small $i$ and $j$ (larger scales), where there are fewer
modes to average over. On the smallest scales we find
significantly positive non-diagonal
elements. These scales are smaller than the smallest ones we shall be using to
reconstruct the matter power spectrum. On larger scales some
of the off-diagonal elements appear to be small but negative. This 
behavior was also seen in the CWPKH covariance matrix, 
and if it is statistically significant it is
probably caused by the differencing needed to compute
$P_F(k)$ from $P_{F,1D}(k)$.
Given that the covariance matrix is noisy and that 
anti-covariance caused by negative elements would decrease error bounds, we
will adopt the conservative position of using only the diagonal elements
in our analysis of the matter power spectrum.

The error bars in Figure~\ref{fluxpk} are much larger than those
on $\xi_F(r)$ in Figure~\ref{fluxxi} because in this case they are
nearly uncorrelated.  The shape of $\deltaf$ remains roughly the
same at all redshifts on large scales, while the relative amount of power
on small scales appears to decrease with decreasing redshift, 
presumably due to increasing non-linearity
and peculiar velocities. The overall amplitude of $\deltaf$ drops
towards lower redshifts, as for $\xi_F(r)$.
This drop is driven by the decrease of
$\taueff$ as the universe expands. 
We will show in \S\ref{sec:evolution} that, once the evolution of $\taueff$ is
taken into account, these $\deltaf$ results yield a marginal
detection of the expected signature of gravitational growth 
of the underlying matter fluctuations.

\section{From flux to mass: method}
\label{sec:method}

\subsection{Overview}

CWKH proposed a method for recovering the linear matter power spectrum $P(k)$
from measurements of the \lya\ forest flux power spectrum, and
CWPHK applied this method to a sample of 19 moderate resolution quasar spectra.
The method that we use to recover $P(k)$ in this paper has evolved from 
that used by CWKH and CWPHK, but it is significantly better.
Specifically, our current method is to assume that
\begin{equation} 
P_F(k) = b^2(k) P(k)
\label{eqn:pkb}
\end{equation}
and that the values of $b(k)$ can be calibrated using numerical
simulations that are tuned to match the observed $\taueff$ and $P_F(k)$.
In this language, the method
used in CWKH and CWPHK assumed $b={\rm constant}$, and CWKH defined $P_F(k)$ 
from the ``Gaussianized'' flux rather than the flux itself.
The $b={\rm constant}$ assumption is reasonably accurate on large scales,
where, at least according to hydrodynamic simulations, the shape of
the flux power spectrum is similar to that of the linear matter power 
spectrum.  This similarity of shape is expected if the matter density and flux
are related by a local transformation (see, e.g., Coles 1993;
Gazta\~{n}aga \& Baugh 1998; Scherrer \& Weinberg 1998;
M00 Appendix C), as they are in the Fluctuating Gunn-Peterson Approximation.
However, the method adopted here is obviously more general, and
it can account for the effects of redshift-space distortions, non-linear
evolution, and thermal broadening, which change the shape of $P_F(k)$.
This improvement in method is justified by the larger size and dynamic
range of our current data set, since with the previous method the accuracy
of our recovered $P(k)$ would be limited by the accuracy of the
$b={\rm constant}$ approximation.  As in the previous method, there are
systematic uncertainties in the transformation from $P_F(k)$ to $P(k)$
because there are uncertainties in the parameter values to adopt
for the calibrating simulations; we will discuss these systematic
uncertainties in \S 5.  
Our current approach is, in some sense, intermediate between that of CWPHK,
who determined the amplitude of $P(k)$ using
simulations with the initial $P(k)$ shape
inferred from the \lya\ forest data themselves, and that of M00, 
who did not attempt an inversion of $P(k)$ but estimated $P(k)$
parameter constraints by scaling the predictions of a hydrodynamic simulation.
However, our approach here also includes
new features not present in either of these previous methods.

\begin{figure}[t]
\PSbox{flow.eps angle=0 voffset=20 hoffset=0 vscale=75 hscale=75}
{3.5in}{3.7in} 
\caption[flow]{
Flowchart for matter power spectrum reconstruction.
\label{flow}
}
\end{figure}

The results of previous investigations (CWPHK; M00; Phillips et al.\ 2001)
imply that the shape of the linear matter power spectrum $P(k)$
on \lya\ forest scales
is in reasonable agreement with that of a low density CDM model.
We therefore adopt this power spectrum shape 
for the ``normalizing simulations'' that we use to calculate the
function $b(k)$.  We obtain outputs from the simulations corresponding
to a number of different $P(k)$ amplitudes.  For each output amplitude,
we create artificial spectra using the FGPA, adjusting the parameter $A$
of equation~(\ref{eqn:tau}) so that the spectra match an observationally
determined value of $\taueff$.  From these spectra, we calculate
the flux power spectrum $P^{\rm sim}_F(k)$, the corresponding $\deltaf$, and
\begin{equation}
b(k) =  \left[\frac{P^{\rm sim}_F(k)}{P^{\rm sim}(k)}\right]^{1/2},
\label{eqn:bdef}
\end{equation}
where $P^{\rm sim}(k)$ is the linear matter power spectrum of the simulation.
The amplitude of the predicted flux power spectrum increases monotonically
with the amplitude of $P(k)$, since stronger density fluctuations produce
stronger fluctuations of \lya\ optical depth.
To decide which $b(k)$ results apply to the observational data,
we choose the simulation output that has $\deltaf$ in best agreement
with the observed $\deltaf$ (interpolating between outputs to get
a finer grid of amplitudes).
We then divide the observed flux power spectrum by the (interpolated) $b^2(k)$ 
corresponding to this output to obtain our observational estimate of the linear
matter power spectrum:
\begin{equation}
P^{\rm obs}(k)= \frac{P^{\rm obs}_{F}(k)}{b^2(k)}.
\label{eqn:pkobs}
\end{equation}
Figure~\ref{flow} summarizes these steps.
We discuss the normalizing simulations and normalization procedure
in more detail in \S\S 4.2 and 4.3, below.

\subsection{Normalizing simulations}
\label{sec:normsim}

The linear matter power spectrum $P(k)$ that we use in our normalizing
simulations is consistent with that of a low density, inflationary CDM
model with a cosmological constant (LCDM for short).
The analytic form we use is taken from the work
of Bardeen $\etal$ (1986):
\begin{equation}
P(q) = B\, \frac{q^n [\ln(1+\alpha_1 q)/\alpha_1 q]^2}
         {[1+\alpha_2 q +(\alpha_3 q)^2+(\alpha_4 q)^3 +
	  (\alpha_5 q)^4]^{\frac{1}{2}}}~,
	  \label{eqn:bbks}
\end{equation}
where $q\equiv k/\Gamma$, and $B$ is a normalization
constant. We use the same coefficients as M00, $\alpha_1=2.205$, 
$\alpha_2=4.05$, $\alpha_3=18.3$, $\alpha_4=8.725$,
and $\alpha_5=8.0$, which were calculated for 
a baryon fraction $\Omega_{b}=0.05$ by Ma (1996).
We set $\Gamma=0.26$ and $n=0.95$.
For the transfer function coefficients of Bardeen et al.\ (1986),
the equivalent $\Gamma$ would be approximately 0.24.
When scaling the simulated spectra to observational units 
(km$\;{\rm s}^{-1}$), we assume a cosmology with $\Omega_{m}(z=0)=0.4$ and 
$\Omega_{\Lambda}(z=0)=0.6$. 

As mentioned earlier, we choose this $P(k)$ shape because previous work
has shown that such a power spectrum 
is consistent with the \lya\ forest results
on the relevant scales (CWPHK; M00; Phillips \etal 2001).
This adoption of a smooth, theoretically motivated initial power
spectrum represents a change in technique from CWPHK, where
the normalizing simulations were run using the $P(k)$ shape measured
from the flux power spectrum as input.
The new approach has the advantage that
errors in the shape of the $P_{F}(k)$ from the normalizing
spectra are not correlated with those in the
observed  $P_{F}(k)$ (which is the case with the previous technique),
and that it is much easier to allow a scale-dependent $b(k)$. 
However, we should emphasize that the power spectrum derived from
the data has very little dependence on the shape of the power spectrum
assumed in the normalizing simulations, since we use the simulations
only to calculate $b(k)$, which should be insensitive to small changes
in the power spectrum shape.  
We justify our choice of $P(k)$ for the normalizing simulations
retrospectively below, by showing that the flux power spectrum
$\deltaf$ derived from these simulations
yields a good fit (with an acceptable $\chi^{2}$) to the observed 
$\deltaf$.

The normalizing simulations themselves are run with a P$^{3}$M N-body code
(Efstathiou \& Eastwood 1981; Efstathiou \etal 1985), with 
the gravitational softening length set to be 0.8 force mesh cells
as high force resolution is not needed.
We run ten simulations with different random phases. Each one
evolves $160^{3}$ particles using a  $256^{3}$ force mesh
in a box 27.77 $\hmpc$ on a side. These parameters yield
the same mass and force resolution as the normalizing simulations in CWPHK, 
which were shown to be adequate by tests in that paper
(see also Figure~\ref{sphtest} below).
The simulations are run with a background EdS cosmology and evolved
so that the expansion factor $a$ increases by a factor 9.0 from the
initial conditions to the most evolved output, in equal steps of
$\Delta a=0.1$. 

Spectra are extracted from the simulation outputs as described in CWKH. 
Densities are converted to real-space optical depths
using the FGPA (eq.~\ref{eqn:tau}).
These optical depth profiles are then used to compute redshift-space
spectra including the effects of peculiar velocities and thermal broadening.
We determine the gas temperature as a function of density assuming a
power-law relation (eq.~\ref{eqn:td}), with fiducial values for the two 
parameters of $T_{0}=15000$ K and $\alpha=0.6$. 
The relatively high temperature (in CWPKH we used $T_{0}=5600K$)
is motivated by evidence that the high-$z$ IGM is hotter than 
we previously assumed (Theuns \etal 1999; Bryan \& Machacek 2000;
Ricotti et al.\ 2000; Schaye et al.\ 2000; McDonald et al.\ 2001).
The value of $\alpha$ in turn determines the value of the
index $\beta=2-0.7\alpha$ in the FGPA.
The effect of $T_0$ is largely degenerate with that of the other
parameters that enter into the combination $A$ of equation~(\ref{eqn:tau}),
but higher $T_0$ does lead to more thermal broadening and thus to
a depression of $\deltaf$ on small scales.
A crucial step of our procedure is to adjust the value of $A$ so
that the spectra extracted for a particular set of simulation outputs
match our adopted observational estimate of $\taueff$.
Physically we can think of this step as fixing the photoionization rate
$\Gamma_{\rm HI}$, which is only weakly constrained by direct measurements,
to reproduce the observed mean opacity given our assumed values of
$T_0$, $\Omega_b$, $h$, and $H(z)/H_0$.
For our fiducial results, we adopt the $\taueff$ value given by PRS,
which is $\taueff=0.349$ at $z=2.72$.
In \S 5, we will discuss the uncertainties in our derived $P(k)$
associated with the uncertainties in the appropriate choices of
$T_0$, $\alpha$, and $\taueff$.  The uncertainty in $\taueff$ turns
out to be the most important, but the uncertainties in $T_0$ and
$\alpha$ are also significant.

\begin{figure}[t]
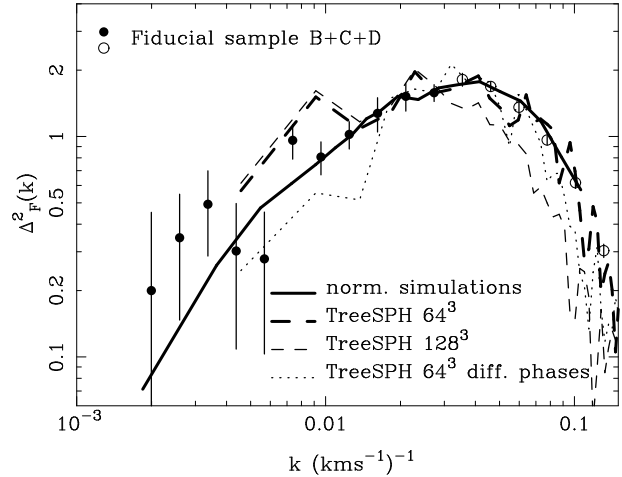

\PSbox{sphtest.ps angle=-90 voffset=245 hoffset=-45 vscale=45 hscale=45}
{3.5in}{2.8in} 
\caption[sphtest]{
The flux power spectrum $\deltaf$ derived from the average of ten
normalizing simulations, each with $160^3$ particles in a $27.77\hmpc$ box,
interpolated to a matter fluctuation amplitude of
$\sigma_{8}(z=0)=0.74$ (thick solid line).
Dotted and dashed lines show results of three TreeSPH simulations with
the same matter fluctuation amplitude.
Two of these simulations (thin dashed and dotted lines) 
have $64^3$ particles in an $11.11\hmpc$
box, the same particle density as our normalizing simulations, and
the difference between them illustrates the effect of cosmic variance
in this small simulation volume.  The third simulation (thick dashed line) 
has the same phases as the first $64^3$ simulation but eight times
more particles.  Points with error bars show $\deltaf$ derived from 
the fiducial observational sample at $z=2.72$; filled points indicate
the scales that we will use for normalization of the matter power spectrum.
\label{sphtest}
}
\end{figure}

We extract 1000 spectra from each box
for a total of 10,000 per output time.
Averaging over a large number of simulations is important to remove 
fluctuations, as the cosmic variance error on the mean $\deltaf$ 
estimated from a small volume can be considerable. 
We can see this cosmic variance in 
Figure~\ref{sphtest}, where we show $\deltaf$ for our normalizing simulations
and for some comparison simulations run with a full cosmological 
hydrodynamic code. The hydrodynamic simulations were run with
parallel TreeSPH (Dav\'{e}, Dubinski \& Hernquist 1997)
by Romeel Dav\'{e} (see Dav\'e \etal 1999) and by Jeffrey Gardner 
(see Gardner \etal, in preparation), and they
include the effects of gas dynamics, shocks, heating of gas
by the UV background, radiative cooling, and star formation.

The LCDM model simulated in the TreeSPH simulations is very close
to the model adopted in our
dissipationless normalizing simulations. The TreeSPH simulations
were output at $z=3$, but in order to compare to our fiducial sample 
($\langle z \rangle= 2.72$), the  $\kms$ length scales were multiplied by 
$\sqrt{3.72/4}$, so that the same
comoving lengths (in this case for EdS scaling, which is
accurate at these redshifts) could be compared against each other.
This use of an earlier output also means that the effective 
mass fluctuation amplitude is lower (equivalent to a model
with $\sigma_{8}=0.74$ rather than the $\sigma_{8}=0.79$
that was actually used). 
We show results from two TreeSPH simulations that have $64^3$ particles in
an $11.11\hmpc$ box and from one simulation with a factor of eight higher
mass resolution, using $128^3$ particles in an $11.11\hmpc$ box.
The former simulations have the same 
particle density as our dissipationless normalizing simulations.
One of the $64^{3}$ simulations has the same phases as
the $128^{3}$ run, and Figure~\ref{sphtest} shows that on large
scales their $\deltaf$ results match well. 
This agreement indicates that at the $64^3$ resolution
the \lya\ forest predictions have 
converged well enough for our normalizing simulations to 
yield the correct amplitude of $\deltaf$, at least on the large
scales where we will normalize the matter 
power spectrum (\S4.3). The other $64^{3}$ run is identical to the
first, except that the initial conditions were generated with different
random phases. The large differences in $\deltaf$ are therefore
due to cosmic variance, and they show that inferences of
the matter $P(k)$ amplitude should rely on normalizing simulations
with a much larger volume (c.f., M00). 
The solid line in Figure~\ref{sphtest} shows $\deltaf$ derived
from our normalizing simulations, interpolated to have the same  
amplitude ($\sigma_{8}=0.74$) as the SPH simulations. 
It lies between the two sets of SPH curves, indicating that the 
combination of dissipationless simulations with the FGPA is accurate
enough for our purposes, to the extent that we can test.
Equally important, the $\deltaf$ curve is smooth, showing that
the total volume sampled ($\sim 150$ times that of the TreeSPH simulation
box) is large enough to eliminate the uncertainty associated with
cosmic variance.

The points with error bars in
Figure~\ref{sphtest} show $\deltaf$ from our fiducial
observational sample, merely for illustrative purposes at this stage. 
The results are roughly consistent with these
$\sigma_8=0.74$ LCDM simulations.
(The quantity $\sigma_8$ is the rms mass fluctuation amplitude in spheres
of comoving radius $8\hmpc$, at $z=0$ unless the redshift is otherwise
specified.)

\subsection{Normalization}
\label{sec:normalization}

\begin{figure}[t]
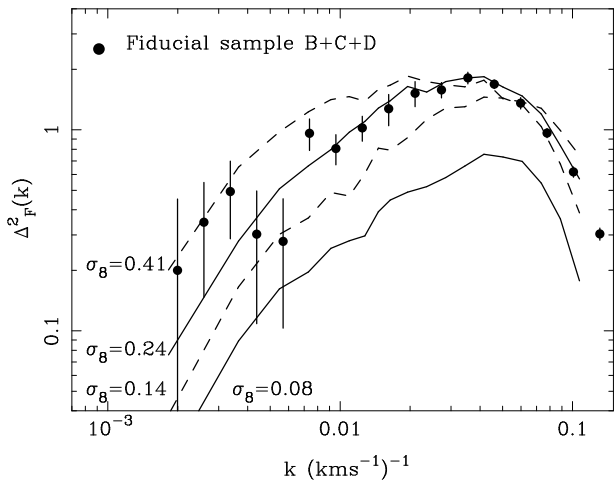

\PSbox{normamp.ps angle=-90 voffset=245 hoffset=-45 vscale=45 hscale=45}
{3.5in}{2.8in} 
\caption[normamp]{
Flux power spectra from four different outputs of the normalizing
simulations, compared to $\deltaf$ of the fiducial observational
sample at $z=2.72$.  The linear matter fluctuation amplitudes 
corresponding to the four outputs are (bottom to top)
$\sigma_8(z=2.72)=0.08$, 0.14, 0.24, 0.41.
(For $\Omega_m=0.4$, $\Omega_\Lambda=0.6$, values at $z=0$ are
larger by a factor of 3.11.)
Filled circles denote
the data points used in determining the best-fit fluctuation
amplitude, which corresponds to $\sigma_8(z=2.72)=0.23$, with a $1\sigma$
fitting uncertainty of $\pm 9\%$.
\label{normamp}
}
\end{figure}

Figure~\ref{normamp} shows the flux power spectrum from four outputs
of our normalizing simulations.  The simulated spectra are scaled to
the same $\taueff$ and the same velocity units at each output, so
the difference in $\deltaf$ just reflects the different amplitude of
the underlying matter fluctuations.
We obtain $\deltaf$ on a finer grid of amplitudes by interpolating
between these outputs, in log space
(the outputs are close enough that interpolating linearly gives essentially
the same result, within $2\%$). 
We then find the amplitude that best matches the observed $\deltaf$
values by $\chi^2$ minimization.
We only consider large scale points, $k\leq 0.0272\invkms$, in this 
amplitude determination,
so that we remain in the regime where our simulation results are
not affected by their finite resolution (Figure~\ref{sphtest}) and 
where the shape of the flux power spectrum is
not sensitive to the adopted temperature of the IGM (see \S5.4 below). 

In our adopted LCDM cosmology, the best-fitting amplitude corresponds
to $\sigma_8=0.23$ at $z=2.72$, or $\sigma_8=0.72$ at $z=0$.
(We will discuss the amplitude in more general terms in 
\S\ref{sec:masspk} and \S\ref{sec:implications}.)
The value of $\chi^{2}$ for the best-fitting $\deltaf$ is
8.5, for 10 degrees of freedom, indicating that our error
bars on $\deltaf$ are realistic and that the LCDM power spectrum shape
is fairly close to the one implied by the observations.
If we change the IGM temperature parameter from our fiducial value of 
$T_0=15,000$ K to $T_0 = 5000$ K while keeping $\alpha=0.6$,
then the fit becomes slightly worse ($\chi^2=10$), but the change
is small because we are restricting the analysis to large scales.
With $T_0$, $\alpha$, and $\taueff$ fixed to their fiducial values,
we find the $1\sigma$ uncertainty in the overall matter fluctuation
amplitude ($\propto \sigma_8$) of the normalizing simulations to be $\pm 9\%$.

\begin{figure}[b]
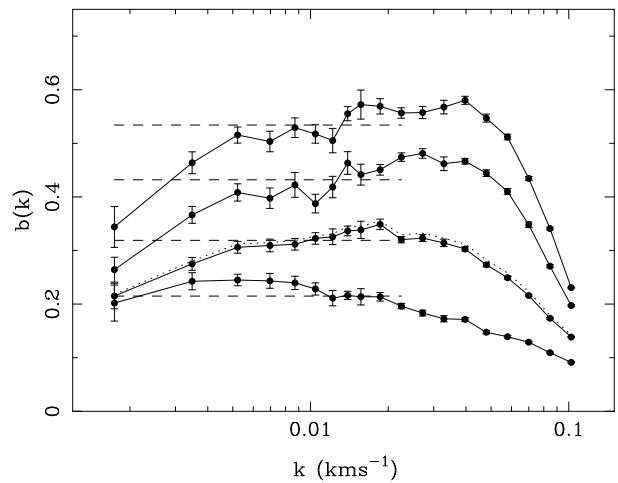

\PSbox{redbias.ps angle=-90 voffset=245 hoffset=-45 vscale=45 hscale=45}
{3.5in}{2.8in} 
\caption[redbias]{
The bias $b(k) \equiv \sqrt{P_F(k)/P(k)}$ between 
the flux power spectrum and the linear matter power spectrum,
measured from the normalizing LCDM simulations,
for different mass fluctuation amplitudes. These are, from
bottom to top, $\sigma_{8}(z=2.72)=0.41, 0.24, 0.14, $ and $0.08$. 
Horizontal dashed lines show 
a $\chi^{2}$ fit to the points with $k< 0.025$,
purely for illustrative purposes.
The dotted curve shows $b(k)$ interpolated to the amplitude
$\sigma_8(z=2.72)=0.23$ that best fits the observed flux power spectrum.
We use this $b(k)$ to infer $P(k)$ from our measured flux power spectrum.
\label{redbias}
}
\end{figure}

Figure~\ref{redbias} shows the biasing function $b(k)$ derived from
the normalizing simulation outputs (see eq.~\ref{eqn:bdef}).
There are two instructive points to note from this figure. 
First, $b(k)$ drops as the matter fluctuation amplitude increases.
Similar behavior can be seen in the one-point
analysis of Gazta\~{n}aga \& Croft (2000), who show that
for low mass fluctuation amplitudes the bias tends to the value 
predicted by perturbation theory.  However, at higher fluctuation
amplitudes, saturation reduces the sensitivity of flux fluctuations
to mass fluctuations: the non-linear mapping of density to flux forces
$F$ into the range zero to one, $\delta_F$ grows more slowly than
$\delta_\rho$, and the bias decreases as $\delta_\rho$ increases.
Second, our large volume simulations show the redshift-space distortion
of the shape of $P_F(k)$ (i.e., a scale-dependent $b(k)$), which was
predicted based on linear theory calculations by Hui (1999) and
McDonald \& Miralda-Escud\'e (1999).
The shape of the distortion follows these predictions qualitatively,
with a suppression on large scales and a boost on intermediate
($k \sim 0.03 \invkms$) scales. 
At higher $k$ we find a substantial suppression of $P_F(k)$,
presumably caused by a combination of thermal broadening, non-linear
effects, and the simulations' finite numerical resolution.

The dotted curve in Figure~\ref{redbias} shows $b(k)$ interpolated
to our best-fit matter fluctuation amplitude.
This is the function that we will use to determine $P^{\rm obs}(k)$
via equation~(\ref{eqn:pkobs}).
The use of normalizing simulations to compute $b(k)$ allows us to
account for the distortion of the shape of $P_F(k)$ caused by
redshift-space distortions and non-linearity.
The distortion in the shape is considerable ($b(k)$ changes by
up to  $\sim 30\%$ between different scales), although the effects are
largest for low $k$, where the statistical uncertainties are already large,
and for high $k$, where we will not attempt to recover the matter 
power spectrum anyway.

There is an overall multiplicative uncertainty in $b(k)$ because
of the range in the amplitudes of normalizing simulations
that yield an acceptable match to our measured flux power spectrum.  
In the neighborhood of our best-fitting fiducial model, the average
value of $b(k)$ in the wavenumber range that we use for normalization
scales as $b \propto \sigma_8^{-0.7}$, so the $\pm 9\%$ 
uncertainty in $\sigma_8$ implies a $\pm 6\%$ uncertainty in $b$.
This in turn contributes a 12\% uncertainty in the overall
amplitude of $P^{\rm obs}(k) = P^{\rm obs}_F(k)/b^2(k)$,
in addition to the error bars on individual points that come
from the jackknife error bars on $P_F(k)$.
Here we are following CWPHK in dividing our error bars into
an overall normalization uncertainty and error bars on individual points. 
This division simplifies our analysis, especially when we consider
the additional uncertainties related to $\taueff$, $T_0$, and $\alpha$;
we will show in \S 5 that these primarily affect the overall amplitude
of $P(k)$ rather than the shape.

\section{Systematic uncertainties in the matter power spectrum}
\label{sec:systematics}

\subsection{Overview}

Our determination of the flux power spectrum $P_F(k)$ in \S 3.3
is essentially a pure measurement.  There are systematic uncertainties
in this measurement associated with continuum fitting, scaling of
pixel sizes and fluxes, inversion from 1D to 3D, and so forth, 
but we have argued in \S 3.3 that these uncertainties are small
compared to the statistical uncertainties of this finite sample.

The inference of the linear matter power spectrum $P(k)$ from $P_F(k)$
requires the biasing function $b(k)$, which we calculate (as 
described in \S 4) using simulations
that incorporate a number of assumptions.  
The uncertainties in these assumptions
are the main source of systematic uncertainties in the derived
matter power spectrum.

Specifically, we compute $b(k)$ from P$^3$M simulations assuming
that the underlying cosmological model is LCDM, that the gas traces
the dark matter in the low density IGM, that all of the gas lies
on the temperature-density relation (eq.~\ref{eqn:td}), that
the parameters $\taueff$, $T_0$, and $\alpha$ have specified values,
that the photoionizing background and temperature-density
relation are spatially uniform, and that metal lines and damping
wings have a negligible impact on $P_F(k)$.
In this section, we will discuss the uncertainties associated
with each of these assumptions in turn.
Because these uncertainties affect the full function $b(k)$, 
they can lead to uncertainties in the shape and amplitude of $P(k)$.
In practice, we will restrict our attention to a range of $k$
for which we expect the
systematic uncertainties in shape to be small compared to the statistical 
uncertainties arising from the finite sample size.
Our principal concern will therefore be the uncertainty
in the overall scaling of $b(k)$, which we will characterize by the ratio
$b_{\rm fid}/b$, where $b$ is the average value of $b(k)$ and 
$b_{\rm fid}$ is the average value of $b(k)$ for our
fiducial normalizing simulations, which have the parameters defined
in \S 4.2 and the matter power spectrum amplitude
$\sigma_8(z=2.72)=0.23$.  The inferred amplitude of $P(k)$ 
is directly proportional to $(b_{\rm fid}/b)^2$.
(Note that higher $b$ implies a lower amplitude, since
we start from the observed flux power spectrum and infer the matter
power spectrum from it.)

We have already concluded in \S 4.3 that the {\it statistical} uncertainty
in $b_{\rm fid}/b$, resulting from the finite size of our data sample, 
is $\pm 6\%$ at the $1\sigma$ level.
We will argue in this section that the main systematic uncertainties
in $b_{\rm fid}/b$ come from the uncertainty in the true value of 
$\taueff$ and the uncertainties in the true values of $T_0$ and $\alpha$.
We will therefore devote most of our effort to quantifying these 
uncertainties and
to showing how our $P(k)$ results should be scaled as new, more precise 
determinations of these parameters become available.

One powerful test for systematic errors is to see whether 
the derived $P(k)$ scales with redshift as it should according to
gravitational instability theory.  
This test is especially important as a way of checking for other
possible sources of fluctuations in the \lya\ forest.
We will discuss this test in \S\ref{sec:evolution}.

\subsection{Cosmological Model}
\label{sec:sysmodel}

The most important assumption underlying our $P(k)$ recovery
method is that structure in the universe formed by gravitational
instability from Gaussian primordial fluctuations.
Gaussian fluctuations are predicted by most versions of inflation,
and there is empirical support for the Gaussian assumption from
many quarters, including microwave background anisotropy statistics
(e.g., Kogut et al.\ 1996), moments and topology
of the galaxy density field (e.g., Bouchet et al.\ 1993; Gazta\~naga 1994;
Canavezes et al.\ 1998), and agreement between the predicted and
observed 1-point flux distribution of the \lya\ forest
(e.g., Rauch \etal 1997; Weinberg \etal 1999b; M00).

Given the Gaussian assumption, the important features of the cosmological
model that we assume for determining $b(k)$ are the shape and amplitude
of $P(k)$ at $z=2.72$, in $\kms$ units.
The amplitude is constrained by matching our 
flux power spectrum data.  The shape of the LCDM $P(k)$ is consistent
with our data and with other data (e.g., Peacock \& Dodds 1994),
and because the quantity we compute from the normalizing simulations is
$b^2(k) = P_F(k)/P(k)$ rather than $P_F(k)$ itself, the results are
not very sensitive to the assumed shape anyway.
The uncertainty in $b(k)$ associated with our adoption of
the LCDM model for the normalizing simulations should therefore be negligible,
unless there is significant non-Gaussianity of the primordial fluctuations
that has somehow escaped detection in the studies cited above.

There is one significant caveat to this statement.  
If the primordial (linear) matter power spectrum is strongly suppressed
on some scale shorter than the scale of non-linearity
(where $\Delta^2(k) \approx 1$, roughly $k_{\rm nl} \sim 0.02 \invkms$
in our fiducial case), then non-linear transfer of power from
large scales to small scales will dominate the growth of $P_F(k)$
on these scales (White \& Croft 2000).  
For $k>k_{\rm nl}$, therefore, our derived $b(k)$ depends on our 
assumption that the matter power spectrum varies smoothly with scale,
as it does in LCDM and other standard variants of the inflationary
CDM scenario.  Models in which the small scale power is truncated
because of warm dark matter (e.g., Sommer-Larsen \& Dolgov 2001)
or broken-scale invariance in inflation (e.g., Kamionkowski \& Liddle 2000)
should be tested directly with numerical simulations against the
measured $P_F(k)$, as in White \& Croft (2000) and Narayanan et al.\ (2000).
For standard variants of inflationary CDM, including models with
CDM and an admixture of massive neutrinos (Croft, Hu, \& Dav\'e 1999a),
one can compare the predicted linear matter power spectrum to our
derived linear matter power spectrum.

\subsection{Simulations}
\label{sec:simsys}

To compute $b(k)$ we use the N-body+FGPA method described in 
\S\ref{sec:intro} and \S\ref{sec:normsim}, rather than full 
hydrodynamic simulations.  Evidence that this approximation is 
adequate for our purposes is provided by CWKH and by Figure~\ref{sphtest}.
The main failing of the N-body approximation is the absence of
shock heating, which in full hydrodynamic simulations pushes some
gas off of the temperature-density relation, reducing its \lya\ optical depth.
However, for the flux power spectrum at these redshifts, shock heating
has little effect --- it occurs in dense regions with small volume 
filling factor, and if the gas is only moderately heated then the
absorption remains saturated even at this higher temperature.
The N-body approximation also ignores the effects
of gas pressure, but these should be unimportant on the scales where
we attempt to recover $P(k)$, though they may become important at
higher $k$.  

Comparison of the two dashed lines in Figure~\ref{sphtest} suggests
that the finite numerical resolution of our normalizing simulations
may start to have a noticeable effect at $k \ga 0.03 \invkms$.
To determine the overall normalization of $b(k)$, we use only
data points with $k<0.03\invkms$, though we continue our calculation
of $P(k)$ to somewhat smaller scales, $k=0.05\invkms$, where
finite simulation resolution could be having a small effect.

As explained in \S\ref{sec:normsim}, we run our simulations with
an Einstein-de Sitter background cosmology for convenience, though
we adopt an LCDM power spectrum and scale comoving $\hmpc$ to
$\kms$ assuming LCDM parameters.  Our use of the EdS background
means that redshift-space distortion effects are computed assuming
$\Omega_m=1$, but since $\Omega_m$ is very close to one in all
cosmological models at high $z$ (and $\Omega_m^{0.6}$ is even closer),
this makes negligible difference to the results
(for further discussion and numerical tests, see CWKH).

We have used ten independent simulation volumes in our estimate
of $b(k)$ (Figure~\ref{redbias}), and there is a small contribution
to the statistical uncertainty in $P(k)$ because of this finite
number of simulations.  We estimate this contribution from the
dispersion in $b(k)$ among the ten realizations and add it in
quadrature to the individual $P(k)$ error bars that result from the finite
number of observed spectra (estimated by the jackknife method
as described in \S\ref{sec:fluxpk}).  This contribution increases the
$1\sigma$ error bars by $\sim 1\%$ on the largest scales
(where the statistical uncertainties are already large) and by
$\sim 15\%$ on the smallest scales at which we calculated $P(k)$,
$0.05 \invkms$.

It is worth reiterating that we use the N-body+FGPA method to 
compute $b(k)$ because it allows us to carry out many large
volume simulations with different cosmological parameters and
IGM parameters.  Large volumes are needed for accurate computation
of $b(k)$, and large numbers of simulations are needed to reduce
the variance in the numerical estimate of $b(k)$.
Our tests imply that the systematic uncertainties introduced by
the use of this approximation and by our finite numerical resolution
are small compared to the other uncertainties in $P(k)$ over the
range of scales where we attempt to derive it (including the uncertainties
that we discuss below).  However, in the future it might become
computationally practical to carry out full hydrodynamic simulations 
in the necessary numbers.  With such simulations, it might be possible
to reduce the systematic uncertainties in $b(k)$ at small scales,
allowing recovery of $P(k)$ over a wider dynamic range.

Recently Gnedin \& Hamilton (2002) have independently run normalizing
simulations to infer the matter power spectrum from our measurement of
the \lya\ flux power spectrum.  Using PM simulations with higher mass
resolution and a smaller volume, and an independent spectral extraction
code, they find nearly identical results when they assume the same cosmology.
They also show that the inferred $b(k)$ is indeed insensitive to the
assumed cosmological model and initial $P(k)$, except for a moderate
increase in the inferred amplitude in open (zero-$\Lambda$) models with
$\Omega_m \la 0.3$, for which $\Omega_m$ is still significantly below
unity at $z\sim 3$.  The good agreement between these independent 
calculations increases our confidence that any systematic errors 
associated with the normalizing simulations are fairly small.
Gnedin \& Hamilton (2002) also show that peculiar velocities induce
correlations in the line-of-sight power spectrum at neighboring $k$
values, which should be taken into account in a full maximum likelihood 
analysis that uses our results.

\subsection{Mean optical depth}
\label{sec:taueff}

As discussed in \S\ref{sec:normsim}, an important input to our
normalizing simulations is the value of the effective mean optical
depth $\taueff$.  This observational constraint allows us to 
fix the parameter $A$ of equation~(\ref{eqn:tau}) for a given
simulation output, which in turn determines the relation between
the mass density field and the \lya\ optical depth.
Although the measurement of the flux power spectrum is 
not sensitive to continuum fitting uncertainties
(see \S\ref{sec:fluxpksystematics}), the measurement of $\taueff$
is very sensitive to continuum determination because a significant
fraction of the mean opacity arises in long stretches of weak
absorption that are close to the unabsorbed continuum.
In principle, the systematic biases of a local continuum fitting
method on $\taueff$ can be calibrated using numerical simulations
(see, e.g., Rauch et al.\ 1997), but it is difficult to do this 
accurately because the simulation boxes are smaller than the scales
over which continua are fitted.  In this paper, therefore, we do
not attempt to determine $\taueff$ from our data but adopt the
value found by PRS (see \S\ref{sec:normsim}), which we check 
below using our filling factor measurements.
An accurate determination of $\taueff$ from a large sample of
HIRES data will be the subject of a future paper.

\begin{figure}[t]
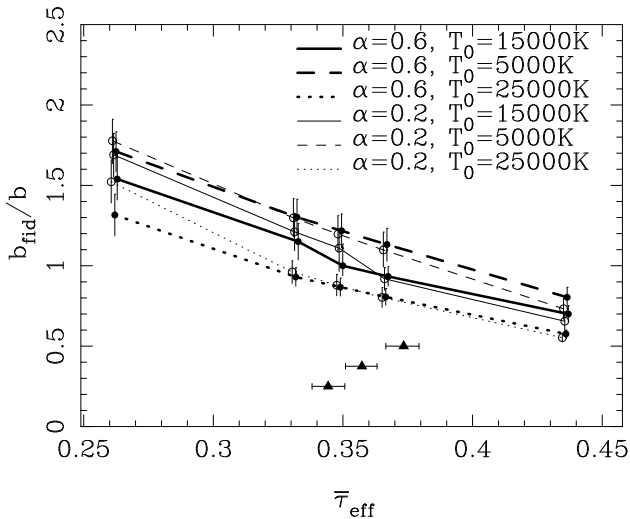

\PSbox{taueffvary.ps angle=-90 voffset=255 hoffset=-45 vscale=45 hscale=45}
{3.5in}{3.1in} 
\caption[taueffvary]{
Influence of the assumed mean optical depth, $\taueff$, and the
assumed parameters of the temperature-density relation, $T_0$ and $\alpha$,
on the inferred amplitude of the matter power spectrum
(which is proportional to $[b_{\rm fid}/b]^{2}$).
The thick solid line with filled circles shows the dependence
of $b_{\rm fid}/b$ on the value of $\taueff$ for our 
fiducial parameters, $T_0=$15000 K, $\alpha=0.6$.
For higher $\taueff$, less matter clustering is required to
match the observed flux power spectrum.
Thick dashed and dotted lines show the effect of changing
$T_0$ to 5000 K and 25000 K, respectively, with $\alpha=0.6$.
Thin lines with open points show results for $\alpha=0.2$ with
the corresponding $T_0$ values.
Points have been slightly displaced in the $x$-direction, for clarity.
Triangles indicate the values of $\taueff$ required to match 
our measured filling factor (see Fig.~\ref{fillfac}) for three
different SPH simulations of the LCDM model, each occupying
an $11.11\hmpc$ box, with horizontal error
bars coming from the $1\sigma$ error bars on the filling factor.
The top two triangles represent two $64^3$-particle simulations with 
different phases, and the lowest represents a $128^3$ simulation with the 
same phases as the lower $64^3$ simulation.
The $y$-axis position of these points is arbitrary.
\label{taueffvary}
}
\end{figure}

We have investigated the dependence of the inferred $P(k)$ amplitude
on $\taueff$ by carrying out our normalization procedure for different
values of $\taueff$.  In Figure~\ref{taueffvary}, the thick solid 
line shows the dependence of $b_{\rm fid}/b$ on $\taueff$ for our
fiducial parameters of the temperature-density relation; note that
we plot the quantity that is proportional to the rms mass fluctuation
amplitude, and hence to $\sqrt{P(k)}$.
Error bars show the $1\sigma$ uncertainty in the mean caused
by our finite number of simulations.
Our fiducial choice of $\taueff=0.349$ at $z=2.72$, based on PRS,
yields $b_{\rm fid}/b=1$ by definition.
A higher $\taueff$ requires a higher value of $A$ in equation~(\ref{eqn:tau}),
which in turn increases the bias between flux and mass.
The inferred power spectrum amplitude is therefore lower when $\taueff$
is higher.
Our results for the fiducial temperature-density relation are 
reasonably well described by the formula
\begin{equation}
{b_{\rm fid} \over b} = \left({\taueff \over 0.349}\right)^{C_\tau}
\label{eqn:ctau}
\end{equation}
with $C_{\tau}=-1.7$, which is accurate to within the measurement
uncertainty of the simulations over the range 
$\taueff=0.26-0.44$ (the full range shown in Figure~\ref{taueffvary}).
Equation~(\ref{eqn:ctau}) can be used to scale the amplitude of
our inferred matter power spectrum in light of new measurements
of $\taueff$, at least if they are not very far from the PRS value.

PRS give a fitting formula for $\taueff(z)$, and their quoted 
uncertainties in the fit parameters imply a $1\sigma$ uncertainty
of approximately $\pm 5\%$ in $\taueff$ at $z=2.72$.
The three central points in Figure~\ref{taueffvary} cover
this range of $\taueff$.  
A 5\% uncertainty in $\taueff$
corresponds to a $\pm 9\%$ uncertainty in the matter
fluctuation amplitude (see equation~\ref{eqn:ctau}).
We therefore adopt $\pm 9\%$ as the contribution of the observational
uncertainty in $\taueff$ to the $1\sigma$ error bar 
on the inferred matter fluctuation amplitude.

In addition to their own internal error estimate, two lines of argument
suggest that the PRS determination of $\taueff$ is not too far from
the true value.  The first is the independent measurement of $\taueff$
by Rauch et al.\ (1997) and M00 from Keck HIRES spectra.
(There are eight spectra in the M00 sample, seven of which are also
in the Rauch et al. sample.)
M00 report $\taueff=0.380\pm 9\%$ at $z=3.0$ (the Rauch et al.\ value
is very similar), with the error bar estimated by bootstrap analysis
of the data.  The PRS formula implies $\taueff=0.448$ at $z=3$,
18\% higher than M00's central value.  The two measurements differ
by slightly more than their estimated $1\sigma$ uncertainties,
but the local continuum fitting approach used for the HIRES data
tends to systematically depress $\taueff$, and correcting for this
effect based on simulated spectra brings the two estimates 
closer together (Rauch et al.\ 1997).
The PRS approach of extrapolating the quasar continuum from redward
of the \lya\ emission line does not suffer from this bias, though
it has systematic uncertainties of its own.
Recently Bernardi et al.\ (2002) have applied an improved version of
the PRS technique to a sample of $\sim 1000$ quasar spectra from the
Sloan Digital Sky Survey, and they find excellent agreement with PRS
(and a much smaller statistical error)
except in a narrow redshift range $3.0 < z < 3.2$, where they find
a $\sim 10\%$ local dip in $\taueff$.  
The good agreement of the PRS and Bernardi et al.\ (2002) determinations
suggests that the error bar we associate with the uncertainty in $\taueff$
may be overly conservative, but one should be cautious until the
difference between the continuum extrapolation approach and the
local continuum fitting approach is completely understood.

The second line of argument is based on the filling factor (FF)
measurement described in \S\ref{sec:ff}.
With the help of simulations, we can ask what what value of
$\taueff$ is compatible with our measurement, a filling factor
of $0.205 \pm 0.004$ for regions of the spectra with $F<0.5$
at $z=2.72$.
The three triangles in Figure~\ref{taueffvary} represent the
values of $\taueff$ for which the three TreeSPH simulations
described previously (see Figure~\ref{sphtest} and the associated
discussion) reproduce the measured FF.
Horizontal error bars represent the $1\sigma$ uncertainty
associated with the $\pm 0.004$ uncertainty in FF.
The upper two points represent the two $64^3$-particle simulations; 
the 4\% difference between them is the effect of cosmic variance
for the $11.11\hmpc$ simulation volume.  The lower point represents
the $128^3$-particle simulation, which has the same phases as the
$64^3$ simulation represented just above it; the factor of eight
increase in particle number reduces $\taueff$ by less than 3\%.
Based on these results, we can conclude that our adopted value
of $\taueff=0.349$ is compatible with our measured FF, a consistency
that would be lost if we changed $\taueff$ by a substantial factor.
The relation between $\taueff$ and the filling factor depends
on the matter power spectrum itself (and on the assumption of primordial
Gaussianity), but the flux power spectra of the TreeSPH simulations
are reasonably close to our measured flux power spectrum
(Figure~\ref{sphtest}), implying that the model adopted in the
simulations should be adequate for calibrating this relation.

As this discussion illustrates, it might be possible to use the
FF measurement itself in place of $\taueff$ when determining the
value of $A$ in the normalizing simulations.  This approach would
remove the dependence of the inferred $P(k)$ amplitude on a
quantity ($\taueff$) that is sensitive to continuum fitting
uncertainties.  We have not followed this route here because 
the computation of FF might be sensitive to the limited resolution
of our normalizing simulations, and because we have not tested
the adequacy of the N-body+FGPA approximation itself for this purpose.
We also have not investigated the influence of $T_0$ and $\alpha$
on the relation between FF and $\taueff$.
However, an approach that uses the filling factor instead of $\taueff$
might become useful in the future, as higher resolution hydrodynamic
simulations become computationally easier.

\subsection{The temperature-density relation}
\label{sec:igm}

There have been several recent attempts to determine the parameters
of the IGM temperature-density relation (a.k.a. ``equation of state'')
by comparing the predicted and observed widths of \lya\ forest absorption
features.  Simulations with standard photoionization heating and
a high reionization redshift do not match the observed line width
distribution (e.g., Theuns et al.\ 1999; Bryan et al.\ 1999),
and several mechanisms have been proposed to resolve this discrepancy
(e.g., Madau \& Efstathiou 1999; Nath, Sethi \&  Shchekinov 1999; 
Abel and Haehnelt 1999). 
Although the widths of most features are dominated by
Hubble flow rather than thermal motions (Weinberg et al.\ 1997a),
the narrowest features occur at velocity caustics and have widths
determined by thermal broadening, so the cutoff in the distribution
of line widths as a function of column density provides a diagnostic
for the temperature-density parameters (Bryan \& Machacek 2000;
Schaye et al. 1999).  Ricotti et al.\ (2000), Schaye et al.\ (2000),
and McDonald et al.\ (2001) have used variations on this theme
to estimate values of $T_0$ and $\alpha$.  
At $z=3$, McDonald et al.\ (2001) find $\alpha \approx 0.3 \pm 0.3$
and $T_0 \approx 18,000$ K (extrapolated from their quoted estimate
of $T$ at overdensity 1.4).  Schaye et al.\ (2000) find slightly
lower temperature ($T_0 \approx 16,000$ K at $z=2.7$) and
Ricotti et al.\ (2000) somewhat higher ($T_0 \approx 25,000$ K at $z=2.75$),
with similar best fit values of $\alpha$.
Zaldarriaga, Hui, \& Tegmark (2001a) obtain results similar to 
those of McDonald et al.\ (2001) with a different technique,
based on the flux power spectrum.
The statistical and systematic uncertainties in these determinations
are still rather large, so we must assess the influence of these
uncertainties on our $P(k)$ determination.
While the relation between the matter and flux power spectra is not strongly
sensitive to $T_0$ or $\alpha$ (see CWKH), there is enough
dependence to influence the inferred $P(k)$ at the level of
precision achievable with our data set.

The influence of $T_0$ and $\alpha$ on $b(k)$ is subtle
because we always adjust the constant $A$ in the FGPA
(eq.~\ref{eqn:tau}) so that the normalizing simulations
match the adopted $\taueff$.
The direct effect of thermal broadening on $P_F(k)$ is 
confined to small scales
(the thermal broadening width $b_{\rm th}$ at 15,000 K is $16 \kms$). 
However, by changing the structure of the flux distribution on small
scales, thermal broadening can alter the value of $A$ required for a given 
matter distribution.
The parameter $\alpha$ has a direct impact on the flux--density
relation in the FGPA (eq.~\ref{eqn:tau}), but this effect
is again mediated by the requirement of matching $\taueff$.

The various lines in Figure~\ref{taueffvary} show the relation
between $b_{\rm fid}/b$ and $\taueff$ for different temperature-density
parameters.  The thick solid line (discussed in \S\ref{sec:taueff})
corresponds to our fiducial choice: $T_0 = 15,000$ K, motivated
roughly by the observational results cited above, and $\alpha=0.6$,
the asymptotic slope that should be approached long after reionization
(Hui \& Gnedin 1997).  The thick dashed and dotted lines have
$T_0=5000$ K and 25,000 K, respectively, with $\alpha=0.6$.
Thin lines correspond to $\alpha=0.2$, with the same temperatures.
Comparison of these lines shows that the inferred $P(k)$ amplitude
is lower for a hotter IGM temperature or a steeper 
temperature-density relation (higher $\alpha$).

We can use the results in Figure~\ref{taueffvary} to parameterize
the dependence of the mass fluctuation amplitude on $T_0$ and $\alpha$,
as we did for the dependence on $\taueff$ in equation~(\ref{eqn:ctau}).
We use a slightly different form in order to ensure 
reasonable behaviour for values of $T_0$ and $\alpha$ close to zero.
For $\taueff=0.349$ and $\alpha=0.6$, the $T_0$ dependence is roughly
\begin{equation}
{b_{\rm fid} \over b} = \left({1+T_0/15000\;\rm{K} \over 2}\right)^{C_T}
\label{eqn:ct0}
\end{equation}
with $C_T \approx -0.5$. 
We measure no statistically significant dependence of our results on
$\alpha$.
 $C_T$ is  
less well determined than $C_\tau$ because we have only a sparse
grid of $T_0$ and $\alpha$ values.  
Since $C_\tau=-1.7$, the power spectrum normalization is 
obviously much more sensitive to $\taueff$ than to $T_0$ or $\alpha$.
However, the fractional uncertainty in $T_0$  is still
larger than the fractional uncertainty in $\taueff$,
so it still makes a significant contribution to the overall
uncertainty in the amplitude of $P(k)$.

\begin{figure}[t]
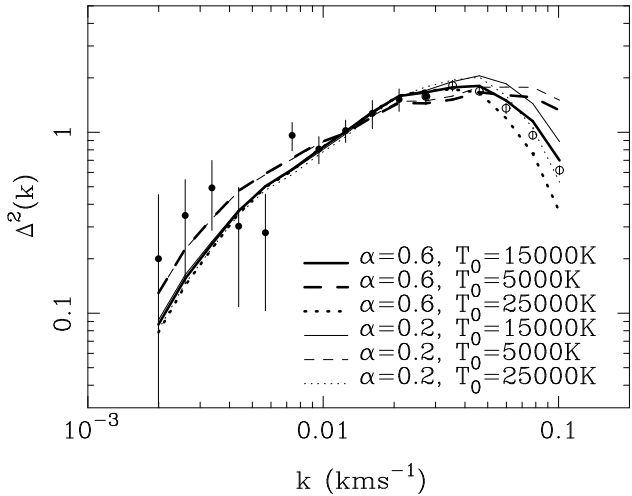

\PSbox{tempvary.ps angle=-90 voffset=250 hoffset=-45 vscale=45 hscale=45}
{3.5in}{3.1in} 
\caption[tempvary]{
Influence of the IGM temperature-density parameters on the
shape of the flux power spectrum in the normalizing simulations.
The six lines have the same meaning as those in Figure~\ref{taueffvary}:
solid, dashed, and dotted for $T_0=15,000$ K, 5000 K, and 25,000 K, 
respectively; thick for $\alpha=0.6$, thin for $\alpha=0.2$.
The flux power spectrum of the fiducial observational sample
is shown by points.
The normalization of the matter power spectrum is chosen separately in each 
case to yield the best fit to the points with $k \leq 0.03 \invkms$
(solid points).
\label{tempvary}
}
\end{figure}

To assign the uncertainty in $b_{\rm fid}/b$, we will assume that
the range of parameter values considered in Figure~\ref{taueffvary},
$T_0=5000-25,000$ K, $\alpha = 0.2 - 0.6$, represents the 95\% 
confidence range on the true values, based on the papers cited
above, on figure 5 of White \& Croft (2000), and on 
Figure~\ref{tempvary} discussed below.
This assumption is probably overly conservative with respect to $T_0$,
but it perhaps underestimates the viable range of $\alpha$, which is a
more difficult parameter to pin down observationally.
{}From the points in Figure~\ref{taueffvary} at $\taueff$ close to $0.349$,
we then estimate that the resulting uncertainty in $b_{\rm fid}/b$ is
$+10\%, -7\%$, at the $1 \sigma$ ($68\%$ confidence) level.

The values of $T_0$ and $\alpha$ also affect the shape of the flux
power spectrum (for a given matter power spectrum) on small scales.
Figure~\ref{tempvary} shows the flux power spectrum $\deltaf$ of
the normalizing simulations for each of the six 
parameter combinations illustrated previously in Figure~\ref{taueffvary}.
In each case we choose the overall normalization to get the best fit
to points with $k \leq 0.03 \invkms$, interpolating between the
simulation outputs.  On large scales, the shape of $\deltaf$ is
 almost independent of $T_0$ and $\alpha$
 (at least with respect to the large obseravtional
error bars).  For $k \ga 0.05 \invkms$, however,
there is a strong dependence, with hotter models having less small
scale power as expected.  These differences in the shape of $\deltaf$
would translate into scale-dependent changes in the function $b(k)$.
We therefore restrict our estimate of the matter power spectrum
to $k < 0.05 \invkms$, so that the inferred shape is insensitive
to the uncertainties in $T_0$ and $\alpha$.

It is clear from Figure~\ref{tempvary} that our fiducial choice of
parameters yields the best match to the observed shape of $\deltaf$
on small scales.  We do not regard this agreement as a solid determination
of $T_0$ (which has more influence on the shape than $\alpha$), since we are 
relying on moderate resolution
N-body simulations rather than high resolution hydrodynamic
simulations and have not explored the tradeoffs between the IGM
parameters and the assumed shape of the matter power spectrum on 
these scales (see White \& Croft 2000).
However, the result is reassuringly consistent with the estimates
cited above.  On the basis of this match to the observed $\deltaf$,
one could argue that we have overestimated the uncertainty in the
$P(k)$ normalization associated with $T_0$ and $\alpha$ --- it may not
be the temperature-density relation itself that matters but only
the sum total of all physical and numerical effects on the small
scale power in the flux, which then determines the value of $A$ that
is required to match the observed $\taueff$ for a given mass distribution.
We will adopt the more conservative position described above,
for which the uncertainty contributed by the temperature-density
relation is similar to that contributed by the finite sample size
and the uncertainty in $\taueff$.  Setting the temperature-density
contribution to zero would reduce our error bar on $b_{\rm fid}/b$
by about 20\%.

By using the N-body + FGPA method for our normalizing simulations,
we implicitly assume that uncertainties in $T_0$ and $\alpha$
influence the inferred matter $P(k)$ only through their effects 
on the flux-overdensity relation~(\ref{eqn:tau}) and through thermal
broadening.  Zaldarriaga et al.\ (2001; hereafter ZHT), who compare results
from a grid of simulations to the 1-d flux power spectrum measurements of M00, 
suggest that gas pressure effects may introduce additional uncertainty in
the inferred shape of $P(k)$ by depressing the flux power spectrum
on small scales.  We believe that such effects are unlikely to be
significant in our analysis, for several reasons.  First, we infer 
the matter $P(k)$ only at $k \leq 0.05\invkms$, while much of the
weight in the ZHT likelihood analysis comes from smaller scale data points,
with $0.05 \invkms \la k \la 0.15 \invkms$.  
Second, our larger number of spectra yields better determination of
the $P(k)$ shape on large scales, where any pressure effects are negligible, 
and we see no change in the inferred $P(k)$ slope as we go from large
scales to small scales (see Figure~\ref{models1} below).
Finally, we suspect that the ZHT approach exaggerates the possible
effects of gas pressure because it parameterizes these effects as a Gaussian
smoothing of arbitrary strength, constrained only by the shape of the
flux power spectrum itself.  Our SPH simulations and the simulations
of Meiksin \& White (2000) using the hydro-PM technique (Gnedin \& Hui 1998)
show no signs of significant gas pressure effects at $k < 0.05 \invkms$
(however, these simulations do not examine scenarios with late heating from
HeII reionization).  More generally, we suspect that gas pressure effects
on the flux power spectrum are always small relative to the effects of
thermal broadening and therefore add little additional uncertainty to
the matter power spectrum. Experiments by McDonald (2002),
using the hydro-PM technique support this point of view.
McDonald (2002) finds that for an uncertainty of 4000K in temperature
(slightly less than our 5000K 1 $\sigma$ error bar), the uncertainty
in $\deltaf$ is less than $2\%$ on the large scales we consider.
Further investigation with hydro-PM and full hydrodynamic simulations
will be required to settle the issue entirely.  For now, we assign
no additional uncertainty to our results associated with gas pressure
effects.

\subsection{UV background and temperature fluctuations}
\label{sec:uvbg}

Our normalizing simulations assume that the UV background (UVBG)
is spatially uniform, i.e., that the photoionization rate $\Gamma$ in 
equation~(\ref{eqn:tau}) is constant.  
Fluctuations in the UVBG could in principle be an additional source
of structure in the \lya\ forest, beyond that due to density fluctuations. 
UVBG fluctuations and their influence on the \lya\ forest have been 
studied theoretically by Zuo (1992), Fardal \& Shull (1993), 
and Haardt \& Madau (1996), among others. 

CWPHK calculated the impact of UVBG inhomogeneities
on the flux power spectrum $\deltaf$, using a simplified model and
the assumption that the background is
produced by a population of clustered quasars.
They found (as expected based on earlier work) that the fluctuations
induced in the \lya\ forest are small in amplitude and occur mainly
on large scales.  
For example, UVBG fluctuations contribute $< 1\%$ of the
signal in $\deltaf$ at $z=2.5$ for $k=2\times 10^{-3} \invkms$. 
The reason for this small impact is simple: through most of the 
volume occupied by the IGM, the photoionization rate reflects
the summed contribution of 
many distant quasars rather than a few nearby quasars.  
The UVBG fluctuations from more numerous
sources (e.g., galaxies) would be smaller still (see the
analytic argument in CWPHK, following Kovner \& Rees 1989).
Based on these results, we expect UVBG fluctuations to have
a negligible impact on our $P(k)$ measurement at the fiducial
redshift $z=2.72$, at least on the scales that we are able to probe with this
data set.

At high redshifts, the universe becomes more optically thick, so that 
the effective number of sources seen by a given point in space is smaller
and fluctuations increase. Simulations of stellar reionization
of hydrogen by Gnedin (2000) make {\it ab initio} predictions for the UVBG 
fluctuations expected in this case, at least down to $z=4$.
Gnedin \& Hamilton (2002) report that the inhomogeneity of the background
in this simulation changes the \lya\ flux power spectrum by less than 1\%,
so it appears that UVBG fluctuations are unimportant for this purpose
even at $z=4$.  However, more complete theoretical calculations that
include the effects of inhomogeneous HeII reionization at lower redshift 
are still desirable.

Another possible source of structure in the \lya\ forest is spatially
coherent variation in the temperature-density relation.
Fluctuations in $T_0$ and $\alpha$ could arise during inhomogeneous
heating of the IGM, for example during HeII reionization.
They would be damped on a Hubble timescale by the 
competition between adiabatic and photoionization heating, which establishes
the temperature-density relation itself
(Hui \& Gnedin 1997). However, if the IGM is heated at relatively
low redshift then significant fluctuations could potentially
survive to the epoch probed by our observations.
We do not know of any direct theoretical predictions of such
temperature fluctuations.

Several lines of empirical argument suggest that UVBG fluctuations
and temperature fluctuations have much less impact on the \lya\ forest
than the density fluctuations that we are trying to measure.
The first 
is the approximate consistency between the inferred matter power spectrum
and other statistical properties of the forest like the 
flux distribution function (Rauch et al.\ 1997; Weinberg et al.\ 1999b; M00),
implying reasonable agreement with the model of density fluctuations
induced by gravitational instability from Gaussian initial conditions.
The second is the constancy of the shape of the flux correlation function
(Figure~\ref{fluxxi}) and flux power spectrum (Figure~\ref{fluxpk}) over
the range of redshifts probed by our five data subsamples, 
$\langle z \rangle = 3.51$ to $\langle z \rangle = 2.13$.
A nearly constant shape is expected if these statistics reflect
structure in the underlying mass distribution, but it would be
quite surprising if UVBG fluctuations or temperature fluctuations
were having a significant impact, since the quasar space density
and the mean opacity to photoionizing radiation vary substantially
over this interval, and temperature fluctuations would be expected
to evolve if they were present.
A third argument is presented by Zaldarriaga, Seljak, \& Hui (2001b),
who have devised a statistical diagnostic for distinguishing
gravitationally induced power from non-gravitational fluctuations.
Applying their method to a spectrum of Q1422+231, they conclude
that non-gravitational fluctuations contribute less than 10\% of
the observed power.
Finally, we will show in \S\ref{sec:evolution} that the evolution
of the amplitude of $\deltaf$ is consistent with the predictions
of gravitational growth of mass fluctuations.  In contrast,
we would expect the contribution of other sources 
to decrease towards lower redshift rather than increase, so they 
would spoil this agreement if they were significant in any 
of our subsamples.
Given the theoretical expectations and these empirical arguments,
we will not assign any additional uncertainty to our $P(k)$ 
estimates to account for UVBG or temperature fluctuations.

\subsection{Metal lines and damping wings}

Metal lines in the \lya\ forest region of spectra are rare
compared to \lya\ lines, so their filling 
factor in spectra is small. Their effect on a measurement
of $\deltaf$ can be checked in two ways, by attempting 
to remove them or by adding simulated lines to spectra. 
The first of these tests was carried out by M00, who showed that
any changes in $\deltaf$ are confined to scales $k > 0.1 \invkms$.
This is as we might expect, given that the metal lines are
rare, sharp features in the absorption spectrum.
Our data are not useful for measurements on these scales because
the power there is dominated by shot noise (see Figure~\ref{noisepk}),
so we do not need to take this effect into account.
CWPHK carried out the second test, adding simulated CIV lines 
to their observed spectra.  They found a small effect
on relatively large scales, at wavenumbers related to harmonics
of the CIV doublet spacing.
For reasonable assumptions about the metallicity of the IGM,
the magnitude of the effect was $\sim 1 \%$, small
enough to be hidden in the cosmic variance noise of the M00 analysis,
and not likely to be a significant factor here.

Damped \lya\ systems are another potential problem, since the
FGPA (eq.~\ref{eqn:tau}) does not allow for the effect of damping wings.
However, the filling factor of damped \lya\ systems is again very
small ($\sim 0.1$ systems per unit redshift), so they are unlikely
to make a significant contribution to $\deltaf$.
Direct empirical evidence for this point comes from CWPHK, who show
that including or excluding damped absorption regions makes little
difference to $\deltaf$ even in a sample {\it selected} to have
damped \lya\ absorbers in each spectrum.

Constancy of the measured shape of $\xi_F(r)$ and $\deltaf$ 
(see Figures~\ref{fluxxi} and~\ref{fluxpk}) provides further
evidence that metal lines and damped \lya\ systems do not alter
the flux power spectrum, since the opacity 
of metal lines and damped systems relative to the low column
density forest varies substantially with redshift and they would
add power primarily on small scales.

\section{Matter power spectrum results}
\label{sec:masspk}

\subsection{The linear matter power spectrum at z=2.72}
\label{sec:masspkfid}

Table~\ref{pkmassfid} presents this paper's primary result,
the linear matter power spectrum $P(k)$ at $\langle z \rangle = 2.72$
estimated from our fiducial data sample.
These values of $P(k)$ are obtained by dividing the values of
$P_F(k)$ in Table~\ref{pktabfid} by (the square of) our fiducial estimate
of $b(k)$ shown by the dotted line in Figure~\ref{redbias}.
We have used only points with $k < 0.05 \kms$, so that the
inferred shape should be insensitive to the assumed values of
the IGM temperature parameters (see \S\ref{sec:igm} and Figure~\ref{tempvary}).
The error bars on individual points represent the same fractional
uncertainty as the error bars on $P_F(k)$ reported in Table~\ref{pktabfid},
except that we have added the small additional contribution from
statistical errors in $b(k)$ caused by the finite number of 
normalizing simulations (see \S\ref{sec:simsys}).
The $P_F(k)$ error bars were derived by applying
a jackknife estimator to 50 separate subsamples drawn
from the fiducial data set, as described in \S\ref{sec:fluxpk}.
The covariance matrix of these errors 
is approximately diagonal on these scales (see Figure~\ref{covhires}), 
and the estimates of off-diagonal terms are noisy, so we do not quote them.
A plot comparing the measured $P(k)$ to CDM model predictions
appears in Figure~\ref{models1}, which
we discuss in \S\ref{sec:implications} below.

\begin{table}[ht]
\centering
\caption[pkmassfid]{\label{pkmassfid} 
The linear matter $P(k)$, 
for the fiducial sample ($\langle z \rangle =2.72$).}
\begin{tabular}{cc}
\hline &\\
k    & $P(k)$    \\
 $(\kms)^{-1}$   & $(\kms)^{-3}$ \\
\hline &\\
 0.00199  & $(4.84 \pm 6.20) \times 10^{8}$ \\
 0.00259 & $(3.14 \pm 1.85 ) \times 10^{8}$ \\
 0.00337 & $(1.67 \pm 0.72 ) \times 10^{8}$ \\
 0.00437 & $(3.90 \pm 2.63 ) \times 10^{7}$ \\
 0.00568& $(1.46 \pm 0.96 ) \times 10^{7}$ \\
 0.00738 &  $(2.28 \pm 0.45 ) \times 10^{7}$ \\
 0.00958 &  $(8.38 \pm  1.58) \times 10^{6}$ \\
 0.0124 &  $(4.64 \pm 0.74 ) \times 10^{6}$ \\
 0.0162&  $(2.48 \pm 0.47 ) \times 10^{6}$ \\
 0.0210&  $(1.35 \pm 0.20 ) \times 10^{6}$ \\
 0.0272&  $(6.86 \pm 0.64 ) \times 10^{5}$ \\
 0.0355&  $(3.83 \pm 0.28 ) \times 10^{5}$ \\
 0.0461&  $(1.90 \pm 0.11 ) \times 10^{5}$ \\
\hline &\\
\end{tabular}
\\
Note: We give the $1 \sigma$ error
on $P(k)$. An additional error should also be assigned to the normalization
of all points, which is $+29\%, -25\%$ in $P(k)$ ($1 \sigma$,
see text). 
\end{table}

While the uncertainty in the shape of $b(k)$ is small over the range
of scales in Table~\ref{pkmassfid}, there is uncertainty in the
amplitude of $b(k)$ contributed by the various sources discussed
in \S\ref{sec:fluxpk} and \S\ref{sec:systematics}.
This normalization error could be treated as a set of off-diagonal
terms in the $P(k)$ error covariance matrix, but we think it is
simpler to regard it as an overall multiplicative uncertainty in the
amplitude of $P(k)$.  The $1\sigma$ uncertainty in the average
value of $b(k)$ contributed by the finite size of the data set
(cosmic variance) is 6\% (\S\ref{sec:normalization}).  The $1\sigma$ 
uncertainty contributed by the error bar on $\taueff$ is 9\%
(\S\ref{sec:taueff}).   The $1\sigma$ uncertainty contributed
by the uncertainty in the IGM temperature parameters is 8.5\%
(\S\ref{sec:igm}).  Summing these error contributions in 
quadrature leads to a $1\sigma$ error bar of $\pm 13.5\%$ on the rms
mass fluctuation amplitude, or $+29\%,\;-25\%$ on the amplitude
of $P(k)$.  At the $1\sigma$ level, therefore, the $P(k)$ values
in Table~\ref{pkmassfid} can be multiplied coherently by 
a factor $F$ in the range $0.75-1.29$.

The cosmic variance uncertainty is estimated directly from our
data, and the range of $T_0$ and $\alpha$ that we have used to 
assign an associated error bar is probably conservative.
The most plausible source for an error in the $P(k)$ amplitude
that is substantially larger than our quoted $1\sigma$ uncertainty
would be an error in our adopted value of $\taueff$ that is larger than
the 5\% $1\sigma$ uncertainty we have assigned based on PRS.
We think that a large departure from the PRS $\taueff$ is unlikely
for the reasons given in \S\ref{sec:taueff}, but we cannot rule
out the possibility.  Another possible source of systematic error
beyond that reflected in our error bar is numerical limitations
of our normalizing simulations --- finite resolution and use of
the N-body+FGPA method in place of full hydrodynamic simulations.
We think that the uncertainty associated with these limitations
is unlikely to be significant for the reasons discussed in 
\S\ref{sec:sysmodel} and \S\ref{sec:simsys}, but confirmation
of our $b(k)$ estimate with large numbers of high resolution
hydrodynamic simulations would be valuable once it becomes
computationally feasible.

As discussed in \S\ref{sec:sysmodel} and White \& Croft (2000),
the main effect that might cause errors in the derived
{\it shape} of $b(k)$ outside those quoted in the Table
is truncation of the true linear $P(k)$ below the scale
of non-linearity, $k \sim 0.02 \invkms$.  

\subsection{Parameterized fits to $P(k)$}
\label{sec:fits}

Over the range of scales that we probe, our derived matter power
spectrum can be adequately described by a power-law or by some
other smooth functional form.  For many applications, it is simpler
to work with such parameterized descriptions than with the
individual $P(k)$ data points.  Following CWPHK, we describe the
amplitude of the power spectrum by quoting its value at 
a pivot wavenumber $k_{p}$, chosen so that the errors in 
amplitude are approximately independent of the errors in the 
logarithmic slope or other shape descriptor.
For this data set (which probes higher wavenumbers than the
CWPHK data set), we adopt $k_p=0.03\invkms$.

For power-law fits, our parameterized form is
\begin{equation}
P(k)=P_p\left(\frac{k}{k_{p}}\right)^{\nu},
\label{eqn:pl}
\end{equation}
where $P_p=P(k_{p})$ and we use $\nu$ to avoid confusion
with the inflationary spectral index $n$ (see equation~\ref{eqn:bbks}).
We determine the parameters by a maximum likelihood fit to
the $P(k)$ data points, using a diagonal $\chi^2$ and assuming that
the relative likelihood ${\cal L}$ obeys $-2\ln{\cal L}=\chi^{2}$.
We account for the additional normalization error on $P_p$ by convolving
the likelihoods in the $P_{p}-\nu$ plane along the $P_p$ direction with 
a likelihood distribution derived from the normalization
error bar.  We assume that the latter distribution is Gaussian in 
the log of the mass fluctuation amplitude (i.e., a Gaussian with specified
fractional error rather than absolute error).
The mass fluctuation amplitude is proportional to $\sqrt{P_p}$.
With these assumptions, the likelihood convolution is then
\begin{equation}
{\cal L}(P_{p},\nu)=\int^{\infty}_{0} {\cal L}'(x,\nu)
\exp\left[-{1 \over 2}\left(\frac{\log\sqrt{P_p}-\log\sqrt{x}}
  {\log(1+\sigma_a)}\right)^{2}\right]dx,
\label{eqn:likconv}
\end{equation}
where ${\cal L}'(P_{p},\nu)$ is the likelihood before the convolution
and $\sigma_a$ is the $1\sigma$ fractional normalization error 
on the fluctuation amplitude ($13.5\%$ for the fiducial sample).
In practice, the normalization error dominates over the uncertainty in
the amplitude of the power-law fit, i.e., ${\cal L}'(x,\nu)$ is effectively
a $\delta$-function.  Since $k_p$ is chosen so that errors in $\nu$ and
$P_p$ are uncorrelated, the result of equation~(\ref{eqn:likconv}) in
the $P_p$ direction is very close to a log-normal distribution in $\sqrt{P_p}$
with dispersion $\sigma_a$.

In place of $P_p$, we quote the dimensionless quantity
\begin{equation}
\Delta^{2}(k_{p}) = {1 \over 2\pi^{2}} k_{p}^{3}P_p,
\end{equation}
the contribution to the mass fluctuation variance from an
interval $d\ln k=1$ about $k_p=0.03\invkms$.
For the fiducial sample we obtain $\Delta^2 (k_p)  = 0.74^{+0.20}_{-0.16}$
and $\nu = -2.43 \pm 0.06$ (see Table~\ref{pltab}).
The absolute value of $\chi^2$ for the best fitting parameter
values is 6.9 for 11 degrees of freedom, indicating that
the power-law shape is an adequate description of the data.

\begin{figure*}[t]
\centering
\vspace{8cm}
\includegraphics{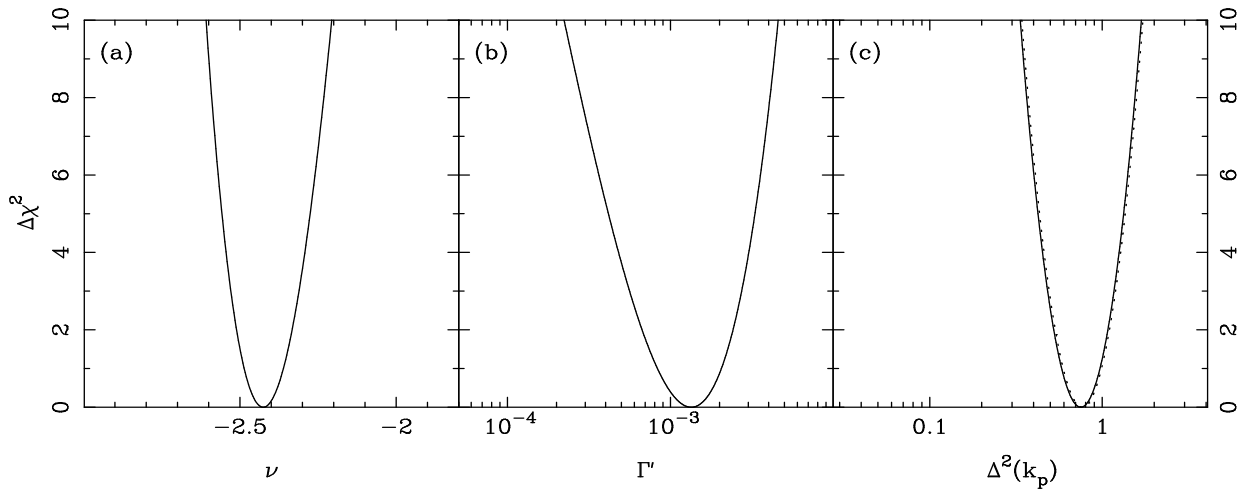}
\caption[dchi21d]{
Constraints on fit parameters for the matter power spectrum of the fiducial
sample.  Panel (a) 
shows $\Delta \chi^2$ vs. $\nu$ for power-law fits,
with $\Delta^2(k_p)$ fixed at its best-fit value.
Panel (b) is similar but for the parameter $\gprime$ of CDM-like fits.
Panel (c) shows $\Delta\chi^2$ vs. $\Delta^2(k_p)$ for
power-law fits with $\nu$ fixed to its best-fit value (solid curve)
and CDM-like fits with $\gprime$ fixed to its best-fit value (dotted
curve, nearly obscured).
\label{dchi21d}}
\end{figure*}

The solid curves in Figure~\ref{dchi21d} show the value of
$\Delta\chi^2$ as a function of $\nu$ (left panel) or $\Delta^2(k_p)$ 
(right panel), in each case with the other parameter fixed at its
best-fit value.  Because the errors on $\nu$ and $\Delta^2(k_p)$ are
nearly independent, one can simply add the $\Delta \chi^2$
associated with each parameter separately to get the value of $\Delta\chi^2$
for any combination of $\nu$ and $\Delta^2(k_p)$.
The $\Delta\chi^2$ curve for $\Delta^2(k_p)$ is well
described by the equation
\begin{equation}
\Delta\chi^2 = \left[{\log(\Delta^2(k_p)/0.74) \over 2\log(1.135)}\right]^2,
\label{eqn:dchi2amp}
\end{equation}
and the $\Delta\chi^2$ curve for $\nu$ is adequately described by
\begin{equation}
\Delta\chi^2 = \left({\nu + 2.43 \over 0.06}\right)^2.
\label{eqn:dchi2nu}
\end{equation}
These formulae can be used to compute joint confidence intervals on 
$\Delta^2(k_p)$ and $\nu$ for cosmological model tests.

As an alternative parameterized description, we fit the fiducial $P(k)$
data with the generic shape predicted by inflationary CDM models,
equation~(\ref{eqn:bbks}), taking as free parameters the amplitude
$\Delta^2(k_p)$ and the shape parameter $\gprime$ specified in $\invkms$
at $z=2.72$.  We set the inflationary spectral index $n=1$.
We use the same pivot wavenumber, $k_p=0.03\invkms$, that we used
for the power-law fit.
The parameters $\gprime$ and $\Delta^2(k_p)$ can be related to
the quantities $\Gamma$ (in $\invhmpc$) and $\sigma_8$ at $z=0$,
as discussed in \S\ref{sec:implications} below.  However, this relation
is cosmology dependent, while the fit in terms of $\gprime$ and
$\Delta^2(k_p)$ is not.  We should note that our data do not yield
a significant detection of curvature of the power spectrum; they constrain
$\gprime$ because the slope at scale $k_p$ depends on the location
of the peak of $P(k)$, which is determined by $\gprime$.

The constraint on $\Delta^2(k_p)$ is essentially identical for 
the power-law and CDM-like fits, as demonstrated by the agreement of the
dotted and solid curves in the right panel of Figure~\ref{dchi21d}.
The constraint on $\gprime$ (middle panel) is asymmetric, since
the slope of the CDM spectrum changes more rapidly towards lower $k$
than towards higher $k$.  
We find $\gprime= 1.3^{+0.7}_{-0.5}\times 10^{-3} \invkms$ at the $1 \sigma$
level and $\gprime=1.3^{+1.6}_{-0.9}\times 10^{-3} $ at $2 \sigma$.
The errors on $\Delta^2(k_p)$ and $\gprime$
are again nearly uncorrelated, so one can add the $\Delta\chi^2$
values from the two 1-dimensional curves to obtain the $\Delta\chi^2$
for a combination of values.
We discuss cosmological constraints based on this fit in 
\S\ref{sec:implications} below.

\subsection{Evolution of $P(k)$}
\label{sec:evolution}

\begin{figure*}[t]
\centering
\vspace{14cm}
\includegraphics{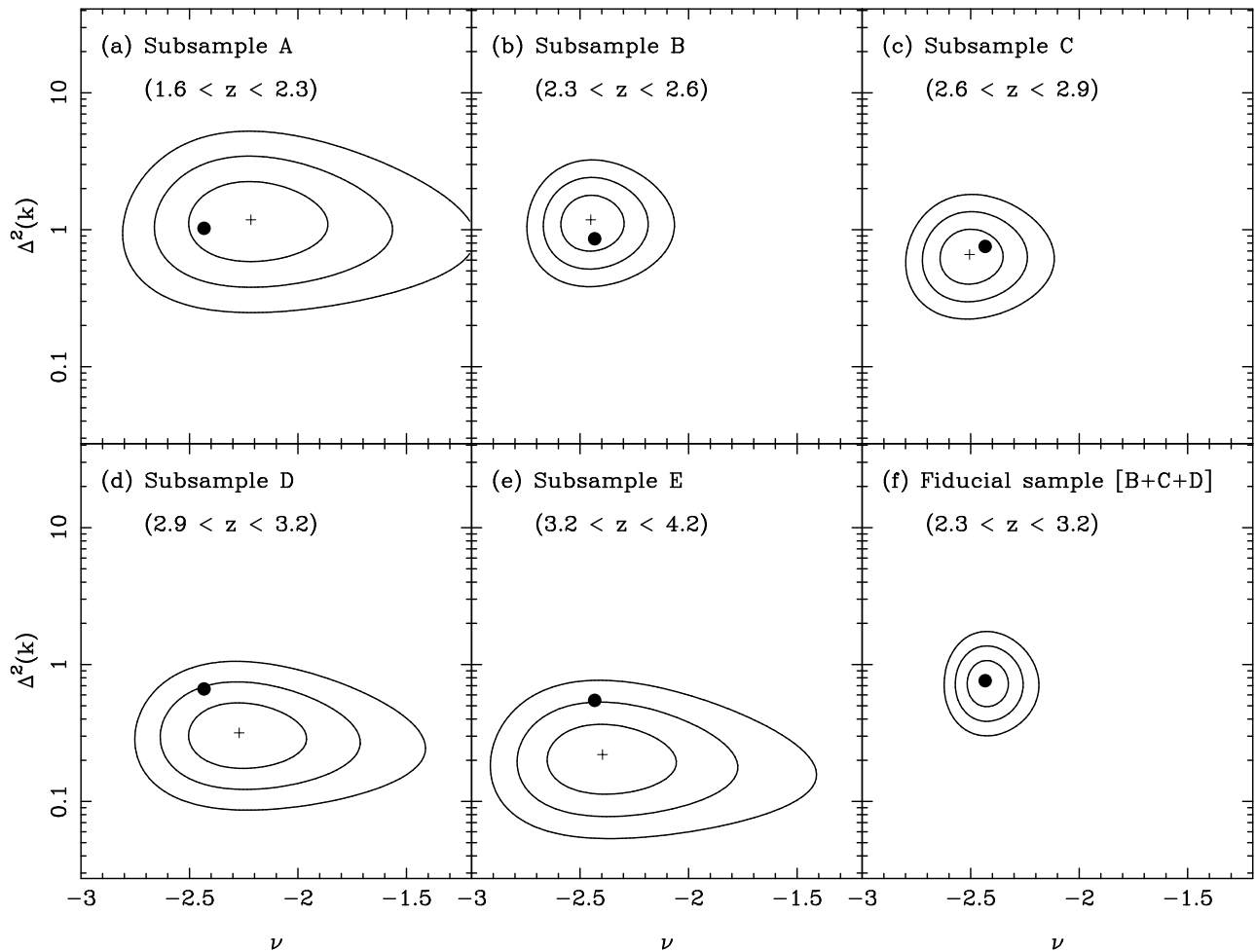}
\caption[plfit]{
Evolution of the matter power spectrum, measured from the data subsamples
described in Table~\ref{subtab}.
In each panel, contours show the 1, 2, and 3$\sigma$ confidence intervals
on the parameters $\Delta^2(k_p)$ and $\nu$ of power-law fits to the
inferred matter power spectrum, and crosses mark the best-fit parameter 
values.  Filled circles show the best-fit parameters derived from the
fiducial sample (see panel f), scaled to the mean redshift
of the subsample according to the gravitational instability prediction.
\label{plfit}}
\end{figure*}

Figure~\ref{fluxpk} demonstrates substantial evolution of the observed flux
power spectrum over the redshift range covered by our data set,
$\langle z \rangle = 3.51$ (subsample E) to $\langle z \rangle = 2.13$
(subsample A).  However, most of this evolution is driven by the change
in the mean optical depth.  In this section, we test whether the 
evolution of the inferred matter $P(k)$ is consistent with the predictions
of gravitational instability theory.

We determine the matter power spectrum for the five subsamples using
the same normalizing simulations described in \S\ref{sec:normsim},
except that at each redshift we use the appropriate scaling 
of comoving distances to $\kms$ and match the mean optical depth
implied by the PRS fitting formula at the subsample's mean
redshift.  These optical depth values are $\taueff=0.192$, 
0.274, 0.355, 0.460, 0.679 for subsamples A to E, respectively.
We fit power-laws to the resulting $P(k)$ data points as we did
for the fiducial sample in \S\ref{sec:fits}.  In computing the
contribution of $\taueff$ to the 
normalization uncertainty, we assume fractional errors in $\taueff$
of 5\%, which we convert to uncertainties in the mass fluctuation
amplitude using equation~(\ref{eqn:ctau}) with 
$C_{\tau}=$ $-0.5$, $-0.75$, $-3.0$, $-1.5$, $-2.0$,
for subsamples A to E.  (These values of $C_{\tau}$, derived
using the simulations, are themselves rather uncertain.)
This procedure is likely to underestimate the true uncertainty in
the $P(k)$ amplitude due to $\taueff$, since the PRS data 
cover only the range $z > 2.5$ and we are using their fitting formula to
extrapolate the behavior of $\taueff$ to lower redshift.
Also, we have not carried out the filling factor analysis for the
individual subsamples, so we do not have this additional supporting
evidence for the adopted values of $\taueff$.
For the normalization uncertainty associated with the IGM temperature
parameters, we have assumed the same fractional error as for the
fiducial sample.  We assume constant values of $T_0$ and $\alpha$
in the normalizing simulations;
incorporating redshift evolution of these parameters
might be worthwhile in the future, as they become better constrained
observationally.

Figure~\ref{plfit} shows likelihood contours of the power-law fits
in the $\Delta^2(k_p)-\nu$ plane, with $k_p=0.03\invkms$ in each case.
The three contours enclose
$68\%$, $95\%$ and $99.7\%$ of the joint probability 
($\Delta\chi^{2}=2.3,6.2,11.8$), and crosses mark the best-fit 
parameter values.  
The bottom right panel shows likelihood contours for the fiducial sample,
which are, of course, tighter than those of the individual subsamples.
The absolute values of $\chi^2$ for the best fitting
parameter values are 14.9, 6.8, 6.9, 10.8, and 7.1 for subsamples
A to E, respectively.  The number of degrees of freedom is 11 in each
case, so the power-law descriptions are statistically acceptable.
Values of the fit parameters, and $1\sigma$ and $2\sigma$ error bars,
appear in Table~\ref{pltab}.

In each panel of Figure~\ref{plfit}, a filled circle marks
the slope and amplitude predicted by scaling the best-fit parameters of
the fiducial sample to the
subsample's mean redshift according to gravitational instability theory.
Since $P(k)$ represents
the linear theory power spectrum, and we do not detect significant
curvature of $P(k)$ over our range of $k$, the predicted slope is the
same at all redshifts.
The scaling of the amplitude includes both the linear growth factor
and the change in comoving scale at fixed $k_p$.  
We compute both effects assuming $\Omega_m=1$, which should be
a good approximation at these redshifts.

There are several features to note from Figure~\ref{plfit} and
Table~\ref{pltab}.  First, the slope measured from each subsample
is consistent with that of the fiducial sample at the $1\sigma$ level,
confirming the expected constancy of the shape of $P(k)$.
Second, the best-fit amplitude grows steadily from $\langle z \rangle = 3.51$
(subsample E) to $\langle z \rangle = 2.47$ (subsample B), before dropping
slightly (by less than $1\sigma$) between $\langle z \rangle = 2.47$ 
and $\langle z \rangle = 2.13$ (subsample A).
The best-fit value of $\Delta^2(k_p)$ is a factor of 5.7 higher 
for subsample A than for subsample E.  
Third, the measured growth is roughly consistent with the gravitational
instability prediction, but not perfectly so.
The scaled amplitude of the fiducial sample is consistent with the
measurements from subsamples A-C at the $\sim 1\sigma$ level, but
it is $\sim 2\sigma$ from the results of subsamples D and E.
We suspect that this marginal discrepancy arises because we have
underestimated the contribution of $\taueff$ uncertainty to the
error bars for the individual subsamples.
The dip in $\taueff$ at $z \sim 3.1$ found by Bernardi et al.\ (2002)
could contribute to the low apparent amplitude of $P(k)$ for subsample D,
since we have assumed smooth evolution of $\taueff$.

Figure~\ref{plfit} provides evidence for growth of the amplitude of
mass fluctuations over the redshift range $z=3.5-2.1$, though it
does not constitute a high-precision confirmation of the
gravitational instability prediction.  A strong application of the
gravitational evolution test will require better determinations
of $\taueff$ as a function of redshift.  It would also benefit from 
a larger set of quasar spectra, since the statistical normalization
error bars for our individual subsamples are fairly large.
Nonetheless, the approximate consistency between our results and
the gravitational instability prediction supports the view
that structure in the \lya\ forest is dominated by structure in the
underlying mass density field.  Other effects, such as ionizing
background fluctuations or temperature fluctuations, would be
expected to decrease towards low redshift, as the IGM becomes more
transparent and moves closer to equilibrium. 
They are therefore likely to predict $\Delta^2(k_p)$ evolution 
opposite in sign to what we observe. 
Future theoretical work that incorporates
radiative transfer will help us understand these predictions
quantitatively for different hydrogen and helium reionization scenarios.
For the time being, we also note that it is unlikely
that the flux power spectrum associated
with other effects should have the same shape as that associated
with mass fluctuations, so the constancy of the measured $P(k)$ 
slope suggests that these effects are not significant even at the highest 
redshifts probed by this sample.

\begin{table}[ht]
\centering
\caption[pltab]{\label{pltab} 
Power-law fit parameters for P(k) (see Equation~\ref{eqn:pl}).}
\begin{tabular}{ccc}
\hline &&\\
Subsample &  $\Delta^{2}(k_{p})\pm1\sigma\pm2\sigma$  &
 $\nu\pm1\sigma\pm2\sigma$  \\
\hline &&\\
A &   $ 1.21^{+0.62}_{-0.43}{}^{+1.62}_{-0.71}$ &
 $-2.22^{+0.22}_{-0.19}{}^{+0.50}_{-0.36}$             \\
B &   $ 1.15^{+0.39}_{-0.29}{}^{+0.97}_{-0.53}$ &
 $-2.45^{+0.10}_{-0.09}{}^{+0.21}_{-0.18}$           \\
C &   $ 0.66^{+0.22}_{-0.17}{}^{+0.55}_{-0.30}$ &
 $-2.51^{+0.10}_{-0.09}{}^{+0.21}_{-0.18}$          \\
D &   $ 0.32^{+0.13}_{-0.10}{}^{+0.33}_{-0.16}$ &
 $-2.27^{+0.19}_{-0.16}{}^{+0.42}_{-0.31}$           \\
E &   $ 0.21^{+0.09}_{-0.07}{}^{+0.24}_{-0.12}$ &
 $-2.40^{+0.21}_{-0.17}{}^{+0.47}_{-0.32}$           \\
Fiducial &   $ 0.74^{+0.20}_{-0.16}{}^{+0.49}_{-0.28}$ &
 $-2.43^{+0.06}_{-0.06}{}^{+0.13}_{-0.12}$          \\
\hline &&\\
\end{tabular}
\end{table}

\subsection{Comparison to previous results}
\label{sec:previous}

We can use results of the power-law fits of \S\ref{sec:fits} to
compare our new measurement of $P(k)$ at $z=2.72$ to the results
of CWPHK at $z=2.5$.  Here we have used a pivot wavenumber $k_p=0.03\invkms$,
whereas CWPHK, working with lower resolution spectra, used $k_p=0.008\invkms$.
We therefore extrapolate our measurement in $k$ using the best fit 
$\gprime$ model (\S\ref{sec:fits}),
as well as scaling the amplitude to $z=2.5$ assuming linear growth
(for an EdS cosmology) between $z=2.72$ and 2.5. We find
$\Delta^{2}(0.008 \invkms)=0.42^{+0.12}_{-0.1}$ (1 $\sigma$ errors),
compared to the CWPHK result of $\Delta^{2}(0.008 \invkms)
=0.57^{+0.26}_{-0.18}$. 
Our new $P(k)$ amplitude is therefore within $1 \sigma$ of 
that of CWPHK. The sign of the difference (our
new amplitude is lightly lower) is likely to be partly due to
the fact that CWPHK used a much 
lower gas temperature in their normalizing simulations, $T_0=6000$ K,
which is probably inconsistent with current observations 
(e.g., McDonald et al.\ 2001).
Here we have used $T_0=15,000$ K, and we have included a contribution
from $T_0$ uncertainty to the error bar on $P(k)$, which CWPHK did not do.
The local logarithmic slope of our model
extrapolation is $\nu=-2.35$ at $k_{p}=0.008 \invkms$, compared to
$\nu=-2.25\pm 0.18$ for CWPHK. 
The slope is thus in good agreement ($0.6\sigma$) with the previous
measurement, and part of the slight difference can be accounted for by
our correction for redshift-space distortions, which was not
incorporated into the CWPHK method.

Other differences between the CWPHK result and ours include the
data sample size, and the correction for smoothing bias.
Tests in \S\ref{sec:smoothingbias} show that smoothing bias can
boost the flux power spectrum by as much as 10\% on small scales 
for $2 $\AA\ resolution data; the CWPHK analysis included some spectra  
with resolution as coarse as $2.3$ \AA.
On the scales where the measurements overlap, the error bars on
our individual $P(k)$ data points are larger than one might have
expected given the error bars found by CWPHK for a smaller data set.
In both cases, these error bars are estimated internally from the data,
and we expect that the estimates from the larger data set are more reliable.
For example, here we have been able to use 50 subsamples of the data set
to generate jackknife error estimates, whereas CWPHK used only 10 subsamples.
Smoothing bias may also have made the CWPHK error bars artificially small.
The good agreement between our current results and those
of CWPHK supports CWPHK's assumption that the statistical uncertainties
in their data dominated over systematic uncertainties associated
with redshift-space distortions or the temperature-density relation.
However, the improvements to the $P(k)$ determination method introduced
here, which account for these systematic effects,
are clearly needed to take advantage of our larger, higher 
resolution data set.  Relative to CWPHK, our current determination
of $P(k)$ has significantly smaller statistical uncertainties and
better quantified and understood systematic uncertainties.

Other estimates of linear matter clustering  from \lya\ forest flux statistics
include those of Nusser \& Haehnelt (2000) and M00.
The former measure the fluctuation amplitude on the Jeans scale,
using an analytic fit to the flux distribution function. 
Using this information
on the smallest scales, they find results that appear
to be $1-2 \sigma$ lower in amplitude that ours.
The interpretation does, however, depend on assumptions that relate the
actual Jeans scale to that derived
using linear perturbation theory (Gnedin \& Hui 1998). 
M00 use a single, hydrodynamic simulation of a $\Lambda$-dominated CDM model
and the flux power spectrum of eight HIRES spectra
to infer the best fitting $P(k)$ amplitude, within the context of that
model.  Guided by the simulation result, they fit
their observed one-dimensional flux power spectrum to infer $P(k)$, finding
$\Delta^{2}(k_{p})=0.72 \pm0.09$ at $k_{p}=0.04 \invkms$,
for $z=3$.  Scaling our results as we did for the CWPHK comparison, we
find an equivalent
$\Delta^{2}(k_{p})=0.73^{+0.20}_{-0.15}$. The slope
we find is $\nu=-2.56$, compared to M00's $\nu=-2.55 \pm 0.10$.
We thus find essentially perfect agreement in both slope 
and rms mass fluctuation amplitude.
Figure~\ref{1dpk} shows good agreement between M00's 1-d flux 
power spectrum and that measured from a subset of our HIRES data
with the same redshift range.
This said, however, there is an important difference in our analyses.
We adopt the PRS
estimate of $\taueff$ instead of the slightly lower value found by M00.
Equation~(\ref{eqn:ctau}) implies that if we scaled our fiducial $\taueff$
(at $z=2.72$) by the ratio of the values found by M00 and PRS at $z=3$,
then we would derive an rms matter fluctuation amplitude 30\% higher
than M00.  The agreement between our results is therefore somewhat
fortuitous. The conversion between flux and mass amplitudes
in each case is different, but this is cancelled out by the
different $\taueff$ values used.
Among the numerous other differences in our procedures,
 one possible concern with the M00 method is that they rely on one 
$10\hmpc$ simulation for their normalization, which could leave a
significant cosmic variance error (see Figure~\ref{sphtest}).
As M00 state explicitly, their quoted error bar applies in the context
of a particular cosmological model and does not include contributions
from uncertainties in $\taueff$, the temperature-density relation,
or the cosmological model.
Given the differences in modeling approach, agreement of these two studies
at the level seen is impressive.

\section{Cosmological implications}
\label{sec:implications}

\begin{figure}[t]
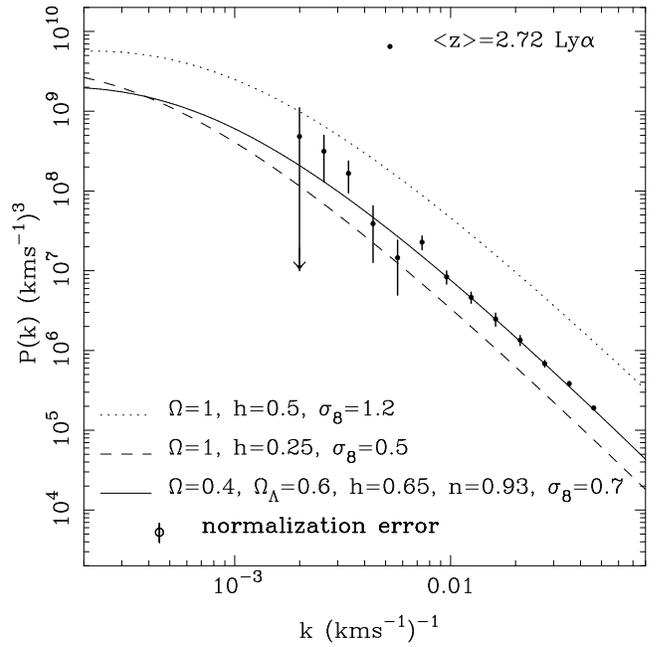

\PSbox{models1.ps angle=-90 voffset=285 hoffset=-70 vscale=50 hscale=50}
{3.5in}{3.8in} 
\caption[models1]{
Comparison of the linear matter power spectrum inferred from the
fiducial \lya\ forest sample to the predictions of three inflationary
CDM models.  Points with error bars show the inferred $P(k)$; at the
$1\sigma$ level these points can be shifted coherently by the 
normalization error bar shown in the lower left.
Dotted and dashed lines show, respectively, COBE-normalized and
cluster normalized CDM models with $\Omega_m=1$.
The solid line shows a low-$\Omega_m$ model with a cosmological 
constant.  While the models represented by the solid and dotted
lines have quite different power spectrum shapes in comoving $\hmpc$ units,
they have similar shapes here because of the different values of
$H(z)/H_0$ for the two cosmologies.
\label{models1}
}
\end{figure}

Figure~\ref{models1} compares our fiducial $P(k)$ measurement 
(from Table~\ref{pkmassfid}) to the linear matter power spectra
of three CDM models.
Note that the data points can be shifted up and down coherently
at the $1\sigma$ level by the amount indicated by the normalization
error bar in the lower left.
The model power spectra are computed using the CDM transfer
function of equation~(\ref{eqn:bbks}).
The amplitude of $P(k)$ is specified by $\sigma_8$,
the rms mass fluctuation in $8\hmpc$ spheres at $z=0$.
The theoretical power spectra have been scaled to redshift $z=2.72$
and converted to $\kms$ units using the parameters of
the corresponding cosmological model.

The highest amplitude model (dashed line) is a COBE-normalized 
(Bennett et al.\ 1996), $\Omega_m=1$ model with scale-invariant
($n=1$) inflationary fluctuations, $h=0.5$, $\Gamma = \Omega_m h =0.5$,
and $\sigma_8=1.2$.
It is ruled out at high significance.  Although this model is well
known to fail other cosmological tests, in particular to produce
excessively massive clusters at $z=0$ (e.g., White, Efstathiou, \& Frenk 1993),
most of those failures reflect the {\it combination} of high fluctuation
amplitude with $\Omega_m=1$, while the failure here is
almost entirely one of the fluctuation amplitude 
(see equation \ref{eqn:sigma8} below).
The lowest amplitude model (dotted line) is an $\Omega_m=1$ model
with $\Gamma=0.25$, a shape closer to that favored by galaxy
large-scale structure (see, e.g., Peacock \& Dodds 1994).
We achieve this low $\Gamma$ by setting $h=0.25$ with $n=1$, though a similar
power spectrum shape would arise with larger $h$ and a tilted ($n<1$)
inflationary spectrum or boosted relativistic neutrino background.
The normalization, $\sigma_8=0.5$, is close to the value required
to match both COBE-DMR anisotropies and the cluster mass function at $z=0$
(see, e.g., Cole et al.\ 1997).
This model predicts a power spectrum amplitude that is too low
to match the \lya\ forest results, by about $2.5\sigma$.

The solid line shows the power spectrum of a flat, $\Omega_m=0.4$, 
$\Omega_\Lambda=0.6$ model, with $h=0.65$ and a slightly tilted
($n=0.93$) inflationary spectrum, and a normalization $\sigma_8=0.70$.
M00 conclude that this model reproduces
the shape and amplitude of their measured flux power spectrum,
and it is clear from Figure~\ref{models1} that it provides an
excellent fit to our inferred matter $P(k)$.  As discussed
in the previous section, our fiducial value of $\taueff$ is different
from M00's, so this near perfect agreement is somewhat fortuitous.
The amplitude is 12.5\% (about $1\sigma$) lower than the amplitude 
$\sigma_8=0.80$ that Eke et al.\ (1996) estimate is needed to match the 
cluster mass function for $\Omega_m=0.4$, 
and it is 18\% lower than the
amplitude $\sigma_8=0.85$ implied by COBE normalization for these
parameters (computed using the CMBFAST code of Zaldarriaga, Seljak,
\& Bertschinger [1998], assuming no tensor contribution).
Thus, the shape of our measured $P(k)$ is similar to that of
currently popular CDM models, and the amplitude is compatible 
with other estimates but on the low side, a point that we will
return to below.

We can make a more systematic comparison to the inflationary CDM 
predictions using the parameterized fit in terms of $\gprime$ and
$\Delta^2(k_p)$ from \S\ref{sec:fits}.
These quantities can be related to the values of $\Gamma$ 
(in $\hmpc$) and $\sigma_8$ at $z=0$, but the relation depends on
cosmological parameters because of the dependence of the linear
growth factor and the Hubble ratio $H(z)/H_0$ on cosmology.
The scaling of $\Gamma$ is quite straightforward,
\begin{equation}
\Gamma = \gprime (1+z)^{-1} H(z) =
  0.16 \left({\gprime \over 1.3 \times 10^{-3}}\right) 
  \left({{\cal H} \over 4.6}\right) \;\invhmpc,
\label{eqn:gamma}
\end{equation}
where we have introduced the notation ${\cal H} \equiv H(z=2.72)/H_0$
and scaled the formula to our best-fit value of $\gprime$ in $\invkms$
and to the value ${\cal H} = 4.6$ appropriate for $\Omega_m=0.4$, 
$\Omega_\Lambda=0.6$.
The Hubble ratio at $z=2.72$ 
is ${\cal H}=4.0$ for $\Omega_m=0.3$, $\Omega_\Lambda=0.7$,
${\cal H}=5.4$ for $\Omega_m=0.4$, $\Omega_\Lambda=0$, and
${\cal H}=7.2$ for $\Omega_m=1$, $\Omega_\Lambda=0$.
Equation~(\ref{eqn:gamma}) is based on the Ma (1996) transfer
function coefficients (eq.~\ref{eqn:bbks}); for the Bardeen et al.\ (1986)
coefficients, 0.16 should be changed to 0.15.
Recent estimates of the galaxy power spectrum from the 2dF Galaxy
Redshift Survey find best-fit parameter combinations $\Omega_m h = 0.20$ and
$\Omega_b/\Omega_m = 0.15$ for a CDM model with scale-invariant
initial fluctuations (Percival et al.\ 2001).
Incorporating Sugiyama's (1995) approximation for the suppression
of small scale power by baryons, these parameter combinations imply 
$\Gamma \approx \Omega_m h \exp[-(2h)^{1/2}\Omega_b/\Omega_m-\Omega_b]
\approx 0.16$.  Thus, for ${\cal H} \approx 4.6$, the shape parameter
inferred from the \lya\ forest $P(k)$ is in excellent agreement with
that inferred from the galaxy power spectrum today, though our
$1\sigma$ error bar on $\gprime$ is nearly 50\%, so the constraint
from equation~(\ref{eqn:gamma}) is not particularly tight.

The constraint on $\gprime$ is derived assuming a scale-invariant
($n=1$) inflationary spectrum, so equation~(\ref{eqn:gamma}) implicitly
incorporates the same assumption.  Since we do not have a significant
detection of curvature in $P(k)$, the effects of $\Gamma$ and $n$ 
are degenerate, and one can only identify $\Gamma$ with the
above parameter combination of CDM models 
for a specified value of $n$.
To a fairly good approximation, one can account for 
the degeneracy by replacing $\Gamma$ with $\Gamma+1.4(1-1/n)$ on the 
left-hand side of equation~(\ref{eqn:gamma}).  For example, a model
with $\Gamma=0.26$ and $n=0.93$ has almost the same $P(k)$ shape as a 
model with $\Gamma=0.16$, $n=1$, on the scales probed by our measurement.

While the scaling of $\gprime$ is determined by the evolution of $H(z)$,
relating $\Delta^2(k_p)$ to $\sigma_8$ is somewhat trickier.
The wavenumber at $z=0$ that corresponds to the same comoving scale as $k_p$
is
\begin{equation}
k_0 = k_p (1+z)^{-1} H(z) = 3.710 \left({{\cal H} \over 4.6}\right) \invhmpc.
\label{eqn:k0}
\end{equation}
Since $\Delta^2$ and $P(k)$ refer to the linear power
spectrum, $\Delta^2(k_0)=D_L^{-2} \Delta^2(k_p)$, where $D_L$ is the
linear growth factor at $z=2.72$ for the cosmological model under 
consideration.  However, the ratio of $\sigma_8$ to $[\Delta^2(k_0)]^{1/2}$
depends on the shape of the power spectrum and on the value of $k_0$,
and there is no exact equation relating the two quantities.
Empirically, we find that the fitting formula
\begin{eqnarray}
\sigma_8 & = &0.82 \left({\Delta^2(k_p) \over 0.74}\right)^{1/2}
  \left({0.322 \over D_L}\right)
  \left({{\cal H}}\over 4.6\right)^{-0.25}\;\times \nonumber \\
& & \left({\Gamma \over 0.15}\right)^{-0.44-0.06\ln(\Gamma/0.15)
                                    +0.05(4.6/{\cal H}-1)}
\label{eqn:sigma8}
\end{eqnarray}
describes our results with an accuracy of 1\% over the parameter range
$3.5 \leq {\cal H} \leq 7.0$ and $0.05 \leq \Gamma \leq 0.5$.
We have scaled equation~(\ref{eqn:sigma8}) to our best-fit value
of $\Delta^2(k_p)$, to $\Gamma=0.15$, and to the values of $D_L$ and ${\cal H}$
that apply for $\Omega_m=0.4$, $\Omega_\Lambda=0.6$.
For specified $D_L$, ${\cal H}$, and $\Gamma$, the $1\sigma$ uncertainty
in our estimate of $\sigma_8$ is $\sim 13.5\%$, the same as our $1\sigma$
uncertainty in $[\Delta^2(k_p)]^{1/2}$ (see \S\ref{sec:fits}).
Because decreasing $\Omega_m$ raises $D_L$ but lowers ${\cal H}$, the
inferred $\sigma_8$ is not very sensitive to the value of $\Omega_m$
in flat cosmological models: for $\Omega_m=0.5$, 0.3, 0.2, the factor
0.82 in equation~(\ref{eqn:sigma8}) changes to 0.84, 0.80, and 0.76,
respectively.
The value of $\sigma_8$ is more sensitive to $\Gamma$, which determines
the ratio of power on the $\sim 8\hmpc$ scale to the power at $k_0$.
Equation~(\ref{eqn:sigma8}) applies for $n=1$, and other $(n,\Gamma)$
combinations that yield the same slope of $P(k)$ on the scale of 
the \lya\ forest data do not necessarily yield the same $\sigma_8$,
as one can see from the example in Figure~\ref{models1}.

\begin{figure*}[t]
\centerline{
\epsfxsize=4.5truein
\epsfbox[55 235 550 720]{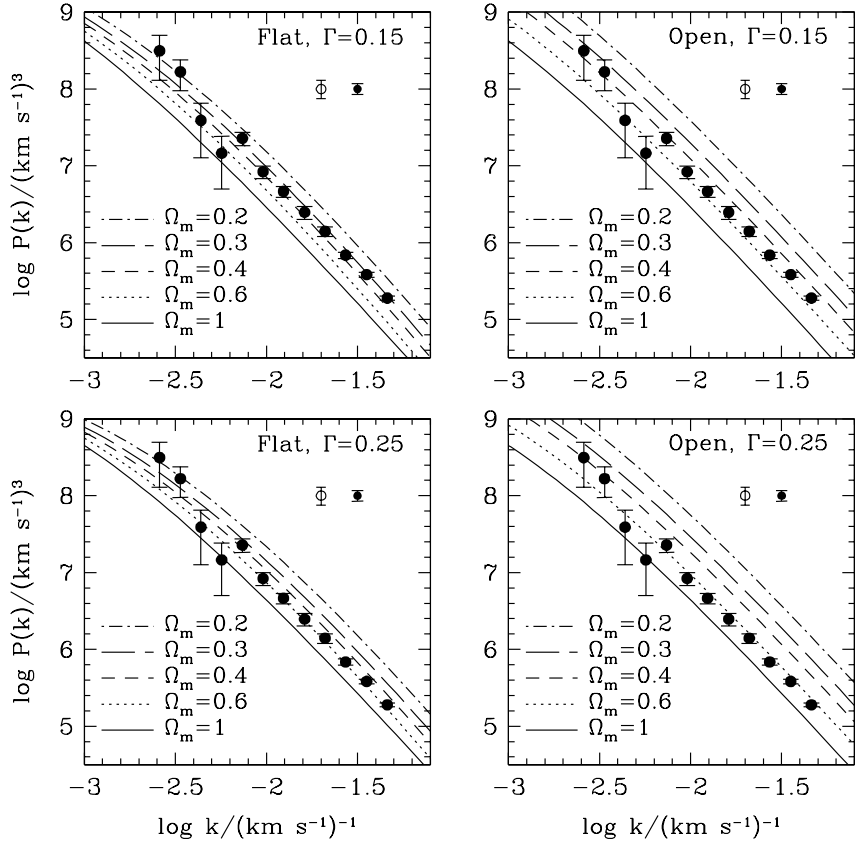}
}
\caption[pkcluster]{
Comparison of the matter $P(k)$ inferred from the \lya\ forest to
the predictions of cluster normalized models with different values of 
$\Omega_m$.  In each panel, points with error bars show our fiducial $P(k)$,
which has 
the $1\sigma$ normalization uncertainty marked on the open circle at the 
upper right.  Model power spectra satisfy the cluster normalization 
constraint of Eke et al.\ (1996): 
$\sigma_8=0.52\Omega_m^{-0.52+0.13\Omega_m}$ for $\Omega_\Lambda=1-\Omega_m$ 
(left panels) and 
$\sigma_8=0.52\Omega_m^{-0.46+0.10\Omega_m}$ for $\Omega_\Lambda=0$ (right
panels).
The error bar on the filled circle at upper right indicates the quoted 
$\sim 15\%$ uncertainty in the cluster normalization of $P(k)$.
Upper panels show power spectra with shape parameter $\Gamma=0.15$,
lower panels $\Gamma=0.25$.
\label{pkcluster}
}
\end{figure*}

Following the lines of Weinberg et al.'s (1999a) analysis of the CWPHK result,
we can constrain the value of $\Omega_m$ by combining our measurement
of $P(k)$ at $z=2.72$ with the constraint on $\Omega_m$ and $\sigma_8$
implied by the cluster mass function or temperature function at $z=0$
(see, e.g., White et al.\ 1993; Eke et al.\ 1996;
Viana \& Liddle 1996; Pierpaoli, Scott, \& White 2001).
Figure~\ref{pkcluster} compares our fiducial power spectrum to those
of cluster-normalized models with $\Gamma=0.15$ (top panels) and
$\Gamma=0.25$ (bottom panels).  For each value of $\Omega_m$, we
determine $\sigma_8$ from the Eke et al.\ (1996) cluster normalization
constraint, then scale the power spectrum back to $z=2.72$.
Both $D_L$ and ${\cal H}$ depend on $\Omega_\Lambda$ as well as $\Omega_m$,
so the scaling is different for flat models (left panels) and
open models (right panels).  The error bar on the open circle in each 
panel shows the $+29\%,-25\%$ $1\sigma$ uncertainty in the \lya\ $P(k)$
normalization, while the error bar on the filled circle next to it shows
the $\sim 15\%$ uncertainty in the cluster normalization (8\% in $\sigma_8$)
quoted by Eke et al.\ (1996).  

For flat models with $\Gamma=0.15$, the best match to the \lya\ $P(k)$
is obtained for $\Omega_m$ between 0.3 and 0.4.  For higher $\Omega_m$, the
amplitude at $z=0$ is lower 
and $D_L(z=2.72)$ is smaller, so the predicted power spectrum is too low.
Conversely, the predicted $P(k)$ amplitude is too high for lower $\Omega_m$.
The implied value of $\Omega_m$ is higher in open models, mainly because
they have a larger value of $D_L(z=2.72)$.
The best-fit $\Omega_m$ is also higher for $\Gamma=0.25$, since in this
case there is less contribution to $\sigma_8$ from large scales, and the
power on the scales probed by the \lya\ data must be higher to compensate.

\begin{figure*}[t]
\centerline{
\epsfxsize=4.5truein
\epsfbox[50 470 550 720]{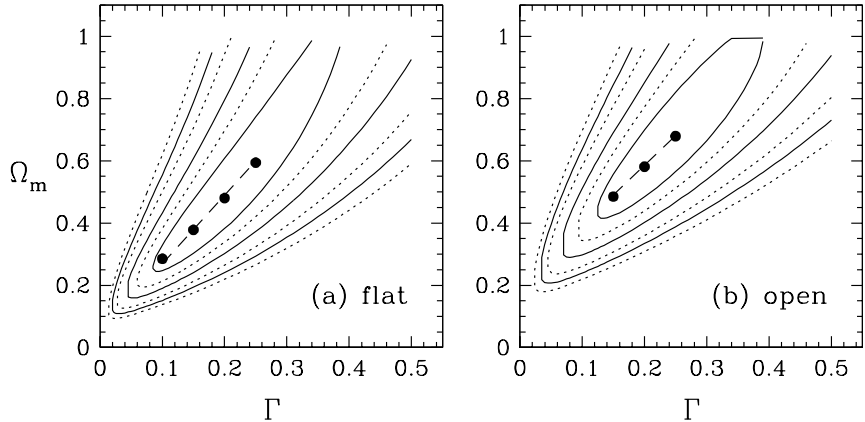}
}
\caption[omega]{
Constraints in the $\Gamma-\Omega_m$ plane from the combination of
cluster normalization with the \lya\ forest $P(k)$, for
models with $\Omega_\Lambda=1-\Omega_m$ (left) and $\Omega_\Lambda=0$ (right).
Points show the best-fit value of $\Omega_m$ at each $\Gamma$;
the dashed lines are the linear fits to these points
listed in eqs.~(\ref{eqn:omegafit1}) and~(\ref{eqn:omegafit2}).
Solid curves show contours of $\Delta\chi^2=1$, 4, 9, giving the 
1, 2, and $3\sigma$ confidence intervals on $\Omega_m$ for specified $\Gamma$.
Dotted curves show contours of $\Delta\chi^2=2.3$, 6.7, 11.8, corresponding
to 68\%, 95\%, and 99.7\% joint confidence levels for 2-parameter fits.
\label{omega}
}
\end{figure*}

To make the $\Omega_m$ constraints more quantitative, we compute
the values of $\Delta^2(k_p)$ and $\gprime$ for different combinations of 
$\Omega_m$ and $\Gamma$, then compute $\Delta\chi^2(\Omega_m,\Gamma)$ 
by comparing to the parameter constraints in \S\ref{sec:fits}
(Figure~\ref{dchi21d}).
We account for uncertainty in cluster normalization itself 
by adding in quadrature our $\sim 13.5\%$ error bar on the rms fluctuation
amplitude and the 8\% error bar quoted by Eke et al.\ (1996).
Specifically, we compute the contribution 
to $\Delta\chi^2$ from $\Delta^2(k_p)$
using equation~(\ref{eqn:dchi2amp}) with 1.135 
replaced by 1.157.  Figure~\ref{omega} shows contours of $\Delta\chi^2$
in the $\Gamma-\Omega_m$ plane for flat models (left) and open models (right).
All models assume an inflationary index $n=1$.
Points show the best-fit value of $\Omega_m$ at each $\Gamma$.
Solid contours show $\Delta\chi^2=1,$ 4, and 9, so they give the
$1$, 2, and $3\sigma$ confidence limits on $\Omega_m$ for a specified
value of $\Gamma$.  Dotted contours show $\Delta\chi^2=2.3$, 6.7, and 11.8,
corresponding to 68\%, 95\%, and 99.7\% confidence intervals on the
joint values of the two parameters.  

For $\Gamma \sim 0.1-0.25$, the best-fit $\Omega_m$ values are well
described by the dashed lines in Figure~\ref{omega},
\begin{eqnarray}
\widehat \Omega_m &=& 0.38 + 2.2(\Gamma-0.15) \qquad \Omega_\Lambda=1-\Omega_m,
  \label{eqn:omegafit1}\\
\widehat \Omega_m &=& 0.49 + 1.9(\Gamma-0.15) \qquad \Omega_\Lambda=0.
  \label{eqn:omegafit2}
\end{eqnarray}
For $\Gamma=0.15$, the value of $\Omega_m$ with $1\sigma$ and $2\sigma$
error bars is $\Omega_m = 0.38^{+0.10+0.23}_{-0.08-0.14}$ for
$\Omega_\Lambda=1-\Omega_m$ and
$\Omega_m = 0.49^{+0.07+0.20}_{-0.07-0.15}$ for $\Omega_\Lambda=0$.
The dependence of $\widehat{\Omega}_m$ on $\Gamma$ is stronger than that in 
Weinberg et al.\ (1999a) because our higher resolution data give their
best $P(k)$ constraint at a scale smaller than that of the CWPHK data.
The best-fit $\Omega_m$ values are also higher than those in
Weinberg et al.\ (1999a), by about $1\sigma$, because of the 
lower power spectrum amplitude implied by our data relative to CWPHK.
If we used the recent Pierpaoli et al.\ (2001) formulation of the
cluster constraint in place of Eke et al.'s, the zero points in
equation~(\ref{eqn:omegafit1}) and~(\ref{eqn:omegafit2}) would be
nearly identical while the $\Gamma$ dependence would be slightly stronger.
However, some authors have recently argued for a substantially
lower normalization of the cluster $(\Omega_m,\sigma_8)$ constraint
(e.g., Bahcall et al.\ 2002; Reiprich \& Bohringer\ 2002;
Seljak 2002; Viana, Nichol, \& Liddle 2002), which would imply
lower values for $\Omega_m$ when combined with the \lya\ $P(k)$.
For example, Seljak's (2002) normalization would imply a best-fit
value $\widehat{\Omega}_m=0.26$ for flat cosmology with $\Gamma=0.15$.

Cosmic shear measurements offer another route to normalizing the
present day matter power spectrum, with a dependence on $\Omega_m$
that is similar to that from cluster normalization
(Bacon et al.\ 2002; Hoekstra, Yee, \& Gladders 2002; Refregier et al.\ 2002;
van Waerbeke et al.\ 2002).  These results agree fairly well with
those of Eke et al.\ (1996), so they lead to similar estimates of 
$\Omega_m$.  For example, if we take the constraint 
$\sigma_8=0.45\Omega_m^{-0.55}$ with a 12\% fractional uncertainty
(Hoekstra et al.\ 2002; we have symmetrized and halved their quoted
95\% error bar to obtain a $1\sigma$ error bar), we obtain 
$\Omega_m = 0.34^{+0.10+0.22}_{-0.07-0.13}$ (flat models)
and
$\Omega_m = 0.46^{+0.07+0.20}_{-0.06-0.14}$ (open models)
for $\Gamma=0.15$. The $\Gamma$ dependences are slightly shallower
than those of equations~(\ref{eqn:omegafit1}) and~(\ref{eqn:omegafit2}), 
with slopes of 1.8 and 1.6, respectively.
The best-fit $\sigma_8$ values in the other papers listed above are
$\sim 10\%$ higher, so they would require $\Omega_m$ values $\sim 20\%$ higher.

A different combination of cosmological parameters is constrained by
combining the \lya\ $P(k)$ measurement with COBE-DMR measurements of
large-angle CMB anisotropies, as done for the CWPHK data by
Phillips et al.\ (2001).  While the cluster normalization comparison
primarily constrains the value of $\Omega_m$, the CMB comparison
depends on all of the parameters that control the COBE $P(k)$ normalization
and the shape of the matter transfer function, since the COBE and
\lya\ forest measurements tie down $P(k)$ at very different scales.
Phillips et al.\ (2001) show that the main parameters of inflationary
CDM models have nearly degenerate effects on the amplitude and slope
of $P(k)$ at the scales probed by the \lya\ forest, and that matching
COBE and the \lya\ forest $P(k)$ leads to constraints of the form
\begin{equation}
\Omega_m h^\alpha n^\beta \Omega_b^{\gamma} = k \pm \epsilon .
\label{eqn:phillips}
\end{equation}
The tabulations of $\alpha$, $\beta$, $\gamma$, $k$, $\epsilon$
in Phillips et al.\ (2001) cannot be immediately adapted to our
new measurement because we have a different mean redshift and,
more importantly, a different $k_p$.
While we have not repeated the full analysis of Phillips et al.\ (2001),
we have calculated the constraints imposed by matching our new
measurement of $\Delta^2(k_p)$ for COBE normalized, flat models
with $\Omega_\Lambda=1-\Omega_m$.
We find
\begin{equation}
\left({\Omega_m \over 0.4}\right)
\left({h \over 0.65}\right)^{1.90}
\left({n \over 0.895}\right)^{2.89}
\left({\Omega_b h^2 \over 0.02}\right)^{-0.25} = 1.0 \pm 0.14 \;(1\sigma)
\label{eqn:constraint1}
\end{equation}
for models with no tensor fluctuations and
\begin{equation}
\left({\Omega_m \over 0.4}\right)
\left({h \over 0.65}\right)^{1.83}
\left({n \over 0.936}\right)^{4.49}
\left({\Omega_b h^2 \over 0.02}\right)^{-0.24} = 1.0 \pm 0.14 \;(1\sigma)
\label{eqn:constraint2}
\end{equation}
for models with the tensor fluctuation amplitude implied
by power-law inflation models, $T/S=7(1-n)$.
We computed these constraints using CMBFAST 
(Zaldarriaga et al.\ 1998) to obtain COBE normalized
CDM transfer functions.  The $1\sigma$ error bar reflects a 10\%
uncertainty in the COBE normalization added in quadrature to our
13.5\% normalization uncertainty in $\left[\Delta^2(k_p)\right]^{1/2}$.

The no-tensor model with the fiducial parameter choices 
of equation~(\ref{eqn:constraint1}) has $\sigma_8=0.79$ at $z=0$. 
The logarithmic slope at $k_p$ is
$\nu = -2.55$, about $1.8\sigma$ below our best-fit slope of $-2.43$.
For the tensor model with the fiducial parameters of 
equation~(\ref{eqn:constraint2}), $\sigma_8=0.74$,
and the slope at $k_p$ is $\nu=-2.51$, $\sim 1.3\sigma$ below our best fit.
The predicted values of $\sigma_8$ and $\nu$ are not sensitive to moderate
changes in the parameter values provided the 
constraints~(\ref{eqn:constraint1}) or~(\ref{eqn:constraint2}) are
satisfied.  

Confirming and strengthening the results of CWPHK and M00,
our estimate of the \lya\ $P(k)$ supports one of the key predictions
of the inflationary CDM scenario, a matter power spectrum that bends
steadily towards $P(k) \propto k^{n-4}$ on small scales.
Specifically, CDM spectra that are normalized to match COBE and
the amplitude of the \lya\ $P(k)$ match our inferred slope at the $1-2\sigma$
level, after traversing $\sim 3-4$ orders of magnitude in length scale
(see Tegmark \& Zaldarriaga 2002 for an impressive illustration).
The \lya\ forest is the only observable probe of the linear matter
power spectrum on these scales, so this is a fundamental and 
(except for the CWPHK and M00 predecessors) new test of the CDM paradigm.
However, as discussed in \S5.2,
we have not investigated models in which the initial, linear power spectrum
is truncated on scales that are small enough to be significantly non-linear
at the observed epoch $z\sim 2.7$, such as models with warm dark matter
or broken scale-invariance of the type discussed by Kamionkowski \& Liddle 
(2000).
Our inference of the linear $P(k)$ will not be correct in such models on
scales $k \ga 0.02$, so they should be tested directly by simulating
the flux power spectrum itself (Narayanan \etal 2000; White \& Croft 2000).

While the slope of the \lya\ $P(k)$ provides a basic test of the
inflationary CDM scenario, the amplitude is more sensitive to the
values of cosmological parameters.  From the results discussed above,
we can see that the parameters required to fit our measured amplitude
are in generally good agreement with those implied by other cosmological
observations, such as the cluster mass function, cosmic shear, 
CMB anisotropies, large scale galaxy clustering, and the Type Ia supernova 
Hubble diagram.  However, the amplitude is on the low side of expectations,
so in combination with the first two observations it tends to favor relatively
high values of $\Omega_m$ ($\sim 0.3-0.4$ instead of $\sim 0.2-0.3$),
which in combination with COBE normalization favors
some degree of tilt ($n<1$) for the inflationary fluctuation spectrum.
The error bars are still too large to draw strong conclusions on this point,
however, since a model with $\Omega_m=0.3$, $\Omega_\Lambda=0.7$, 
$n=1$, and $h=0.65$ agrees perfectly with the COBE constraint
(eq.~\ref{eqn:constraint1}) and is within $1\sigma$ of the 
cluster/cosmic shear constraint.  
For reference, COBE-normalized LCDM models with $n=1$ inflationary spectra and 
$\Omega_b h^2=0.02$ predict $\Delta^2(k_p)=0.81$, 1.14, 1.63 for
$(\Omega_m,h)=(0.3,0.65)$, $(0.3,0.7)$, $(0.4,0.65)$, respectively,
which can be compared to 
our measured value of $\Delta^2(k_p)=0.74^{+0.20+0.49}_{-0.16-0.28}$.

The analyses discussed here ---
model comparisons and combinations with cluster or COBE constraints ---
give only a few illustrations of the cosmological
applications of our $P(k)$ measurement.
For example, the \lya\ $P(k)$ can also constrain the mass
of light neutrinos through their influence on the power spectrum shape
(Croft et al.\ 1999a).
A more general approach is to incorporate the constraints 
on $\Delta^2(k_p)$ and $\nu$ or $\gprime$ into global 
analyses that consider many observational constraints simultaneously,
as done with the CWPHK data by, e.g., Novosyadlyj et al.\ (2000)
and Wang et al.\ (2000).  
Wang, Tegmark, \& Zaldarriaga (2001) and Tegmark \& Zaldarriaga (2002)
have incorporated the results reported here into such global analyses.
The \lya\ forest complements other cosmological constraints
because it probes fluctuations in a regime of lengthscale and
redshift that is difficult to approach in any other way.

\section{Summary}
\label{sec:summary}

We have analyzed a sample of 53 quasar spectra covering the \lya\
forest in the redshift range $z=2-4$, focusing on a fiducial sample
with range $z=2.3-2.9$ and mean absorption redshift 
$\langle z \rangle = 2.72$.  The HIRES and LRIS spectra contribute
about equally to the total path length $\Delta z \approx 25$ of the
fiducial sample, but the HIRES spectra allow us to measure structure
down to smaller scales, where there are more independent modes.
We have measured the flux filling factor, the flux correlation function,
and the flux power spectrum, for the fiducial sample and for five
subsamples of the data set that cover separate ranges in redshift.
The results, illustrated in Figures~\ref{fillfac}, \ref{fluxxi},
and \ref{fluxpk} and tabulated in Tables~\ref{xitabfid}, \ref{pktabfid},
\ref{xitabz}, and \ref{pktabz}, can be compared directly to the
predictions of numerical simulations or analytic models of the \lya\ 
forest for different cosmologies.  The analysis in \S\ref{sec:statistics}
implies that the uncertainties in these measurements are dominated 
by the finite size of the data set rather than systematic uncertainties.
Our measurement of the 1-dimensional flux power spectrum agrees
well with that of M00 (see Figure~\ref{1dpk}), but our error bars
are smaller on scales $k \la 0.1 \invkms$ because of the larger
number of spectra used.

We have recovered the matter power spectrum $P(k)$ from the 3-d flux
power spectrum $P_F(k)$ using a modified version of the method
introduced by CWKH, as described in \S\ref{sec:method}.
The central assumptions of this method are that structure in the
universe formed by gravitational instability from Gaussian initial
conditions and that the diffuse intergalactic gas responsible for
the \lya\ forest traces the underlying mass distribution in the
relatively simple way found in hydrodynamic cosmological simulations.
Specifically, we use numerical simulations to calibrate the function
$b(k) = \sqrt{P_F(k)/P(k)}$, requiring that these simulations
reproduce the observed $P_F(k)$ and the observed value of
the mean opacity $\taueff$ (which we take from PRS).
Relative to CWKH and CWPHK, the key improvement in the method is
allowing for scale dependence of $b(k)$ (instead of $b={\rm constant}$),
thereby accounting more accurately for the effects of
redshift-space distortions, non-linearity, and thermal broadening.
This improvement in method is warranted by the greater statistical
precision of this data set.

We have restricted our measurement of $P(k)$ to a range of scales
where we expect the statistical uncertainty in individual $P(k)$ data
points to dominate over uncertainty in the {\it shape} of $b(k)$.
The statistical uncertainty in $P_F(k)$ leads to a 6\% $1\sigma$
fractional uncertainty in the overall {\it amplitude} of $b(k)$,
and hence to a 6\% normalization uncertainty in $\sigma \propto \sqrt{P(k)}$,
for the fiducial sample.  There are also systematic uncertainties
in the amplitude of $b(k)$ associated with uncertainties in $\taueff$
and the parameters $T_0$ and $\alpha$ of the IGM temperature-density
relation (see \S\ref{sec:taueff} and \S\ref{sec:igm}).
We estimate that the $\taueff$ and $T_0,\alpha$ uncertainties
contribute 9\% and 8.5\% fractional errors, respectively, to the
amplitude of $b(k)$.  Adding the statistical and systematic error bars in
quadrature, we obtain a $1\sigma$ error bar of 13.5\% on the
rms fluctuation amplitude of the fiducial sample, or $+29\%,-25\%$
on the normalization of $P(k)$.  We also determine $P(k)$ for the
five redshift subsamples, with error bars for each subsample comparable
to the error bar for the totality of the data in CWPHK.

The matter power spectrum deduced from the fiducial sample
(\S\ref{sec:masspkfid}) can be adequately described by a
two-parameter fit (\S\ref{sec:fits}) that gives the logarithmic
slope $\nu$ or CDM shape parameter $\gprime$ and the dimensionless
amplitude $\Delta^2(k_p) \equiv k_p^3 P(k_p)/2\pi^2$ at a wavenumber
$k_p = 0.03 \invkms$ ($2\pi/k \sim 1-2\hmpc$ comoving, 
see eq.~\ref{eqn:k0}).  Best-fit values and $1\sigma$ uncertainties are 
$\nu = -2.43 \pm 0.06$, 
$\gprime = 1.3^{+0.7}_{-0.5} \times 10^{-3} \invkms$,
$\Delta^2(k_p) = 0.74^{+0.20}_{-0.16}$,
and the errors in $\Delta^2(k_p)$ are uncorrelated with those
in $\nu$ or $\gprime$.
These parameter constraints (and the likelihood functions 
in Figure~\ref{dchi21d}) can be incorporated into global likelihood
analyses of cosmological parameter values
(e.g., Novosyadlyj et al.\ 2000; Wang et al.\ 2000; Wang et al.\ 2001).
Extrapolating to the redshift and wavenumbers of the CWPHK and M00
measurements, we find good agreement in the logarithmic slope and
an rms fluctuation amplitude that is $\sim 15\%$ lower than that
of CWPHK (a difference of $<1 \sigma$), and the same as that of M00
 (\S\ref{sec:previous}).
The slope measurement confirms the generic prediction of inflationary
CDM models of a $P(k)$ slope bending asymptotically towards $k^{n-4}$
on small scales.  The parameters $\gprime$ and $\Delta^2(k_p)$ can
be related to the quantities $\Gamma$ and $\sigma_8$ at $z=0$, in a
cosmology dependent way (see eqs.~\ref{eqn:gamma} and~\ref{eqn:sigma8}).
For $\Omega_m=0.4$, $\Omega_\Lambda=0.6$, $\Gamma = 0.16^{+0.09}_{-0.06}$,
in good agreement with measurements of the galaxy power spectrum
shape from the 2dF Galaxy Redshift Survey (Percival et al.\ 2001).
The shape parameter $\Gamma$ and 
the inflationary index $n$ have nearly degenerate effects on the shape of 
$P(k)$ over the range of our measurement, and the values of $\gprime$ and
$\Gamma$ quoted above assume $n=1$.  The effective combination of
parameters constrained by our measurement of the $P(k)$ shape is,
to a fair approximation, $\Gamma+1.4(1-1/n)$.

When we measure $\Delta^2(k_p)$ and the logarithmic slope $\nu$ 
for the five redshift subsamples,
we find consistent values of $\nu$ at each redshift and growth of
$\Delta^2(k_p)$ that is compatible at the $1-2\sigma$ level with
the predictions of gravitational instability (\S\ref{sec:evolution}).
These results do not constitute strong confirmation of the predicted 
gravitational growth of $P(k)$, but they do imply that fluctuations in
the \lya\ forest do not have a major contribution from sources other
than density fluctuations (e.g., large-scale variations in the ionizing
background or IGM temperature).  Other mechanisms would be unlikely
to mimic the shape of the matter power spectrum, and their contribution
is likely to decrease over the redshift interval $z=4$ to $z=2$ rather
than increase.

The power spectrum of the fiducial sample is well matched by an inflationary
CDM model with $\Omega_m=0.4$, $\Omega_\Lambda=0.6$, $h=0.65$, $n=0.93$,
and $\sigma_8=0.7$ (see Figure~\ref{models1}).  
The measured amplitude is in good agreement with the predictions of 
CDM models with parameter values favored by independent observations,
though it lies on the low side of the expected range of values.
To obtain consistency between our measurement and Eke et al.'s (1996)
cluster mass function constraint on the present day power spectrum
normalization, we require
$\Omega_m = 0.38^{+0.10+0.23}_{-0.08-0.14}$ 
($1\sigma,2\sigma$) for models with $\Omega_\Lambda=1-\Omega_m$ and
$\Gamma=0.15$, and higher values of $\Omega_m$ for open models
or higher $\Gamma$ (see eqs.~\ref{eqn:omegafit1} and~\ref{eqn:omegafit2},
Figure~\ref{omega}).  Combination with Hoekstra et al.'s (2002)
cosmic shear constraint yields a similar result,
$\Omega_m = 0.34^{+0.10+0.22}_{-0.07-0.13}$.
COBE-normalized, flat CDM models match our measured $P(k)$ amplitude
for $\Omega_m\approx 0.3$, $h=0.65$, and $n=1$, or for higher (lower) 
$\Omega_m$ if $h$ or $n$ are slightly lower (higher) 
(see eqs.~\ref{eqn:constraint1} and~\ref{eqn:constraint2}).
Models with $\Omega_m$ significantly below 0.3 are difficult
to reconcile simultaneously with the \lya\ $P(k)$, COBE, and 
the Eke et al.\ (1996) estimate of the cluster normalization constraint.

There are several ways to reduce the systematic uncertainty in 
inferring $P(k)$ from our flux power spectrum measurement.
The most straightforward is to improve the measurement of $\taueff$,
in order to check the PRS value adopted here
(which may be systematically in error) and reduce the $\taueff$
error bar.  The results in \S\ref{sec:taueff} can be used to 
scale our estimated $P(k)$ and its error bar for new determinations
of $\taueff$.  Tighter constraints on $T_0$ and $\alpha$ would also reduce
the $P(k)$ error bar, and the impact of new determinations can be
judged approximately using the results in \S\ref{sec:igm}.
Finally, replacing our normalizing simulations with higher resolution,
full hydrodynamic simulations would remove one source of systematic
uncertainty in our results and might allow extension of the $P(k)$
recovery to smaller scales.  Such an approach is computationally demanding,
but it might become feasible in the near future, especially if the
hydrodynamic simulations are designed specifically to model the \lya\
forest rather than galaxy formation.

Larger data samples can improve the precision of the $P_F(k)$ measurement
and extend its dynamic range in wavenumber and in redshift.  At high $k$,
moderate numbers of high-resolution spectra are enough to yield high
statistical precision.  On large scales, one needs many spectra to reduce
cosmic variance, but these do not need to be high resolution.
The LRIS sample used here illustrates the power of 10-m class telescopes
for probing large-scale structure in the \lya\ forest; quadrupling
this sample would require only a few nights of observing.
The Sloan Digital Sky Survey (York et al.\ 2000) will obtain thousands
of high-redshift quasar spectra with $\sim 2.5$\AA\ resolution, which
should be an ideal sample for measuring $P_F(k)$ on large scales.
With a high enough surface density of quasars, one can measure the
3-dimensional power spectrum by correlating flux on different lines
of sight, obtaining a new diagnostic for redshift-space distortions
and cosmological geometry (see McDonald 2002).
Such 3-dimensional analyses may be the
best way to extend $P_F(k)$ measurements to very large scales, since
they provide more baselines on these scales and should be less
sensitive to continuum fitting errors.

The shape of $P_F(k)$ on small scales depends on the thermal state
of the IGM as well as the underlying density fluctuations.
In this regime, measurements of $P_F(k)$ and its evolution may
provide useful tests of inhomogeneous reionization models or simulations
that incorporate metal enrichment and IGM feedback from galactic winds.
The most valuable {\it cosmological} tests will
come from improving the precision of $P(k)$ on the scales measured
here and extending the analyses to larger scales so as to constrain the shape
of $P(k)$ over a wider dynamic range.  Combination of the \lya\
forest power spectrum with CMB anisotropies and cluster masses already
yields interesting new constraints on cosmological parameter values.
Over the next year or two, anticipated improvements in these and other
measures of cosmic structure should zero in on a small allowed
region in the parameter space of simple cosmological models
that incorporate a power-law inflationary spectrum, cold dark matter
with a standard baryon component, and a cosmological constant.
Alternatively, the tightening of complementary constraints could
show that these models are not yet complete, and that the real 
universe incorporates more complicated physics of inflation, dark matter,
or vacuum energy.

\bigskip
\acknowledgments
We thank David Sprayberry for assistance with the LRIS observations
and Fred Chaffee for allotting some discretionary time to complete
the LRIS sample.
We thank Patrick McDonald,
Jasjeet Bagla, John Peacock, and John Phillips for useful discussions
and comments.
We thank Romeel Dav\'{e} and Jeffrey Gardner for permission to show results
from the TreeSPH calculations in Figure~\ref{sphtest}, and George Efstathiou
for permission to use the P$^{3}$M N-body code.
We thank Nickolay Gnedin for helpful exchanges on the subject
of normalizing simulations, and for discussions which led to
the discovery of an error in an earlier
version of this work.
This work was supported by NASA Astrophysical Theory Grants NAG5-3111,
NAG5-3922, and NAG5-3820,
by NASA Long-Term Space Astrophysics Grant NAG5-3525,
and by NSF grants AST-9802568, ASC 93-18185, and AST-9803137.

\appendix

\section{Tabulated flux statistics at 5 redshifts}

\begin{table*}[ht]
\centering
\caption[xitabz]{\label{xitabz} 
The flux correlation function, $\xi_{F}(r)$,
for the different redshift subsamples (see Table ~\ref{subtab})}
\begin{tabular}{cccccc}
\hline &&\\
r  &   Subsample  A  & Subsample  B  &Subsample C &Subsample D &Subsample E \\
    ($\kms$) & $\xi_{F}(r)$& $\xi_{F}(r)$&$\xi_{F}(r)$&$\xi_{F}(r)$&
$\xi_{F}(r)$ \\
\hline &&\\
 11.4 &  $(9.8 \pm 1.1) \times 10^{-2}$  & $ 0.162 \pm 0.010$  & 
$0.181 \pm 0.013$ & $0.220 \pm 0.015$ &
$0.355 \pm 0.022$ \\
 14.9 &  $(9.2 \pm 1.1) \times 10^{-2} $ & $ 0.152 \pm 0.009$  &
$0.172 \pm 0.013$ & $0.209 \pm 0.014 $&
$0.334 \pm 0.021$  \\
 19.4 &  $(9.1 \pm 1.0) \times 10^{-2}$  & $ 0.150 \pm 0.011$  &
$0.165 \pm 0.013$ & $0.200 \pm 0.014$ &
$0.318 \pm 0.020$ \\
 25.3 &  $(8.3 \pm 1.0) \times 10^{-2}$  & $ 0.137 \pm 0.009 $ &
$0.154 \pm 0.013$ & $0.186 \pm 0.013$ &
$0.295 \pm 0.018$ \\
 32.9 &  $(7.2 \pm 0.9) \times 10^{-2}$  & $ 0.129 \pm 0.010 $ &
$0.139 \pm 0.012$ & $0.166 \pm 0.012$ &
$0.262 \pm 0.017$ \\
 42.9 &  $(6.6 \pm 0.8) \times 10^{-2}$  & $ 0.112 \pm 0.009 $ &
$0.122 \pm 0.013$ & $0.141 \pm 0.011$ &
$0.225 \pm 0.015$ \\
 56.0 &  $(5.3 \pm 0.8) \times 10^{-2}$  &  $(9.6 \pm 0.9) \times 10^{-2}$ &
$0.101 \pm 0.011$ & $0.120 \pm 0.011$ &
$0.187 \pm 0.014$ \\
 72.9 &  $(4.2 \pm 0.7) \times 10^{-2}$  & $(7.8 \pm 0.8) \times 10^{-2}$  &
$(8.3 \pm 1.1) \times 10^{-2}$ & $(9.4 \pm 1.0) \times 10^{-2}$ &
$0.151 \pm 0.012$ \\
 95.0 &  $(3.1 \pm 0.6) \times 10^{-2}$  & $(6.0 \pm 0.7) \times 10^{-2}$  &
$(6.5 \pm 1.1) \times 10^{-2}$ &$(7.1 \pm 0.9) \times 10^{-2}$ &
$0.120 \pm 0.011$ \\
 124  &  $(2.1 \pm 0.6) \times 10^{-2}$  & $(4.5 \pm 0.7) \times 10^{-2}$  &
$(5.0 \pm 1.1) \times 10^{-2}$ &$(5.3 \pm 0.8) \times 10^{-2}$ &
$(9.5 \pm 1.0) \times 10^{-2}$ \\
 161  &  $(1.4 \pm 0.5) \times 10^{-2}$  & $(3.4 \pm 0.6) \times 10^{-2}$  &
$(3.8 \pm 1.1) \times 10^{-2}$ &$(4.2 \pm 0.8) \times 10^{-2}$ &
$(7.5 \pm 0.8) \times 10^{-2}$ \\
 210  &  $(9.9 \pm 3.5) \times 10^{-3}$  & $(2.2 \pm 0.6) \times 10^{-2}$  &
$(2.8 \pm 1.1) \times 10^{-2}$ &$(3.6 \pm 0.7) \times 10^{-2}$ &
$(5.6 \pm 0.8) \times 10^{-2}$ \\
 274  &  $(7.2 \pm 3.0) \times 10^{-3}$  & $(1.5 \pm 0.3) \times 10^{-2}$  &
$(2.4 \pm 0.6 ) \times 10^{-2}$&$(2.0 \pm 0.5) \times 10^{-2}$ &
$(4.1 \pm 0.9) \times 10^{-2}$ \\
 357  &  $(1.4 \pm 3.0) \times 10^{-3}$  & $(1.1 \pm 0.3) \times 10^{-2}$  &
$(1.9 \pm 0.6) \times 10^{-2}$&$(9.9 \pm 4.1) \times 10^{-3}$ &
$(3.1 \pm 1.0) \times 10^{-2}$ \\
 466  &  $(-6.0 \pm 28.5) \times 10^{-4}$ & $(5.9 \pm 2.8) \times 10^{-3}$ &
$(1.3 \pm 0.6) \times 10^{-2}$&$(6.6 \pm 4.4) \times 10^{-3}$ &
$(2.0 \pm  0.8) \times 10^{-2}$ \\
 607  &  $(3.2 \pm 2.7) \times 10^{-3}$  & $(1.9 \pm 2.7) \times 10^{-3}$  &
$(9.0 \pm 5.4) \times 10^{-3}$&$(4.9 \pm 3.9) \times 10^{-3}$ &
$(1.1 \pm 0.8) \times 10^{-2}$ \\
 791  &  $(2.9 \pm 2.5) \times 10^{-3}$  & $(1.4 \pm 2.5) \times 10^{-3}$  &
$(7.2 \pm 5.0) \times 10^{-3}$ &$(3.0 \pm 3.0) \times 10^{-3}$ &
$(1.7 \pm 0.8) \times 10^{-2}$ \\
 1030 &  $(2.2 \pm 2.2) \times 10^{-3}$  & $(6.4 \pm 21.7) \times 10^{-4}$ &
$(2.5 \pm 4.9) \times 10^{-3}$ &$(3.3 \pm 3.5) \times 10^{-3}$ &
$(1.7 \pm 0.8) \times 10^{-2}$ \\
 1340 &  $(-2.7 \pm 1.7) \times 10^{-3}$ & $(-1.1 \pm 2.2) \times 10^{-3}$ &
$(-5.6 \pm 43.1) \times 10^{-4}$ &$(9.9 \pm 39.7) \times 10^{-4}$ &
$(1.5 \pm 0.8) \times 10^{-2}$ \\
 1750 &  $(-1.3 \pm 22.4) \times 10^{-4}$ & $(-2.8 \pm 1.9) \times 10^{-3}$ &
$(1.9 \pm 3.2) \times 10^{-3}$ & $(-6.3 \pm 3.3) \times 10^{-3}$ &
$(8.1 \pm 9.2) \times 10^{-3}$ \\
\hline &&\\
\end{tabular}
\end{table*}

\begin{table*}[ht]
\centering
\caption[pktabz]{\label{pktabz} 
The one-dimensional flux power spectrum, 
for the different redshift subsamples (see Table ~\ref{subtab})}
\begin{tabular}{cccccc}
\hline &\\
   &   Subsample  A  & Subsample  B  &Subsample C &Subsample D &Subsample E \\
k  &  $P_{F,1D}(k)$&$P_{F,1D}(k)$&$P_{F,1D}(k)$&$P_{F,1D}(k)$&$P_{F,1D}(k)$ \\
 $(\kms)^{-1}$  &$(\kms)^{-1}$ &$(\kms)^{-1}$ &$(\kms)^{-1}$&$(\kms)^{-1}$
&$(\kms)^{-1}$ \\
\hline &\\
 0.00199 & $15.9\pm7.5$& $42.1\pm7.3$& $42.3\pm6.8$&
$46.6\pm15.0$&$50.1\pm10.3$ \\
 0.00259 & $18.6\pm5.3$& $31.9\pm4.0$& $38.8\pm4.8$&
$37.5\pm9.1$&$70.8\pm13.9$ \\
 0.00337 & $17.5\pm6.0$& $27.5\pm2.9$& $33.9\pm4.3$&
$26.7\pm5.1$&$43.4\pm9.8$ \\
 0.00437 & $9.19\pm4.30$&$25.3\pm2.9$& $22.1\pm2.7$&
$28.2\pm4.8$&$57.2\pm13.8$ \\
 0.00568& $14.9\pm4.0$&  $24.5\pm2.8$& $24.4\pm2.2$&
$30.7\pm4.5$&$52.4\pm10.2$ \\
 0.00738& $12.9\pm3.2$&  $20.4\pm2.6$& $20.8\pm3.1$&
$19.8\pm3.7$&$40.2\pm4.0$ \\
 0.00958& $11.0\pm1.5$& $17.0\pm2.5$& $17.6\pm1.7$&
$26.0\pm5.1$&$31.1\pm5.5$ \\
 0.0124& $8.15\pm1.32$& $13.4\pm1.0$& $14.0\pm1.0$&
$16.3\pm2.2$&$25.6\pm3.3$ \\
 0.0162& $6.37\pm1.42$& $10.4\pm0.8$& $11.7\pm1.3$&
$13.6\pm1.1$&$23.9\pm4.0$ \\
 0.0210& $5.71\pm0.63$& $6.65\pm0.52$&$8.51\pm0.62$&
$10.8\pm1.1$&$15.7\pm1.5$ \\
 0.0272& $3.40\pm0.52$& $5.09\pm0.35$&$5.96\pm0.43$&
$9.02\pm0.71$&$12.3\pm1.0$ \\
 0.0355& $2.18\pm0.22$& $3.34\pm0.21$&$3.78\pm0.27$&
$5.44\pm0.54$&$8.37\pm0.63$ \\
 0.0461& $1.25\pm0.25$& $1.69\pm0.12$&$2.38\pm0.18$&
$2.83\pm0.35$&$5.47\pm0.48$ \\
 0.0598& $0.642\pm0.153$&$0.904\pm0.059$&$1.13\pm0.076$&
$1.75\pm0.14$&$3.16\pm0.29$ \\
 0.0777& $0.370\pm0.110$&$0.446\pm0.040$&$0.541\pm0.0455$&
$0.758\pm0.131$&$1.42\pm0.16$ \\
 0.101&  $0.173\pm0.066$&$0.212\pm0.025$&$0.215\pm0.0211$&
$0.338\pm0.097$&$0.637\pm0.135$ \\
 0.131&  $0.107\pm0.043$&$0.105\pm0.013$&$0.0938\pm0.0143$&
$0.166\pm0.081$&$0.328\pm0.157$ \\
 0.170& $0.0705\pm0.0376$&$0.0642\pm0.0124$&$0.0574\pm0.0119$&
$0.114\pm0.077$&$0.201\pm0.141$ \\
 0.221& $0.0593\pm0.0340$&$0.0473\pm0.0111$&$0.0409\pm0.0112$&
$0.0986\pm0.0826$&$0.142\pm0.114$ \\
 0.287& $0.0469\pm0.0230$&$0.0382\pm0.0099$&$0.0290\pm0.0098$&
$0.0628\pm0.0517$&$0.0983\pm0.0844$ \\
\hline &\\
\end{tabular}
\end{table*}

\end{document}